\documentclass[a4paper,10pt]{article}
\usepackage{amssymb}
\usepackage[dvips]{graphicx}
\usepackage[usenames]{color}
\usepackage{amsmath}
\usepackage{fullpage}
\usepackage{boxedminipage}
\usepackage{listings}
\usepackage{minitoc}
\usepackage{wrapfig}
\usepackage{latexsym}

\makeatletter \@addtoreset{equation}{section} \makeatother

\input epsf

\begin{document}

\renewcommand{\[}{\begin{equation}} \renewcommand{\]}{\end{equation}} %
\renewcommand{\>}{\rangle}

\begin{titlepage}

    \thispagestyle{empty}
    \begin{flushright}
        \hfill{CERN-PH-TH/132}
    \end{flushright}

    \vspace{55pt}
    \begin{center}
        { \huge{\textbf{$d=4$ Attractors, Effective Horizon Radius\\\vspace{10pt}and Fake Supergravity}}}\vspace{25pt}

        \vspace{5pt}

         {\Large{\bf Sergio Ferrara$^{\diamondsuit\clubsuit}$, Alessandra Gnecchi$^{\spadesuit\diamondsuit}$ and\ Alessio Marrani$^{\heartsuit\clubsuit}$}}

        \vspace{20pt}

        {$\diamondsuit$ \it Physics Department,Theory Unit, CERN, \\
        CH 1211, Geneva 23, Switzerland\\
        \texttt{sergio.ferrara@cern.ch}}

        \vspace{10pt}

        {$\clubsuit$ \it INFN - Laboratori Nazionali di Frascati, \\
        Via Enrico Fermi 40,I-00044 Frascati, Italy\\
        \texttt{marrani@lnf.infn.it}}

         \vspace{10pt}

        {$\spadesuit$ \it Dipartimento di Fisica, Universit\`{a} di Pisa\\
        and INFN - Sezione di Pisa,\\
        Largo B. Pontecorvo 3, I-56127 Pisa, Italy\\
        \texttt{gnecchi@df.unipi.it}}

        \vspace{10pt}

        {$\heartsuit$ \it Museo Storico della Fisica e\\
        Centro Studi e Ricerche ``Enrico Fermi"\\
        Via Panisperna 89A, I-00184 Roma, Italy}

\end{center}

\vspace{55pt}

\begin{abstract}
We consider extremal black hole attractors (both BPS and non-BPS) for $%
\mathcal{N}=3$ and $\mathcal{N}=5$ supergravity in $d=4$ space-time
dimensions.

Attractors for \textit{matter-coupled} $\mathcal{N}=3$ theory are
similar to attractors in $\mathcal{N}=2$ supergravity
\textit{minimally coupled} to Abelian vector multiplets.

On the other hand, $\mathcal{N}=5$ attractors are similar to
attractors in $\mathcal{N}=4$ \textit{pure} supergravity, and in
such theories only $\frac{1}{\mathcal{N}}$-BPS non-degenerate
solutions exist.

All the above mentioned theories have a simple interpretation in the \textit{%
first order (fake supergravity)} formalism. Furthermore, such
theories do not have a $d=5$ uplift.

Finally we comment on the \textit{``duality''} relations among the
attractor solutions of $\mathcal{N}\geq2$ supergravities sharing the
same full bosonic sector.

\end{abstract}

\end{titlepage}
\newpage\tableofcontents\newpage

\section{\protect\bigskip \label{Introduction}Introduction}

The \textit{Attractor Mechanism} \cite{ferrara1}--\nocite
{ferrara2,strominger2}\cite{FGK} is an important dynamical phenomenon in the
theory of gravitational objects, which naturally appears in modern theories
of gravity, such as supergravity, superstrings \cite{maldacena}--\nocite
{schwarz1,schwarz2}\cite{gasperini} or \textit{M-theory } \cite
{witten,schwarz3}.

Even if such a phenomenon was originally shown to occur for $\frac{1}{2}$%
-BPS extremal black holes (BHs) in $\mathcal{N}=2$, $d=4$ ungauged
supergravity coupled to Abelian vector multiplets, it has a more general
validity, since it may take place also for non-BPS extremal BHs,
irrespective of whether the underlying gravitational theory is endowed with
local supersymmetry or not \cite{Sen-old1}--\nocite
{GIJT,Sen-old2,K1,TT,G,GJMT,Ebra1,K2,Ira1,Tom,
BFM,FKlast,Ebra2,BFGM1,rotating-attr,K3,Misra1,Lust2,Morales,Astefa,CdWMa,
DFT07-1,BFM-SIGRAV06,Cer-Dal-1,ADFT-2,Saraikin-Vafa-1,Ferrara-Marrani-1,
TT2,
ADOT-1,ferrara4,CCDOP,Misra2,Astefanesei,Anber,Myung1,Ceresole,BMOS-1,Hotta,
Gao,Sen-review,Belhaj1,Gaiotto1,GLS1,ANYY1,bellucci2,Cai-Pang,Vaula,
Li,Saidi2,Saidi3,Saidi4,FHM,Unattractor,Vaula2}\cite{Trigiante} (for further
developments, see also \textit{e.g.} \cite{OSV}--\nocite{OVV,ANV}\cite{GSV}%
). Moreover, such a phenomenon also exists in \ higher space-time dimensions
for black $p$-branes coupled to scalar fields, provided certain constraints
are met.

In theories with $\mathcal{N}>2$ local supersymmetry, extremal BH attractors
with regular horizon geometry and non-vanishing classical Bekenstein-Hawking
entropy exhibit a new feature. This indicates that the Hessian matrix of a
suitably defined \textit{effective BH potential} $V_{BH}$ may present, in
contrast with the $\mathcal{N}=2$ case, ``flat'' directions even for ($\frac{%
1}{\mathcal{N}}$-)BPS configurations. This is for instance the case when the
scalar manifold is a locally symmetric space, as it holds for all $\mathcal{N%
}>2$, $d=4$ supergravities. A general analysis of extremal BH attractors in $%
\mathcal{N}=2$ symmetric special K\"{a}hler geometry was performed in \cite
{BFGM1}, and the related moduli spaces were discovered and classified in
\cite{ferrara4}. For $\mathcal{N}>2$ supergravities similar results were
obtained in \cite{ADFT} and \cite{bellucci2}. In \cite{Ferrara-Marrani-1} it
was further observed that ``flat'' directions for $\mathcal{N}>2$ (both $%
\frac{1}{\mathcal{N}}$-BPS and non-BPS) attractors, as well as for $\mathcal{%
N}=2$ non-BPS attractors, are closely related to the fact that in $\mathcal{N%
}=2$ ungauged supergravity the hypermultiplets' scalars do not participate
in the Attractor Mechanism. As a consequence, the moduli space of $\frac{1}{%
\mathcal{N}}$-BPS attractors in $\mathcal{N}>2$ supergravities is a
quaternionic manifold, spanned by the left-over would-be hypermultiplets'
scalar degrees of freedom in the supersymmetry reduction of the original
theory down to $\mathcal{N}=2.$

The corresponding orbits of electric and magnetic BH charges, supporting the
critical points of $V_{BH}$ which determine the attractor scalar
configurations on the BH event horizon, have also been classified in \cite
{BFGM1} and \cite{bellucci2}. The non-compactness of the stabilizer of such
orbits (with the only exception of $\mathcal{N}=2$ BPS orbits) is
responsible for the existence of ``flat'' directions of the BH potential at
its corresponding critical points.

Most of the supergravities based on symmetric scalar manifolds have the
property that the classical BH entropy, as given by the Bekenstein-Hawking
entropy-area formula \cite{hawking2}, is expressed in terms of the square
root of the absolute value of a(n \textit{unique}) invariant $\mathcal{I}_{4}
$ of the relevant representation of the $U$-duality group. Such an invariant
is \textit{quartic} in electric and magnetic BH charges:
\begin{equation}
S_{BH}=\frac{A}{4}=\pi \left. V_{BH}\right| _{\partial V_{BH}=0}=\pi \sqrt{%
\left| \mathcal{I}_{4}\right| }.  \label{BH-entropy-I4}
\end{equation}
$\mathcal{I}_{4}$, which is \textit{moduli-independent}, can also be written
in terms of the \textit{``dressed''} charges, \textit{i.e.} in terms of the
\textit{moduli-dependent} \textit{central charge matrix} and \textit{matter
charges}, in a(n \textit{unique}) combination, such that the overall
dependence on moduli drops out. However, a peculiar class of $d=4$
supergravities exists, such that the \textit{unique} $U$-duality invariant
(and thus moduli-independent) combination of \textit{``dressed''} charges
turns out to be a \textit{perfect square} of a \textit{quadratic} expression
in the skew-eigenvalues of the relevant \textit{central charge matrix}.
Namely, this holds for \textit{pure }$\mathcal{N}=4$ \cite{CSF}\textbf{\ }%
and $\mathcal{N}=5$ \cite{N=5-Ref}\textbf{\ }supergravity.

Furthermore, another class of $d=4$ theories exists, so that the unique $U$%
-invariant $\mathcal{I}_{2}$ is \textit{quadratic} in BH charges, yielding:
\begin{equation}
S_{BH}=\frac{A}{4}=\pi \left. V_{BH}\right| _{\partial V_{BH}=0}=\pi \left|
\mathcal{I}_{2}\right| .  \label{BH-entropy-I2}
\end{equation}
$\mathcal{I}_{2}$ is also given by a \textit{quadratic} expression in terms
of the \textit{``dressed''} charges. Such a class of theories is given by $%
\mathcal{N}=2$ supergravity \textit{minimally coupled} to Abelian vector
multiplets \cite{Luciani}, and by $\mathcal{N}=3$ supergravity coupled to
matter (Abelian vector) multiplets \cite{N=3-Ref}.

For both the peculiar class of theories admitting $\mathcal{I}_{4}$ as a
\textit{perfect square} of the skew-eigenvalues of the \textit{central
charge matrix} and the supergravities admitting $\mathcal{I}_{2}$, there is
a very simple alternative expression for the classical Bekenstein-Hawking
entropy in terms of a \textit{(square) effective horizon radius} $R_{H}$.
This turns out to be \textit{moduli-independent}, and dependent \textit{only}
on the set of magnetic and electric BH charges, shortly indicated as $\left(
p,q\right) $. The formula for the entropy of the extremal BH in these cases
reads:
\begin{eqnarray}
S_{BH} &=&\frac{A}{4}=\pi \left. V_{BH}\right| _{\partial V_{BH}=0}\equiv
\pi R_{H}^{2}\left( p,q\right) =\left\{
\begin{array}{l}
\pi \sqrt{\left| \mathcal{I}_{4}\right| } \\
or \\
\pi \left| \mathcal{I}_{2}\right|
\end{array}
\right. =  \notag \\
&=&\pi \left[ r_{H}^{2}\left( \varphi _{\infty },p,q\right) -\frac{1}{2}%
G_{ab}\left( \varphi _{\infty }\right) \Sigma ^{a}\left( \varphi _{\infty
},p,q\right) \Sigma ^{b}\left( \varphi _{\infty },p,q\right) \right] .
\label{Ita-Rom-1}
\end{eqnarray}
$r_{H}$ is the radius of the unique \textit{(event) horizon} of the extremal
BH, $\Sigma ^{a}$ denoting the set of \textit{scalar charges} asymptotically
associated to the scalar field $\varphi ^{a}$, and $G_{ab}$ is the covariant
metric tensor of the scalar manifold (in the real parametrization). Notice
that the first line of Eq. (\ref{Ita-Rom-1}) only contains the definition of
$R_{H}^{2}$ itself, whereas the second line of the very same Eq. expresses
it through a \textit{moduli-independent} combination of \textit{%
moduli-dependent} quantities, holding only for the aforementioned $d=4$
supergravity theories.

Actually, $R_{H}^{2}$ can be expressed as a suitable integral in terms of a
\textit{square} \textit{effective radius} $R^{2}$, as follows:
\begin{eqnarray}
R_{H}^{2} &=&\left. R^{2}\left( r,\varphi _{\infty },p,q\right) \right| _{r=%
\sqrt{\frac{1}{2}G_{ab}\left( \varphi _{\infty }\right) \Sigma ^{a}\left(
\varphi _{\infty },p,q\right) \Sigma ^{b}\left( \varphi _{\infty
},p,q\right) }}^{r=r_{H}\left( \varphi _{\infty },p,q\right) };
\label{Tue-3} \\
&&  \notag \\
R^{2}\left( r,\varphi _{\infty },p,q\right)  &\equiv &r^{2}-\frac{1}{2}%
G_{ab}\left( \varphi _{\infty }\right) \Sigma ^{a}\left( \varphi _{\infty
},p,q\right) \Sigma ^{b}\left( \varphi _{\infty },p,q\right) \leqslant r^{2},
\label{Tue-4}
\end{eqnarray}
where $r$ is the usual radial coordinate, and in the last inequality the
positive definiteness of $G_{ab}$ was exploited.

It is worth pointing out that the second line of Eq. (\ref{Ita-Rom-1}) and
Eqs. (\ref{Tue-3}) and (\ref{Tue-4}) are generalizations of the formul\ae\ %
holding in the so-called \textit{Maxwell-Einstein-axion-dilaton system}
(similarly also in the \textit{non-extremal case}, see \textit{e.g.} \cite
{K3} and \cite{Garfinkle}; see also the treatment, and in particular Eqs.
(2.7), (2.8) and (2.15), of \cite{FHM}). Notice that in the first of Refs.
\cite{K3} the variable $R$ is named \textit{physical radial coordinate} (see
\textit{e.g.} Eq. (72) therein). Clearly, as the second line of Eq. (\ref
{Ita-Rom-1}), also the definition (\ref{Tue-4}) of $R^{2}\left( r,\varphi
_{\infty },p,q\right) $ holds only for the aforementioned $d=4$ supergravity
theories.

Within the \textit{first order (fake supergravity) formalism} \cite
{Fake-Refs}, recently used to describe non-BPS attractor flows of $d=4$
extremal BHs \cite{Cer-Dal-1,ADOT-1}, the quantities appearing in the second
line of Eq. (\ref{Ita-Rom-1}) can easily be expressed in terms of a \textit{%
real ``fake superpotential''} $\mathcal{W}\left( \varphi ,p,q\right) $ as
follows (see Eqs. (\ref{CERN!-1}) and (\ref{CERN!-2}) below, respectively):
\begin{eqnarray}
r_{H}\left( \varphi _{\infty },p,q\right)  &=&\mathcal{W}\left( \varphi
_{\infty },p,q\right) ;  \label{Ita-Rom-2} \\
&&  \notag \\
\Sigma ^{a}\left( \varphi _{\infty },p,q\right)  &=&2G^{ab}\left( \varphi
_{\infty }\right) \left( \partial _{b}\mathcal{W}\right) \left( \varphi
_{\infty },p,q\right) .  \label{Ita-Rom-3}
\end{eqnarray}
An explicit expression for $\mathcal{W}$ can be given for the supergravity
theories mentioned above \cite{ADOT-1}. It is worth noticing here that for $%
\frac{1}{\mathcal{N}}$-BPS \textit{non-degenerate} attractor flows, it
simply holds $\mathcal{W}\left( \varphi ,p,q\right) =\left| \mathcal{Z}%
\right| \left( \varphi ,p,q\right) $, where $\left| \mathcal{Z}\right| $ is
the biggest absolute value of the skew-eigenvalues of the \textit{central
charge matrix} $Z_{AB}$, saturating the BPS bound \cite{BPS}.

Eq. (\ref{Ita-Rom-1}) would seem to yield a \textit{moduli-dependent}
expression for $R_{H}^{2}$, but, as we prove explicitly in the present
paper, for the class of $d=4$ ungauged supergravities under consideration it
just transpires that the dependence on moduli drops out in the combination $%
r_{H}^{2}-\frac{1}{2}G_{ab}\Sigma ^{a}\Sigma ^{b}$, when Eqs. (\ref
{Ita-Rom-2}) and (\ref{Ita-Rom-3}) are taken into account. Summarizing, such
a phenomenon happens in the following theories:

\begin{itemize}
\item  $\mathcal{N}=2$ supergravity \textit{minimally coupled} to Abelian
vector multiplets \cite{Luciani}, whose scalar manifold is endowed with a
symmetric special K\"{a}hler geometry, with the completely symmetric rank-3
tensor $C_{ijk}=0$, and with $U$-invariant \textit{quadratic} in BH charges.
For such a theory, in \cite{FHM} Eq. (\ref{Ita-Rom-1}) has been proved to
hold for both $\frac{1}{2}$-BPS and non-BPS ($Z=0$) attractor flows

\item  $\mathcal{N}=3$ supergravity coupled to matter (Abelian vector)
multiplets \cite{N=3-Ref}\textbf{, }with $U$-invariant \textit{quadratic} in
BH charges

\item  $\mathcal{N}=4$ \textit{pure} supergravity \cite{CSF}, with $U$%
-invariant \textit{quartic} in BH charges

\item  $\mathcal{N}=5$ supergravity \cite{N=5-Ref}, with $U$-invariant
\textit{quartic} in BH charges.
\end{itemize}

It is worth pointing out that $\mathcal{N}=2$ supergravity \textit{minimally
coupled} to one Abelian vector multiplet, corresponding to the $\left(
U\left( 1\right) \right) ^{6}\rightarrow \left( U\left( 1\right) \right) ^{2}
$ gauge truncation of $\mathcal{N}=4$ \textit{pure} supergravity, is nothing
but the so-called \textit{Maxwell-Einstein-axion-dilaton system}, studied in
\cite{K3,Garfinkle} and recently discussed in \cite{ADFT} and in \cite{FHM}.
As stated above, the formula (\ref{Ita-Rom-1}) indeed holds true (actually,
with suitable changes, also in the \textit{non-extremal} case).

Furthermore, it is interesting to notice that \textit{all} the above
mentioned theories are \textit{all} the $\mathcal{N}\geqslant 2$, $d=4$
supergravities based on \textit{symmetric} scalar manifolds which do \textit{%
not} admit an uplift\footnote{%
Throughout all the treatment of the present paper, by ``uplift to $d=5$'' we
mean the dimensional uplift to a $d=5$ Poincar\'{e} supergravity theory,
having the same massless degrees of freedom of the original $d=4$
supergravity.} to $d=5$ space-time dimensions \cite{GST}.\bigskip

The present paper is organized as follows.\medskip

In Sect. \ref{First-Order} we briefly intoduce the fundamentals of the
\textit{first order (fake supergravity) formalism} for the \textit{%
non-degenerate} attractor flows (both BPS and non-BPS) of extremal BHs in $%
d=4$ space-time dimensions.

Sect. \ref{N=2-d=4-quadratic} is thus devoted to a detailed study of $%
\mathcal{N}=2$, $d=4$ supergravity \textit{minimally coupled} to Abelian
vector multiplets. In Subsect. \ref{N=2-d=4-quadratic-AEs} the related
\textit{Attractor Equations} are explicitly solved, for both the classes of
\textit{non-degenerate} critical points of $V_{BH}$: the $\frac{1}{2}$-BPS
one (Subsubsect. \ref{1/2-BPS-Attractors}) and the non-BPS $Z=0$ one, this
latter with related moduli space (Subsubsect. \ref{non-BPS-Z=0-Attractors}).
By exploiting the \textit{first order (fake supergravity) formalism}, in
Subsects. \ref{N=2-d=4-quadratic-BPS-Attractor-Flow} and \ref
{N=2-d=4-quadratic-non-BPS-Z=0-Attractor-Flow} the \textit{ADM}
(Arnowitt-Deser-Misner) \textit{mass} $M_{ADM}$ \cite{arnowitt}, \textit{%
covariant scalar charges} $\Sigma _{i}$ and \textit{(square) effective
horizon radius} $R_{H}^{2}$ are explicitly computed respectively for $\frac{1%
}{2}$-BPS and non-BPS $Z=0$ attractor flows, proving that the second line of
Eq. (\ref{Ita-Rom-1}) holds true. This latter result, already proved in \cite
{FHM}, generalizes the findings of \cite{K3,Garfinkle}, also holding in the
\textit{non-extremal case}.

Sect. \ref{N=3,d=4} deals with $\mathcal{N}=3$, $d=4$ supergravity coupled
to matter (Abelian vector) multiplets. In Subsect. \ref{N=3,d=4,AES} the
related \textit{Attractor Equations} are explicitly solved, for both the
classes of \textit{non-degenerate} critical points of $V_{BH}$: the $\frac{1%
}{3}$-BPS one (Subsubsect. \ref{1/3-BPS-Attractors}) and the non-BPS $%
Z_{AB}=0$ one (Subsubsect. \ref{N=3,non-BPS,Z=0-Attractors}), both with
related moduli space. Once again, by using the \textit{first order (fake
supergravity) formalism}, in Subsects. \ref{N=3,d=4-BPS} and \ref
{N=3-d=4-non-BPS} the \textit{ADM mass} $M_{ADM}$, \textit{covariant scalar
charges} $\Sigma _{i}$ and \textit{(square) effective horizon radius} $%
R_{H}^{2}$ respectively for $\frac{1}{3}$-BPS and non-BPS $Z_{AB}=0$
attractor flows are explicitly computed, proving that the second line of Eq.
(\ref{Ita-Rom-1}) holds true also for such a theory. This, as the \textit{%
minimally coupled} $\mathcal{N}=2$ supergravity, has a \textit{unique} $U$%
-invariant \textit{quadratic} in BH charges.

Comments on the invariance properties of BH entropy in \textit{minimally
coupled} $\mathcal{N}=2$, as well as in \textit{matter-coupled} $\mathcal{N}%
=3$,\textit{\ ungauged} $d=4$ supergravity are given in Sect. \ref
{Invariance-Props}.

Next, Sect. \ref{N=5,d=4} deals with $\mathcal{N}=5$, $d=4$ supergravity,
which does \textit{not} allow for matter coupling and whose field content
thus only consists of the gravity multiplet (\textit{pure} theory). In
Subsect. \ref{N=5-d=4-AEs} the related \textit{Attractor Equations} are
explicitly solved for the unique class of \textit{non-degenerate} critical
points of $V_{BH}$, namely the $\frac{1}{5}$-BPS one. In Subsubsect. \ref
{1/5-BPS-Attractors} such a class is studied, along with the related moduli
space and Bekenstein-Hawking classical BH entropy \cite{hawking2}. This
latter is proportional to the unique $U$-invariant $\mathcal{I}_{4}$ of $%
\mathcal{N}=5$ supergravity, whose \textit{quartic} expression in terms of
the BH charges is also explicitly derived. Through the formal machinery
presented in Sect. \ref{First-Order}, in Subsect. \ref{N=5,d=4,BPS} the
\textit{ADM mass} $M_{ADM}$, \textit{covariant scalar charges} $\Sigma _{i}$
and \textit{(square) effective horizon radius} $R_{H}^{2}$ for the $\frac{1}{%
5}$-BPS attractor flow are explicitly given, proving that (the second line
of) Eq. (\ref{Ita-Rom-1}) holds true also for such a theory. This is
somewhat surprising because, as mentioned above, $\mathcal{N}=5$
supergravity, in contrast to the \textit{minimally coupled} $\mathcal{N}=2$
and \textit{matter-coupled} $\mathcal{N}=3$ cases, has a \textit{unique} $U$%
-invariant \textit{quartic}, rather than \textit{quadratic}, in BH charges.

So, in Sect. \ref{N=4,d=4-pure} the extremal BH attractors in $\mathcal{N}=4$%
, $d=4$ \textit{pure} supergravity are revisited. In Subsect. \ref
{N=4-d=4-pure-AEs} the resolution of the corresponding \textit{Attractor
Equation} \cite{ADFT} is reviewed for the unique class of \textit{%
non-degenerate} critical points of $V_{BH}$, namely the $\frac{1}{4}$-BPS
one. Its corresponding Bekenstein-Hawking classical BH entropy \cite
{hawking2} is given by the unique $U$-invariant $\mathcal{I}_{4}$ of $%
\mathcal{N}=4$ \textit{pure} supergravity, which is also reported. By using
the formul\ae\ of Sect. \ref{First-Order}, in Subsect. \ref{N=4-d=4-pure-BPS}
the \textit{ADM mass} $M_{ADM}$, \textit{covariant axion-dilaton charge} $%
\Sigma _{s}$ and \textit{square effective horizon radius} $R_{H}^{2}$ for
the $\frac{1}{4}$-BPS attractor flow are explicitly given, proving that the
second line of Eq. (\ref{Ita-Rom-1}) holds true also for such a theory. Also
such a result is rather surprising, for the same reason mentioned above:
\textit{pure} $\mathcal{N}=4$ supergravity, as $\mathcal{N}=5$ theory and in
contrast to the \textit{minimally coupled} $\mathcal{N}=2$ and \textit{%
matter-coupled} $\mathcal{N}=3$ cases, has a \textit{unique} $U$-invariant
\textit{quartic}, rather than \textit{quadratic}, in BH charges.

However, as pointed out in the Introduction, \textit{pure} $\mathcal{N}=4$%
\textbf{\ }and $\mathcal{N}=5$ supergravities are peculiar theories, because
their \textit{unique} (\textit{moduli-independent}) $U$-duality invariant,
\textit{quartic} in BH charges, when expressed as a (\textit{unique})
combination of \textit{``dressed''} (\textit{moduli-dependent}) charges,
turns out to be a \textit{perfect square} of a \textit{quadratic} expression
in the \textit{skew-eigenvalues} $\mathcal{Z}_{1}$ and $\mathcal{Z}_{2}$ of
the relevant \textit{central charge matrix}. Such a key feature is studied
in Sect. \ref{N=4,5,d=4}.

In Sect. \ref{Relations} we consider \textit{all} \textit{ungauged} $%
\mathcal{N}\geqslant 2$, $d=4$ supergravities sharing the same bosonic
sector, and thus with the same number of fermion fields, but with \textit{%
different supersymmetric completions}. Beside the well-known case of the
\textit{``duality'' }between $\mathcal{N}=2$ $J_{3}^{\mathbb{H}}$ (\textit{%
matter-coupled}) and $\mathcal{N}=6$ (\textit{pure}) supergravity (see \cite
{BFGM1} and Refs. therein), other two cases exist, namely:

\begin{itemize}
\item  the \textit{``duality''} exhibited by $\mathcal{N}=2$ supergravity
\textit{minimally coupled} to $3$ Abelian vector multiplets, and $\mathcal{N}%
=3$ supergravity coupled to $1$ matter (Abelian vector) multiplet

\item  the \textit{``duality''} between $\mathcal{N}=2$ supergravity coupled
to $6$ Abelian vector multiplets, with scalar manifold given by the
symmetric reducible special K\"{a}hler manifold $\frac{SU\left( 1,1\right) }{%
U\left( 1\right) }\times \frac{SO\left( 2,6\right) }{SO\left( 2\right)
\times SO\left( 6\right) }$, and $\mathcal{N}=4$ supergravity coupled to $2$
matter (Abelian vector) multiplets.
\end{itemize}

It is here worth commenting that such \textit{``dualities''} constitute
evidence against the conventional wisdom that bosonic interacting theories
have an \textit{unique} supersymmetric extension. The sharing of the same
bosonic backgrounds with different supersymmetric completions implies its
\textit{``dual''} interpretation with respect to the
supersymmetry-preserving properties. Consistent with local supersymmetry,
the number of fermion fields is the same in both theories, but with \textit{%
different} spin/field contents, simply related by the interchange among spin-%
$\frac{1}{2}$ (\textit{gaugino}) and spin-$\frac{3}{2}$ (\textit{gravitino})
fields.

Sect. \ref{Conclusion} contains some comments, outlook and directions for
further developments.

Finally, Appendix I concludes the paper. It presents $\mathcal{N}=4$, $d=4$
\textit{ungauged} supergravity coupled to $1$ matter (Abelian vector)
multiplet (upliftable to the $\mathcal{N}=4$, $d=5$ \textit{pure} theory).
This constitutes a counterexample of a theory with \textit{unique} (\textit{%
moduli-independent}) $U$-duality invariant \textit{quartic} in BH charges
which, when expressed as a combination of \textit{``dressed''} (\textit{%
moduli-dependent}) charges, does \textit{not} turn out to be a \textit{%
perfect square} of a \textit{quadratic} expression in the \textit{%
skew-eigenvalues} of the \textit{central charge matrix} and in the \textit{%
matter charge(s)}. As a consequence, the explicit expression of $R_{H}^{2}$
given by (the second line of) Eq. (\ref{Ita-Rom-1}) does \textit{not} hold
for such a theory, as well as for all other $d=4$ (ungauged) supergravities
not explicitly mentioned above and in the treatment given below.

\section{\label{First-Order}Fake Supergravity Formalism and Effective
Horizon Radius\newline
for $d=4$ Extremal Black Holes}

We recall some facts about the \textit{first order (fake supergravity)
formalism} \cite{Fake-Refs} for static, spherically symmetric,
asymptotically flat dyonic \textit{extremal} (\textit{i.e.} with $c=0$) BHs
in $d=4$, introduced in \cite{Cer-Dal-1} and \cite{ADOT-1} (see also \cite
{FHM}).

Let us start with the general formula for the (positive definite) \textit{BH
effective potential} of $d=4$ supergravities:
\begin{equation}
V_{BH}=\frac{1}{2}Z_{AB}\overline{Z}^{AB}+Z_{I}\overline{Z}^{I},
\label{CERN-4}
\end{equation}
where $Z_{AB}=Z_{\left[ AB\right] }$ ($A,B=1,...,\mathcal{N}$) is the
\textit{central charge matrix}, and $Z_{I}$ ($I=1,...,n$) are the \textit{%
matter charges}, $n\in \mathbb{N}$ being the number of matter multiplets (if
any) coupled to the gravity multiplet. Equivalently, in the \textit{first
order formalism }(see Eq. (2.23) of \cite{Cer-Dal-1}):
\begin{equation}
V_{BH}=\mathcal{W}^{2}+4G^{i\overline{j}}\left( \partial _{i}\mathcal{W}%
\right) \overline{\partial }_{\overline{j}}\mathcal{W=W}^{2}+4G^{i\overline{j%
}}\left( \nabla _{i}\mathcal{W}\right) \overline{\nabla }_{\overline{j}}%
\mathcal{W},
\end{equation}
where $\mathcal{W}$ is the moduli-dependent so-called \textit{first order
fake superpotential}, and $\nabla $ denotes the relevant covariant
differential operator.

An alternative expression for $V_{BH}$ can be given as follows (see Eq.
(5.7) of \cite{BFM}):
\begin{equation}
V_{BH}=e^{\mathcal{G}}\left[ 1+G^{i\overline{j}}\left( \partial _{i}\mathcal{%
G}\right) \overline{\partial }_{\overline{j}}\mathcal{G}\right] =e^{\mathcal{%
G}}\left[ 1+G^{i\overline{j}}\left( \nabla _{i}\mathcal{G}\right) \overline{%
\nabla }_{\overline{j}}\mathcal{G}\right] ,
\end{equation}
where now
\begin{equation}
\mathcal{W}\equiv e^{\mathcal{G}/2}.
\end{equation}

By recalling Eq. (65) of \cite{FHM} and Eqs. (84) and (114) of \cite{FHM}
(which in turn can be traced back to Eq. (29) of \cite{ADOT-1}), in the same
framework the \textit{covariant scalar charges} and the \textit{squared ADM
mass} \cite{arnowitt} can respectively be written as follows\footnote{%
Here and in all our analysis we assume all functions of moduli to be
sufficiently regular, in order to allow one to perform smoothly the \textit{%
radial asymptotical (}$\tau \rightarrow 0^{-}$\textit{) and near horizon (}$%
\tau \rightarrow -\infty $)\textit{\ limits}.}:
\begin{eqnarray}
\Sigma _{i} &=&2\lim_{\tau \rightarrow 0^{-}}\nabla _{i}\mathcal{W}%
=2\lim_{\tau \rightarrow 0^{-}}\partial _{i}\mathcal{W};  \label{CERN!-1} \\
&&  \notag \\
M_{ADM}^{2} &=&r_{H}^{2}=\lim_{\tau \rightarrow 0^{-}}\left[ V_{BH}-4G^{i%
\overline{j}}\left( \partial _{i}\mathcal{W}\right) \overline{\partial }_{%
\overline{j}}\mathcal{W}\right] =\lim_{\tau \rightarrow 0^{-}}\mathcal{W}%
^{2},  \label{CERN!-2}
\end{eqnarray}
where $\tau \equiv \left( r_{H}-r\right) ^{-1}$. Thence, one can introduce
the \textit{(square)} \textit{effective horizon radius} \label{ADOT-1,FHM}
(recall the notation $R_{+,c=0}=R_{-,c=0}\equiv R_{H}$; see the treatment of
\cite{FHM}):
\begin{eqnarray}
R_{H}^{2} &\equiv &\lim_{\tau \rightarrow -\infty }V_{BH}=\left.
V_{BH}\right| _{\partial V_{BH}=0,V_{BH}\neq 0}=\lim_{\tau \rightarrow
-\infty }\mathcal{W}^{2}=\left. \mathcal{W}^{2}\right| _{\partial \mathcal{W}%
=0,\mathcal{W}\neq 0}=  \notag \\
&=&\frac{A_{eff}\left( p,q\right) }{4\pi }=\frac{S_{BH}\left( p,q\right) }{%
\pi },  \label{Frascati-1}
\end{eqnarray}
where $\left( p,q\right) $ denotes the set of magnetic and electric BH
charges, $A_{eff}$ (simply named $A$ in the Introduction) is the \textit{%
effective area} of the BH (\textit{i.e.} the area of the surface pertaining
to $R_{H}$), $S_{BH}$ is the classical BH entropy, and the \textit{%
Bekenstein-Hawking entropy-area formula} \cite{hawking2} has been used.

Whenever allowed by the symmetric nature of the scalar manifold, $R_{H}^{2}$
can thus be expressed in terms of a suitable power of the (generally unique)
invariant of the relevant representation of the $U$\textit{-duality group }$%
G $, determining the symplectic embedding of the vector field strengths. In $%
d=4$ $S_{BH}$ is homogeneous of degree two in $\left( p,q\right) $, and only
two possibilities arise:
\begin{equation}
R_{H}^{2}=\left| \mathcal{I}_{2}\left( p,q\right) \right| ,~\mathit{or}%
~R_{H}^{2}=\sqrt{\left| \mathcal{I}_{4}\left( p,q\right) \right| },
\end{equation}
where $\mathcal{I}_{2}$ and $\mathcal{I}_{4}$ respectively denote $U$%
-invariants \textit{quadratic} and \textit{quartic} in BH charges.

By exploiting the $\tau $\textit{-monotonicity} of $\mathcal{W}$ (which is
indeed an example of $\mathit{C}$\textit{-function} \cite{GJMT} for \textit{%
extremal} BHs) \cite{ADOT-1}:
\begin{equation}
\frac{d\mathcal{W}\left( z\left( \tau \right) ,\overline{z}\left( \tau
\right) ;p,q\right) }{d\tau }\geqslant 0,
\end{equation}
the following inequality (holding for $c=0$) can be obtained \cite{FHM}:
\begin{eqnarray}
M_{ADM}^{2}\left( z_{\infty },\overline{z}_{\infty },p,q\right)
&=&\lim_{\tau \rightarrow 0^{-}}\left[ V_{BH}-4G^{i\overline{j}}\left(
\partial _{i}\mathcal{W}\right) \overline{\partial }_{\overline{j}}\mathcal{W%
}\right] =  \notag \\
&=&\lim_{\tau \rightarrow 0^{-}}\mathcal{W}^{2}\equiv r_{H}^{2}\left(
z_{\infty },\overline{z}_{\infty },p,q\right)  \notag \\
&\geqslant &R_{H}^{2}\left( p,q\right) =\lim_{\tau \rightarrow -\infty }%
\mathcal{W}^{2}=\lim_{\tau \rightarrow -\infty }V_{BH},  \label{CERN-night-2}
\end{eqnarray}
where the radius $r_{H}$ of the BH \textit{event horizon} was introduced.
More concisely:
\begin{equation}
r_{H}^{2}\left( z_{\infty },\overline{z}_{\infty },p,q\right) \geqslant
R_{H}^{2}\left( p,q\right) ,~\forall \left( z_{\infty },\overline{z}_{\infty
}\right) \in \mathcal{M}_{\infty },  \label{CERN-night-1}
\end{equation}
holding in the whole \textit{asymptotical scalar manifold} $\mathcal{M}%
_{\infty }$.

In the \textit{minimally matter-coupled} $\mathcal{N}=2$, $d=4$ supergravity
based on the sequence of symmetric special K\"{a}hler manifolds (\textit{%
complex Grassmannians}) $\frac{SU\left( 1,n\right) }{SU\left( n\right)
\times U(1)}$ \cite{Luciani} (see also the treatment of \cite{FHM}), as well
as in $\mathcal{N}=3$, \textit{pure} $\mathcal{N}=4$ and $\mathcal{N}=5$, $%
d=4$ supergravity, it is possible to specialize further the inequality (\ref
{CERN-night-1}). Indeed, for such theories it holds that (recall Eq. (\ref
{Ita-Rom-1}), as well as Eqs. (\ref{Tue-3}) and (\ref{Tue-4}))
\begin{eqnarray}
R_{H}^{2}\left( p,q\right) &\equiv &\frac{S_{BH}\left( p,q\right) }{\pi }=
\notag \\
&=&r_{H}^{2}\left( z_{\infty },\overline{z}_{\infty },p,q\right) -G_{i%
\overline{j}}\Sigma ^{i}\overline{\Sigma }^{\overline{j}}=r_{H}^{2}\left(
z_{\infty },\overline{z}_{\infty },p,q\right) -4\lim_{\tau \rightarrow
0^{-}}G^{i\overline{j}}\left( \partial _{i}\mathcal{W}\right) \overline{%
\partial }_{\overline{j}}\mathcal{W},  \label{CERN-night-3}
\end{eqnarray}
where in the last step Eq. (\ref{CERN!-1}) was used. Eq. (\ref{CERN-night-3}%
), clearly yielding the inequality (\ref{CERN-night-1}) by the presence of
non-vanishing scalar charges and the (strict) positive definiteness of $G_{i%
\overline{j}}$, is nothing but a \textit{many-moduli generalization} of the
formula holding for the so-called \textit{(axion-)dilaton extremal BH} \cite
{K3}. The crucial feature, expressed by Eq. (\ref{CERN-night-3}) and shared
by the aforementioned supergravities, is the \textit{disappearance} of the
dependence on the \textit{asymptotical moduli} $\left( z_{\infty },\overline{%
z}_{\infty }\right) $ in the combination of quantities $r_{H}^{2}-G_{i%
\overline{j}}\Sigma ^{i}\overline{\Sigma }^{\overline{j}}$, which separately
do depend on moduli\footnote{$r_{H}\left( z_{\infty },\overline{z}_{\infty
},p,q\right) $ is the radius of the BH event horizon, which is the unique
geometrical horizon for \textit{extremal} BHs (in which $c=0\Leftrightarrow
r_{-}=r_{+}\equiv r_{H}$; for a recent treatment, see \textit{e.g.} \cite
{FHM}). It depends on the dyonic BH charges $p^{\Lambda }$ and $q_{\Lambda }$%
, but, in presence of \textit{non-vanishing} scalar charges, also on the
asymptotical scalar fields $\left( z_{\infty },\overline{z}_{\infty }\right)
$.
\par
In order to make contact with the \textit{Attractor Mechanism}, and thus to
characterize $r_{H}$ as the \textit{fixed point} of the scalar radial
dynamics (in the considered static, spherically symmetric and asymptotically
flat \textit{extremal} BH background), one has to evaluate $r_{H}$ at the
peculiar \textit{geometrical locus} in the (asymptotical) moduli space
defined by the (\textit{non-degenerate}) criticality condition of $V_{BH}$ (%
\textit{i.e.} by $\partial V_{BH}=0$, with $\left. V_{BH}\right| _{\partial
V_{BH}=0}\neq 0$). Eq. (\ref{CERN!-1}) yields that
\begin{equation*}
\Sigma ^{i}\left( z_{H}\left( p,q\right) ,\overline{z}_{H}\left( p,q\right)
,p,q\right) =0~~\forall i,
\end{equation*}
where $\left( z_{H}\left( p,q\right) ,\overline{z}_{H}\left( p,q\right)
\right) $ are defined by
\begin{equation*}
\left[ \partial V_{BH}\left( z,\overline{z},p,q\right) \right] _{\left( z,%
\overline{z}\right) =\left( z_{H}\left( p,q\right) ,\overline{z}_{H}\left(
p,q\right) \right) }\equiv 0.
\end{equation*}
Thus, Eqs. (\ref{CERN!-2}) and (\ref{Frascati-1}) (or Eq. (\ref{CERN-night-3}%
)) consistently yield that
\begin{equation*}
r_{H}\left( z_{H}\left( p,q\right) ,\overline{z}_{H}\left( p,q\right)
,p,q\right) =R_{H}\left( p,q\right) .
\end{equation*}
}.

As a generalization of the formula holding (also in the \textit{non-extremal
case}) in the \textit{Maxwell-axion-dilaton supergravity }(see \textit{e.g.}
\cite{K3,Garfinkle}, and also \cite{FHM}), in \cite{FHM} Eq. (\ref
{CERN-night-3}) was proved to hold in the \textit{extremal case} for the
whole sequence of $\mathcal{N}=2$, $d=4$ supergravity \textit{minimally
coupled} to Abelian vector multiplets \cite{Luciani}, in terms of the
(unique) invariant $\mathcal{I}_{2}$ of the $U$-duality group $G=SU\left(
1,n\right) $, which is \textit{quadratic} in charges:
\begin{equation}
R_{H}^{2}\left( p,q\right) =r_{H}^{2}\left( z_{\infty },\overline{z}_{\infty
},p,q\right) -4\lim_{\tau \rightarrow 0^{-}}G^{i\overline{j}}\left( \partial
_{i}\mathcal{W}\right) \overline{\partial }_{\overline{j}}\mathcal{W}=\left|
\mathcal{I}_{2}\left( p,q\right) \right| .  \label{CERN-Thu-1}
\end{equation}
We will report such results in Subsects. \ref
{N=2-d=4-quadratic-BPS-Attractor-Flow} and \ref
{N=2-d=4-quadratic-non-BPS-Z=0-Attractor-Flow}.

Thence, by exploiting the \textit{first order formalism} for $d=4$\textit{\
extremal} BHs outlined above, we will show that \textit{the same} happens
for the following $d=4$ supergravities:

\begin{itemize}
\item  $\mathcal{N}=3$ (\textit{matter-coupled}) \cite{N=3-Ref}, as
intuitively expected by the strict similarity with the so-called \textit{%
minimally coupled} $\mathcal{N}=2$ theory (Subsects. \ref{N=3,d=4-BPS} and
\ref{N=3-d=4-non-BPS}\textbf{)};

\item  $\mathcal{N}=5$\textbf{\ }\cite{N=5-Ref}, with $\left| \mathcal{I}%
_{2}\right| $ replaced by $\sqrt{\left| \mathcal{I}_{4}\right| }$ (Subsect.
\ref{N=5,d=4,BPS});

\item  \textit{pure }$\mathcal{N}=4$ \cite{CSF}, with $\left| \mathcal{I}%
_{2}\right| $ replaced by $\sqrt{\left| \mathcal{I}_{4}\right| }$ (Subsect.
\ref{N=4-d=4-pure-BPS}).
\end{itemize}

Let us here note that while $\mathcal{N}=5$ theory \textit{cannot be }%
coupled to matter, in the case $\mathcal{N}=4$ \textit{matter coupling} is
allowed, but Eq. (\ref{CERN-night-3}) holds \textit{only} in $\mathcal{N}=4$
\textit{pure} supergravity. Having a(n unique) $U$-invariant $\mathcal{I}%
_{4} $ \textit{quartic} in charges, the aforementioned $\mathcal{N}=4$ and $%
\mathcal{N}=5$ theories are pretty different from the \textit{minimally
coupled} $\mathcal{N}=2$ and $\mathcal{N}=3$, $d=4$ supergravity, as we will
point out in the treatment below.

It is here worth pointing out that in the \textit{non-extremal case} (%
\textit{i.e.} $c\neq 0$) the expression generalizing Eq. (\ref{CERN-night-3}%
), namely
\begin{eqnarray}
R_{+}^{2}\left( z_{\infty },\overline{z}_{\infty },p,q\right) &\equiv &\frac{%
S_{BH,c\neq 0}\left( z_{\infty },\overline{z}_{\infty },p,q\right) }{\pi }%
\equiv R_{+}^{2}\left( z_{\infty },\overline{z}_{\infty },p,q\right) =
\notag \\
&=&r_{+}^{2}\left( z_{\infty },\overline{z}_{\infty },p,q\right) -G_{i%
\overline{j}}\Sigma ^{i}\overline{\Sigma }^{\overline{j}}
\end{eqnarray}
can be only \textit{guessed}, but at present cannot be rigorously proved.
Indeed, for static, spherically symmetric, asymptotically flat dyonic
\textit{non-extremal} BHs a \textit{first order formalism} is currently
unavailable, so there is no way to compute the scalar charges (beside the
\textit{direct integration} of the Eqs. of motion of the scalars, as far as
we know at present feasible only for the (axion-)dilaton BH \cite{K3}, and -
partially - for $stu$ model \cite{GLS1}).\setcounter{equation}0

\section{\label{N=2-d=4-quadratic}$\mathcal{N}=2$ \textit{Minimally Coupled}
Supergravity}

We consider $\mathcal{N}=2$, $d=4$ ungauged supergravity \textit{minimally
coupled} \textit{(mc)} \cite{Luciani} to $n_{V}$ Abelian vector multiplets,
in the case in which the scalar manifold is given by the sequence of
homogeneous symmetric \textit{rank-}$1$ special K\"{a}hler manifolds
\begin{equation}
\mathcal{M}_{\mathcal{N}=2,mc,n}=\frac{G_{\mathcal{N}=2,mc,n}}{H_{\mathcal{N}%
=2,mc,n}}=\frac{SU(1,n)}{SU(n)\times U(1)},~dim_{\mathbb{R}}=2n,~n=n_{V}\in
\mathbb{N}.
\end{equation}
The $1+n$ vector field strengths and their duals, as well as their
asymptotical fluxes, sit in the \textit{fundamental }$\mathbf{1+n}$ \
representation of the $U$-duality group $G_{\mathcal{N}=2,mc,n}=SU\left(
1,n\right) $, in turn embedded in the symplectic group\footnote{%
In all our analysis we consider the (semi)classical limit of \textit{%
continuous} (\textit{unquantized}), large BH charges.} $Sp\left( 2+2n,%
\mathbb{R}\right) $.

The general analysis of the Attractor Equations, BH charge orbits and
attractor moduli spaces of such a theory has been performed in \cite{BFGM1}
and \cite{ferrara4}.

By fixing the K\"{a}hler gauge such that $X^{0}=1$ and in a suitable system
of local symplectic \textit{special} coordinates, the geometry of $\mathcal{M%
}_{\mathcal{N}=2,mc,n}$ is determined by the \textit{holomorphic
prepotential function}:
\begin{equation}
\mathcal{F}(z)\equiv -\frac{i}{2}\left[ 1-\left( z^{i}\right) ^{2}\right] .
\end{equation}

The K\"{a}hler potential of $\mathcal{M}_{\mathcal{N}=2,mc,n}$ can be
computed to be ($\Lambda =0,1,...,n$ throughout all the present Section, and
$\left| z\right| ^{2}\equiv \sum_{i=1}^{n_{V}}\left| z^{i}\right| ^{2}$)
\begin{equation}
K\left( z,\overline{z}\right) =-log\left[ i\left( \overline{X}^{\Lambda
}F_{\Lambda }-X^{\Lambda }\overline{F}_{\Lambda }\right) \right] =-log\left[
2\left( 1-\left| z\right| ^{2}\right) \right] ,  \label{K}
\end{equation}
yielding the \textit{metric constraint} $1-\left| z\right| ^{2}>0$, and the
covariant and contravariant metric tensors to be respectively ($G_{i%
\overline{j}}\left( z,\overline{z}\right) G^{i\overline{k}}\left( z,%
\overline{z}\right) =\delta _{\overline{j}}^{\overline{k}}$):
\begin{eqnarray}
G_{i\overline{j}}\left( z,\overline{z}\right) &=&\frac{\left( 1-\left|
z\right| ^{2}\right) \delta _{i\overline{j}}+\overline{z}^{\overline{i}}z^{j}%
}{\left( 1-\left| z\right| ^{2}\right) ^{2}}=2e^{K}\delta _{i\overline{j}%
}+4e^{2K}\overline{z}^{\overline{i}}z^{j}; \\
&&  \notag \\
G^{i\overline{j}}\left( z,\overline{z}\right) &=&\left( 1-\left| z\right|
^{2}\right) \left( \delta ^{i\overline{j}}-z^{i}\overline{z}^{\overline{j}%
}\right) =\frac{1}{2}e^{-K}\left( \delta ^{i\overline{j}}-z^{i}\overline{z}^{%
\overline{j}}\right) ;
\end{eqnarray}

From its very definition (see \textit{e.g.} \cite{4}, and Refs. therein),
the covariantly holomorphic $\mathcal{N}=2$, $d=4$ \textit{central charge
function} can be computed to be
\begin{equation}
Z=e^{K/2}W=e^{K/2}\left[ q_{0}+ip^{0}+\left( q_{i}-ip^{i}\right) z^{i}\right]
=\frac{1}{\sqrt{2}}\frac{1}{\sqrt{1-|z|^{2}}}\left[ q_{0}+ip^{0}+\left(
q_{i}-ip^{i}\right) z^{i}\right] ,  \label{Z}
\end{equation}
where $W$ is the $\mathcal{N}=2$, $d=4$ \textit{superpotential} (also named
\textit{holomorphic} \textit{central charge function}, with K\"{a}hler
weights $\left( 2,0\right) $).

On the other hand, the so-called \textit{matter charges} read
\begin{eqnarray}
Z_{i} &\equiv &D_{i}Z=\partial _{i}Z+\frac{1}{2}\left( \partial _{i}K\right)
Z=e^{K/2}\left[ \partial _{i}W+\left( \partial _{i}K\right) W\right] =
\notag \\
&&  \notag \\
&=&\frac{1}{\sqrt{2}\left( 1-\left| z\right| ^{2}\right) ^{3/2}}\left[
(q_{i}-ip^{i})(1-\left| z\right| ^{2})+(q_{0}+ip^{0})\overline{z}^{\overline{%
i}}+(q_{j}-ip^{j})z^{j}\overline{z}^{\overline{i}}\right] .  \label{DiZ}
\end{eqnarray}
Here, $D$ denotes the $U\left( 1\right) $-K\"{a}hler and $H_{\mathcal{N}%
=2,mc,n}$-covariant differential operator. Due to the global vanishing of
the $C_{ijk}$-tensor of special K\"{a}hler geometry, there are only two ($%
U\left( 1\right) $-K\"{a}hler-) and $H_{\mathcal{N}=2,mc,n}$-invariants,
namely
\begin{eqnarray}
\alpha _{1} &\equiv &\left| Z\right| ;  \label{tango1} \\
&&  \notag \\
\alpha _{2} &\equiv &\sqrt{Z_{i}\overline{Z}^{i}}=\sqrt{G^{i\overline{j}%
}Z_{i}\overline{Z}_{\overline{j}}}=\sqrt{G^{i\overline{j}}\left(
D_{i}Z\right) \overline{D}_{\overline{j}}\overline{Z}},  \label{tango2}
\end{eqnarray}
both \textit{(homogeneous) of degree }$1$ in BH charges $\left( p,q\right) $
(in particular, square roots of quantities \textit{quadratic} in $\left(
p,q\right) $). By a suitable rotation of $U\left( n\right) $, the vector $%
Z_{i}$ of \textit{matter charges} can be chosen real and pointing in a given
direction, \textit{e.g.}
\begin{equation}
Z_{i}=\sqrt{Z_{j}\overline{Z}^{j}}\delta _{i1}=\alpha _{2}\delta _{i1}.
\end{equation}

As recalled at the start of the next Subsection, only ($\frac{1}{2}$-)BPS
and non-BPS ($Z=0$) attractor flows are \textit{non-degenerate} (\textit{i.e.%
} corresponding to \textit{large} BHs, see below) \cite{BFGM1}, and the
corresponding (squared) \textit{first order fake superpotentials} are (\cite
{ADOT-1}; recall Eq. (\ref{salsa-1}) and (\ref{salsa-2}), respectively)
\begin{eqnarray}
\mathcal{W}_{\left( \frac{1}{2}-\right) BPS}^{2} &=&\left| Z\right|
^{2}=\alpha _{1}^{2}=  \notag \\
&&  \notag \\
&=&\frac{\left[ q_{0}+ip^{0}+\left( q_{i}-ip^{i}\right) z^{i}\right] \left[
q_{0}-ip^{0}+\left( q_{j}+ip^{j}\right) \overline{z}^{\overline{j}}\right] }{%
2\left( 1-|z|^{2}\right) };
\end{eqnarray}
\begin{eqnarray}
\mathcal{W}_{non-BPS(,Z=0)}^{2} &=&G^{i\overline{j}}\left( D_{i}Z\right)
\overline{D}_{\overline{j}}\overline{Z}=\alpha _{2}^{2}=  \notag \\
&&  \notag \\
&=&\frac{1}{2\left( 1-\left| z\right| ^{2}\right) ^{2}}\left( \delta ^{i%
\overline{j}}-z^{i}\overline{z}^{\overline{j}}\right) \cdot  \notag \\
&&  \notag \\
&&\cdot \left[ (q_{i}-ip^{i})(1-\left| z\right| ^{2})+(q_{0}+ip^{0})%
\overline{z}^{\overline{i}}+(q_{r}-ip^{r})z^{r}\overline{z}^{\overline{i}}%
\right] \cdot  \notag \\
&&  \notag \\
&&\cdot \left[ (q_{j}+ip^{j})(1-\left| z\right|
^{2})+(q_{0}-ip^{0})z^{j}+(q_{n}+ip^{n})\overline{z}^{\overline{n}}z^{j}%
\right] ,  \label{W-non-BPS}
\end{eqnarray}
where use of Eqs. (\ref{Z}) and (\ref{DiZ}) was made.

\subsection{\label{N=2-d=4-quadratic-AEs}Attractor Equations and their
Solutions}

The \textit{BH effective potential} can be written as
\begin{equation}
V_{BH}=\left| Z\right| ^{2}+G^{i\overline{j}}\left( D_{i}Z\right) \overline{D%
}_{\overline{j}}\overline{Z}=\alpha _{1}^{2}+\alpha _{2}^{2}.
\label{VBH-N=2-quadr.}
\end{equation}
The $\mathcal{N}=2$, $d=4$ \textit{Attractor Eqs.} in the case of \textit{%
minimal coupling} to Abelian vector multiplets are nothing but the \textit{%
criticality conditions} for such an $H_{\mathcal{N}=2,mc,n}$-invariant (and
K\"{a}hler-gauge-invariant) quantity. Such criticality conditions are
satisfied for two classes of critical points \label{BFGM1}:

\begin{itemize}
\item  ($\frac{1}{2}$-)BPS:
\begin{equation}
D_{i}Z=0~\forall i=1,...,n\Leftrightarrow \alpha _{2}=0,~Z\neq 0;
\label{1/2-BPS}
\end{equation}
%\newline

\item  non-supersymmetric (non-BPS with $Z=0$) :
\begin{equation}
D_{i}Z\neq 0~(\text{\textit{at least~}for~some~}i),~Z=0\Leftrightarrow
\alpha _{1}=0.  \label{non-BPS-Z=0}
\end{equation}
\end{itemize}

It is worth counting here the degrees of freedom related to Eqs. (\ref
{1/2-BPS}) and (\ref{non-BPS-Z=0}). The $\frac{1}{2}$-BPS criticality
conditions\ (\ref{1/2-BPS}) are $n$ \textit{complex} independent ones, thus
\textit{all} scalars are stabilized by such conditions. On the other hand,
there is only one \textit{complex} non-BPS $Z=0$ criticality condition\ (\ref
{non-BPS-Z=0}). This fact paves the way to the possibility to have a \textit{%
moduli space} of non-BPS $Z=0$ attractors, spanned by the $n-1$ \textit{%
complex} scalars unstabilized by Eq. (\ref{non-BPS-Z=0}); this actually
holds true \cite{ferrara4}, as it will be explicitly found below for the
first time (see Subsection \ref{non-BPS-Z=0-Attractors}).

\subsubsection{$\frac{1}{2}$-BPS Attractors\label{1/2-BPS-Attractors}}

An \textit{algebraic}, equivalent approach to the direct resolution of the $%
n $ \textit{complex} $\frac{1}{2}$-BPS criticality conditions (\ref{1/2-BPS}%
) is based on the resolution of the \textit{special K\"{a}hler geometry
identities} evaluated along the \textit{geometrical locus} in $\mathcal{M}_{%
\mathcal{N}=2,quadr.,n}$ defined by the constraints (\ref{1/2-BPS}). By
following such an approach, the electric and magnetic BH charges are
constrained as follows \cite{ferrara2}:
\begin{equation}
\left\{
\begin{array}{c}
p^{\Lambda }=ie^{K/2}(\overline{Z}X^{\Lambda }-Z\overline{X}^{\Lambda }); \\
\\
q_{\Lambda }=ie^{K/2}(\overline{Z}F_{\Lambda }-Z\overline{F}_{\Lambda }).
\end{array}
\ \right.
\end{equation}
Summing such two sets of symplectic-covariant Eqs., one gets
\begin{equation}
X^{\Lambda }q_{\Sigma }-p^{\Lambda }F_{\Sigma }=ie^{K/2}Z(\overline{X}%
^{\Lambda }F_{\Sigma }-X^{\Lambda }\overline{F}_{\Sigma }),
\label{sat-night-1}
\end{equation}
in which the scalars $z^{i}$ and $\overline{z}^{\overline{i}}$ and the
\textit{central charge function} $Z$ are understood to be evaluated at the
BH horizon. Thence, we can proceed to solve for the scalars, \textit{%
stabilized} at the BH horizon in terms of the BH charges; by rewriting Eqs. (%
\ref{sat-night-1}) in components, one achieves the following result:
\begin{eqnarray}
\left( \Lambda ,\Sigma \right) &=&\left( 0,0\right) :q_{0}+ip^{0}=2\
e^{K/2}Z;  \notag \\
\left( \Lambda ,\Sigma \right) &=&\left( 0,i\right) :q_{i}-iz^{i}p^{0}=-\
e^{K/2}Z(z^{i}+\overline{z}^{\overline{i}});  \notag \\
\left( \Lambda ,\Sigma \right) &=&\left( i,0\right)
:z^{i}q_{0}+ip^{i}=e^{K/2}Z(z^{i}+\overline{z}^{\overline{i}});  \notag \\
\left( \Lambda ,\Sigma \right) &=&\left( i,i\right)
:z^{i}(q_{i}-ip^{i})=-2e^{K/2}Z|z|^{2}.  \label{system-1}
\end{eqnarray}
The \textit{decoupling} of such $2n_{V}+2$ \textit{real} algebraic Eqs. in
terms of the $n_{V}$ \textit{complex} unknowns $z^{i}$ (the two additional
real degrees of freedom residing in the \textit{homogeneity} of degree $1$
of the system (\ref{system-1}) in BH charges) allows for an effortless
resolution, yielding the following explicit expression of the $n$ complex
scalars determining the $\frac{1}{2}$\textit{-BPS attractor scalar horizon
configurations}:
\begin{equation}
z_{BPS}^{i}=-\frac{\left( q_{i}+ip^{i}\right) }{q_{0}-ip^{0}},~\forall
i=1,...,n.  \label{1/2-BPS-stabilized}
\end{equation}
Notice that \textit{all} $n_{V}$ complex scalars $z^{i}$ are stabilized in
terms of the BH charges, and thus, as well known, no classical moduli space
for $\frac{1}{2}$-BPS attractors exists at all. By recalling Eqs. (\ref
{BH-entropy-I2}) and (\ref{VBH-N=2-quadr.}), and plugging Eqs. (\ref
{1/2-BPS-stabilized}) into Eqs. (\ref{Z}) and (\ref{DiZ}), one obtains that
\begin{equation}
\frac{S_{BH,BPS}}{\pi }=\frac{A_{H,BPS}}{4}=\left. V_{BH}\right|
_{BPS}=\alpha _{1,BPS}^{2}=\left| Z\right| _{BPS}^{2}=\mathcal{I}_{2}>0,
\end{equation}
where $\mathcal{I}_{2}$ is the (unique) invariant of the \textit{%
fundamental/anti-fundamental} $\left( \mathbf{1+n},\overline{\mathbf{1+n}}%
\right) $ \ representation of the $U$-duality group $G_{\mathcal{N}=2,mc,n}$
(not \textit{irreducible} with respect to $G_{\mathcal{N}=2,mc,n}$ itself),
\textit{quadratic} in BH charges (see Eq. (\ref{ven2}) below):
\begin{equation}
\mathcal{I}_{2}=\frac{1}{2}\left[ q_{0}^{2}-q_{i}^{2}+\left( p^{0}\right)
^{2}-\left( p^{i}\right) ^{2}\right] =\frac{1}{2}\left( q^{2}+p^{2}\right) ,
\label{I2-N=2-quadr.-BH-charges}
\end{equation}
where $q^{2}\equiv \eta ^{\Lambda \Sigma }q_{\Lambda }q_{\Sigma }$ and $%
p^{2}\equiv \eta _{\Lambda \Sigma }p^{\Lambda }p^{\Sigma }$, $\eta ^{\Lambda
\Sigma }=\eta _{\Lambda \Sigma }$ being the $\left( 1+n\right) $-dim.
Lorentzian metric with signature $\left( +,-,...,-\right) $ (see the
discussion in Sect. \ref{Invariance-Props}). In terms of the dressed charges
$Z$ and $D_{i}Z$, the (\textit{only apparently} moduli-dependent) expression
of $\mathcal{I}_{2}$ reads (see \textit{e.g.} \cite{BFGM1}):
\begin{equation}
\mathcal{I}_{2}=\left| Z\right| ^{2}-G^{i\overline{j}}\left( D_{i}Z\right)
\overline{D}_{\overline{j}}\overline{Z}=\alpha _{1}^{2}-\alpha _{2}^{2}.
\label{I2-N=2-quadr.-dressed-charges}
\end{equation}
It can be explicitly checked that $z_{BPS}^{i}$ given by Eqs. (\ref
{1/2-BPS-stabilized}) satisfy the \textit{metric constraint} $1-\left|
z^{i}\right| ^{2}>0$ (yielded by Eq. (\ref{K})).

It is well known that $\frac{1}{2}$-BPS critical points of $V_{BH}$ are
\textit{stable}, \textit{at least} as far as the metric $G_{i\overline{j}}$
of the special K\"{a}hler scalar manifold is positive definite (at such
points); indeed, the $2n_{V}\times 2n_{V}$ (covariant) Hessian matrix $%
\mathcal{H}_{BPS}^{V_{BH}}$ of $V_{BH}$ at its $\frac{1}{2}$-BPS critical
points has rank $2n_{V}$, and it reads \cite{FGK}:
\begin{eqnarray}
\mathcal{H}_{BPS}^{V_{BH}} &=&2\left| Z\right| _{BPS}^{2}\left(
\begin{array}{ccc}
0 &  & G_{i\overline{j}} \\
&  &  \\
G_{j\overline{i}} &  & 0
\end{array}
\right) _{BPS}=\left[ q_{0}^{2}-q_{k}^{2}+\left( p^{0}\right) ^{2}-\left(
p^{k}\right) ^{2}\right] \cdot  \notag \\
&&  \notag \\
&&  \notag \\
&&\cdot \left(
\begin{array}{ccc}
0 &  & e^{K_{BPS}}\delta _{i\overline{j}}+e^{2K_{BPS}}\overline{z}_{i,BPS}z_{%
\overline{j},BPS} \\
&  &  \\
e^{K_{BPS}}\delta _{j\overline{i}}+e^{2K_{BPS}}\overline{z}_{j,BPS}z_{%
\overline{i},BPS} &  & 0
\end{array}
\right) ,  \notag \\
&&  \label{wed-3}
\end{eqnarray}
where use of the stabilization Eqs. (\ref{1/2-BPS-stabilized}) was made.

\subsubsection{Non-BPS ($Z=0$) Attractors and their Moduli Space\label%
{non-BPS-Z=0-Attractors}}

As yielded by Eqs. (\ref{non-BPS-Z=0}), non-BPS ($Z=0)$ attractor solutions
are given by $D_{i}Z\neq 0$ for \textit{at least} some $i\in \left\{
1,...,n_{V}\right\} $, and by the vanishing of the central charge $Z$, which
in the considered theory reads as follows (within the \textit{metric
constraint} $1-\left| z\right| ^{2}>0$; recall Eq. (\ref{Z})):
\begin{equation}
Z=0\Leftrightarrow q_{0}+ip^{0}=-\left( q_{i}-ip^{i}\right) z_{non-BPS}^{i}.
\label{wed-1}
\end{equation}
As noticed above,this is one \textit{complex} Eq. in terms of $n_{V}$
\textit{complex} unknowns $z^{i}$, thus at most \textit{only one} of them
will be stabilized in terms of the BH charges. Indeed, one can choose,
without any loss of generality, to solve Eq. (\ref{wed-1}) for $z^{1}$,
getting ($\widehat{i}=2,...,n_{V}$):
\begin{equation}
z_{non-BPS}^{1}=-\frac{\left( q_{\widehat{i}}-ip^{\widehat{i}}\right)
z_{non-BPS}^{\widehat{i}}}{q_{1}-ip^{1}}-\frac{q_{0}+ip^{0}}{q_{1}-ip^{1}}.
\label{wed-2}
\end{equation}
The remaining scalars $z^{\widehat{i}}$ are \textit{not stabilized} at the
considered (\textit{non-degenerate}) non-BPS $Z=0$ critical points of $%
V_{BH} $. As known from group-theoretical arguments (see Table 3 of \cite
{ferrara4}), such scalars span a moduli space given by the \textit{rank-}$1$
symmetric special K\"{a}hler manifold
\begin{equation}
\mathcal{M}_{\mathcal{N}=2,mc,n,non-BPS}=\frac{SU\left( 1,n-1\right) }{%
SU\left( n-1\right) \times U\left( 1\right) }=\mathcal{M}_{\mathcal{N}%
=2,mc,n-1},dim_{\mathbb{R}}=2\left( n-1\right) .
\end{equation}
The unique element of the sequence $\mathcal{M}_{\mathcal{N}=2,mc,n}$, $n\in
\mathbb{N}$, in which the non-BPS $Z=0$ attractors have no associated moduli
space is the $n=1$ case (the so-called $t^{2}$ model), in which all \textit{%
non-degenerate} critical points of $V_{BH}$ are stable, with no ``flat''
directions at all.

The existence of $n-1$ ``flat'' directions at all orders in the (covariant)
differentiation of $V_{BH}$ at its \textit{non-degenerate} non-BPS $Z=0$
critical points in the considered theory can be realized also by the
following argument.

Firstly, it can be explicitly computed that the application of an odd number
of covariant differential operators on $V_{BH}$ always yields a vanishing
result (here the tilded indices can be either holomorphic or
anti-holomorphic; $m\in \mathbb{N}$ throughout):
\begin{equation}
\left( D_{\widetilde{i}_{1}}D_{\widetilde{i}_{2}}...D_{\widetilde{i}%
_{2m-1}}V_{BH}\right) _{non-BPS}=0.
\end{equation}
Thence, the $2n_{V}\times 2n_{V}$ (covariant) Hessian matrix $\mathcal{H}%
_{non-BPS}^{V_{BH}}$ of $V_{BH}$ at its non-BPS $Z=0$ critical points can be
computed to be:
\begin{equation}
\mathcal{H}_{non-BPS}^{V_{BH}}=2\left(
\begin{array}{ccc}
0 &  & \left( D_{i}Z\right) \overline{D}_{\overline{j}}\overline{Z} \\
&  &  \\
\left( D_{j}Z\right) \overline{D}_{\overline{i}}\overline{Z} &  & 0
\end{array}
\right) _{non-BPS},
\end{equation}
and it has thus rank $2$, with $2$ strictly positive and $2n_{V}-2$
vanishing real eigenvalues (\textit{massless} ``Hessian modes'').

In order to investigate the persistence of such $2n_{V}-2$ \textit{massless}
``Hessian modes'' to higher order in the covariant differentiation of $%
V_{BH} $, one can define a \textit{``putative'' mass matrix} $\mathcal{H}%
_{m}^{V_{BH}}$ for scalars, such that $\mathcal{H}_{m=0}^{V_{BH}}=\mathcal{H}%
^{V_{BH}}$ (covariant Hessian matrix of $V_{BH}$), in the following way:
\begin{equation}
\mathcal{H}_{m}^{V_{BH}}\equiv \left( D_{\widetilde{i}}D_{\widetilde{j}}D_{%
\widetilde{i}_{1}}...D_{\widetilde{i}_{2m}}V_{BH}\right) Z^{\widetilde{i}%
_{1}}...Z^{\widetilde{i}_{2m}},
\end{equation}
where $Z^{\widetilde{i}}$ denotes the relevant contravariant \textit{matter
charge}. It can be thus calculated that
\begin{equation}
\mathcal{H}_{m,non-BPS}^{V_{BH}}=2^{2m}\left( V_{BH,non-BPS}\right) ^{m}%
\mathcal{H}_{non-BPS}^{V_{BH}}.
\end{equation}
Therefore, regardless of $m$ the \textit{``putative'' mass matrix} $\mathcal{%
H}_{m}^{V_{BH}}$ has rank $2$, with $2$ strictly positive and $2n_{V}-2$
vanishing real eigenvalues, and these latter thus span a moduli space.

By recalling Eq. (\ref{DiZ}) and plugging Eq. (\ref{wed-1}) into the \textit{%
matter charges} $D_{i}Z$, one obtains:
\begin{equation}
\left. D_{i}Z\right| _{non-BPS}=\frac{q_{i}-ip^{i}}{\sqrt{2}\sqrt{1-\left|
z\right| _{non-BPS}^{2}}}.
\end{equation}
By such a result, $\mathcal{H}_{non-BPS}^{V_{BH}}$ can be rewritten as
follows:
\begin{eqnarray}
\mathcal{H}_{non-BPS}^{V_{BH}} &=&\frac{1}{1-\left| z\right| _{non-BPS}^{2}}%
\left(
\begin{array}{ccc}
0 &  & \left( q_{i}-ip^{i}\right) \left( q_{j}+ip^{j}\right) \\
&  &  \\
\left( q_{j}-ip^{j}\right) \left( q_{i}+ip^{i}\right) &  & 0
\end{array}
\right) .  \notag \\
&&
\end{eqnarray}
Due to Eqs. (\ref{wed-1}) and (\ref{wed-2}), in general the \textit{matter
charges} $D_{i}Z$ and $\mathcal{H}^{V_{BH}}$ are not stabilized in terms of
the BH charges at the considered non-BPS $Z=0$ critical points, but
nevertheless this does \textit{not} affect the \textit{moduli-independence}
of the BH entropy. Indeed, by recalling Eqs. (\ref{BH-entropy-I2}) and (\ref
{VBH-N=2-quadr.}), and plugging Eqs. (\ref{non-BPS-Z=0}) and (\ref{wed-1})
into Eqs. (\ref{Z}) and (\ref{DiZ}), one obtains that
\begin{eqnarray}
\frac{S_{BH,non-BPS}}{\pi } &=&\frac{A_{H,non-BPS}}{4}=\left. V_{BH}\right|
_{non-BPS}=\alpha _{2,non-BPS}^{2}=  \notag \\
&&  \notag \\
&=&\left[ G^{i\overline{j}}\left( D_{i}Z\right) \overline{D}_{\overline{j}}%
\overline{Z}\right] _{non-BPS}=\left[ G^{i\overline{j}}\left( \partial
_{i}Z\right) \overline{\partial }_{\overline{j}}\overline{Z}\right]
_{non-BPS}=  \notag \\
&&  \notag \\
&=&-\mathcal{I}_{2}>0,
\end{eqnarray}
where $\mathcal{I}_{2}$ is the (unique) \textit{quadratic} $G_{\mathcal{N}%
=2,mc,n}$-invariant given by Eqs. (\ref{I2-N=2-quadr.-BH-charges}) and(\ref
{I2-N=2-quadr.-dressed-charges}).

Thus, in $\mathcal{N}=2$, $d=4$ supergravity \textit{minimally coupled} to
Abelian vector multiplets, the BH charges supporting \textit{non-degenerate}
critical points of $V_{BH}$ are split in two branches: the ($\frac{1}{2}$%
-)BPS one, defined by $\mathcal{I}_{2}>0$, and the non-BPS ($Z=0$) one,
corresponding to $\mathcal{I}_{2}<0$.

\subsection{\label{N=2-d=4-quadratic-BPS-Attractor-Flow}Black Hole
Parameters for $\frac{1}{2}$-BPS Flow}

By using the explicit expressions of $\mathcal{W}_{BPS}^{2}$ given by Eq. (%
\ref{W-BPS}), using the differential relations of special K\"{a}hler
geometry of $\mathcal{M}_{\mathcal{N}=2,mc,n}$ (see \textit{e.g.} \cite{4},
and Refs. therein), and exploiting the \textit{first order (fake
supergravity) formalism} discussed in Sect. \ref{First-Order}, one
respectively obtains the following expressions of the \textit{(square) ADM
mass}, \textit{covariant scalar charges} and \textit{(square) effective
horizon radius} for the $\frac{1}{2}$-BPS attractor flow\footnote{%
Throughout the whole paper, for \textit{all} the considered functions $%
f\left( z,\overline{z},p,q\right) $ we assume:
\begin{equation*}
\left( f\left( z,\overline{z},p,q\right) \right) _{\infty }\equiv \lim_{\tau
\rightarrow 0^{-}}f\left( z\left( \tau \right) ,\overline{z}\left( \tau
\right) ,p,q\right) =f\left( z_{\infty },\overline{z}_{\infty },p,q\right) .
\end{equation*}
Furthermore, we assume $f\left( z,\overline{z},p,q\right) $ to be \textit{%
smooth} enough to split the \textit{asymptotical limit} of a product into
the product of the \textit{asymptotical limits} of the factors.} \cite{FHM}:
\begin{eqnarray}
r_{H,BPS}^{2}\left( z_{\infty },\overline{z}_{\infty },p,q\right)
&=&M_{ADM,BPS}^{2}\left( z_{\infty },\overline{z}_{\infty },p,q\right) =%
\mathcal{W}_{BPS}^{2}\left( z_{\infty },\overline{z}_{\infty },p,q\right) =
\notag \\
&&  \notag \\
&=&\lim_{\tau \rightarrow 0^{-}}\left| Z\right| ^{2}\left( z\left( \tau
\right) ,\overline{z}\left( \tau \right) ,p,q\right) =  \notag \\
&&  \notag \\
&=&\frac{\left[ q_{0}+ip^{0}+\left( q_{i}-ip^{i}\right) z_{\infty }^{i}%
\right] \left[ q_{0}-ip^{0}+\left( q_{j}+ip^{j}\right) \overline{z}_{\infty
}^{\overline{j}}\right] }{2\left( 1-|z_{\infty }|^{2}\right) };
\label{thu-1}
\end{eqnarray}
\begin{eqnarray}
\Sigma _{i,BPS}\left( z_{\infty },\overline{z}_{\infty },p,q\right)
&=&2\lim_{\tau \rightarrow 0^{-}}\left( \partial _{i}\mathcal{W}%
_{BPS}\right) \left( z\left( \tau \right) ,\overline{z}\left( \tau \right)
,p,q\right) =  \notag \\
&&  \notag \\
&=&\frac{1}{M_{ADM,BPS}\left( z_{\infty },\overline{z}_{\infty },p,q\right) }%
\lim_{\tau \rightarrow 0^{-}}\left( \overline{Z}D_{i}Z\right) \left( z\left(
\tau \right) ,\overline{z}\left( \tau \right) ,p,q\right) =  \notag \\
&&  \notag \\
&=&\frac{1}{\sqrt{2}\left( 1-\left| z_{\infty }\right| ^{2}\right) ^{3/2}}%
\sqrt{\frac{q_{0}-ip^{0}+\left( q_{j}+ip^{j}\right) \overline{z}_{\infty }^{%
\overline{j}}}{q_{0}+ip^{0}+\left( q_{k}-ip^{k}\right) z_{\infty }^{k}}}\cdot
\notag \\
&&  \notag \\
&&  \notag \\
&&\cdot \left[ (q_{i}-ip^{i})(1-\left| z_{\infty }\right|
^{2})+(q_{0}+ip^{0})\overline{z}_{\infty }^{\overline{i}}+(q_{r}-ip^{r})z_{%
\infty }^{r}\overline{z}_{\infty }^{\overline{i}}\right] ;  \label{euro1}
\end{eqnarray}
\begin{eqnarray}
R_{H,BPS}^{2} &=&\lim_{\tau \rightarrow 0^{-}}\left[
\begin{array}{l}
\mathcal{W}_{BPS}^{2}\left( z\left( \tau \right) ,\overline{z}\left( \tau
\right) ,p,q\right) + \\
\\
-4G^{i\overline{j}}\left( z\left( \tau \right) ,\overline{z}\left( \tau
\right) \right) \left( \partial _{i}\mathcal{W}_{BPS}\right) \left( z\left(
\tau \right) ,\overline{z}\left( \tau \right) ,p,q\right) \cdot \\
\\
\cdot \left( \overline{\partial }_{\overline{j}}\mathcal{W}_{BPS}\right)
\left( z\left( \tau \right) ,\overline{z}\left( \tau \right) ,p,q\right)
\end{array}
\right] =  \notag \\
&&  \notag \\
&&  \notag \\
&=&\mathcal{I}_{2}\left( p,q\right) =V_{BH,BPS}=\frac{S_{BH,BPS}\left(
p,q\right) }{\pi }.  \label{Lun-3}
\end{eqnarray}

Eq. (\ref{Lun-3}) proves Eq. (\ref{CERN-Thu-1}) for the $\frac{1}{2}$-BPS
attractor flow of the $\mathcal{N}=2$, $d=4$ supergravity \textit{minimally
coupled} to $n\equiv n_{V}$ Abelian vector multiplets.

Notice that in the extremality regime ($c=0$) the \textit{effective horizon
radius} $R_{H}$, and thus $A_{H}$ and the Bekenstein-Hawking entropy $S_{BH}$
are \textit{independent} on the particular vacuum or ground state of the
considered theory, \textit{i.e.} on $\left( z_{\infty }^{i},\overline{z}%
_{\infty }^{\overline{i}}\right) $, but rather they depend \textit{only} on
the electric and magnetic charges $q_{\Lambda }$ and $p^{\Lambda }$, which
are \textit{conserved} due to the overall $\left( U(1)\right) ^{n+1}$
gauge-invariance. The independence on $\left( z_{\infty }^{i},\overline{z}%
_{\infty }^{\overline{i}}\right) $ is of crucial importance for the
consistency of the \textit{microscopic state counting interpretation} of $%
S_{BH}$, as well as for the overall consistency of the macroscopic
thermodynamic picture of the BH. However, it is worth recalling that the ADM
mass $M_{ADM}$ generally does depend on $\left( z_{\infty }^{i},\overline{z}%
_{\infty }^{\overline{i}}\right) $ \textit{also in the extremal case}, as
yielded by Eq. (\ref{thu-1}) for the considered $\frac{1}{2}$-BPS attractor
flow.

Furthermore, Eq. (\ref{thu-1}) yields that the $\frac{1}{2}$-BPS attractor
flow of the $\mathcal{N}=2$, $d=4$ supergravity \textit{minimally coupled}
to $n\equiv n_{V}$ Abelian vector multiplets does \textit{not }saturate the
\textit{marginal stability bound} (see \cite{Marginal-Refs} and \cite{GLS1}).

\subsection{\label{N=2-d=4-quadratic-non-BPS-Z=0-Attractor-Flow}Black Hole
Parameters for Non-BPS $\ $($Z=0$) Flow}

By using the explicit expressions of $\mathcal{W}_{non-BPS}^{2}$ given by
Eq. (\ref{W-non-BPS}), using the differential relations of special
K\"{a}hler geometry of $\mathcal{M}_{\mathcal{N}=2,mc,n}$ (see \textit{e.g.}
\cite{4} and Refs. therein), and exploiting the \textit{first order (fake
supergravity) formalism} discussed in Sect. \ref{First-Order}, one
respectively obtains the following expressions of the \textit{(square) ADM
mass}, \textit{covariant scalar charges} and \textit{(square) effective
horizon radius} for the non-BPS $Z=0$ attractor flow \cite{FHM}:

\begin{eqnarray}
r_{H,non-BPS}^{2}\left( z_{\infty },\overline{z}_{\infty },p,q\right)
&=&M_{ADM,non-BPS}^{2}\left( z_{\infty },\overline{z}_{\infty },p,q\right) =%
\mathcal{W}_{non-BPS}^{2}\left( z_{\infty },\overline{z}_{\infty
},p,q\right) =  \notag \\
&&  \notag \\
&=&\lim_{\tau \rightarrow 0^{-}}\left[ G^{i\overline{j}}\left( D_{i}Z\right)
\overline{D}_{\overline{j}}\overline{Z}\right] \left( z\left( \tau \right) ,%
\overline{z}\left( \tau \right) ,p,q\right) =  \notag \\
&&  \notag \\
&=&\frac{1}{2\left( 1-\left| z_{\infty }\right| ^{2}\right) ^{2}}\left(
\delta ^{i\overline{j}}-z_{\infty }^{i}\overline{z}_{\infty }^{\overline{j}%
}\right) \cdot  \notag \\
&&  \notag \\
&&\cdot \left[ (q_{i}-ip^{i})(1-\left| z_{\infty }\right|
^{2})+(q_{0}+ip^{0})\overline{z}_{\infty }^{\overline{i}}+(q_{r}-ip^{r})z_{%
\infty }^{r}\overline{z}_{\infty }^{\overline{i}}\right] \cdot  \notag \\
&&  \notag \\
&&\cdot \left[ (q_{j}+ip^{j})(1-\left| z_{\infty }\right|
^{2})+(q_{0}-ip^{0})z_{\infty }^{j}+(q_{n}+ip^{n})\overline{z}_{\infty }^{%
\overline{n}}z_{\infty }^{j}\right] ;  \notag \\
&&  \label{Sat-aft-1}
\end{eqnarray}
\begin{eqnarray}
\Sigma _{i,non-BPS}\left( z_{\infty },\overline{z}_{\infty },p,q\right)
&=&2\lim_{\tau \rightarrow 0^{-}}\left( \partial _{i}\mathcal{W}%
_{non-BPS}\right) \left( z\left( \tau \right) ,\overline{z}\left( \tau
\right) ,p,q\right) =  \notag \\
&&  \notag \\
&=&\frac{1}{M_{ADM,non-BPS}\left( z_{\infty },\overline{z}_{\infty
},p,q\right) }\lim_{\tau \rightarrow 0^{-}}\left( \overline{Z}D_{i}Z\right)
\left( z\left( \tau \right) ,\overline{z}\left( \tau \right) ,p,q\right) =
\notag \\
&&  \notag \\
&=&\frac{1}{\sqrt{2}}\frac{1}{\left( 1-|z_{\infty }|^{2}\right) }\cdot
\notag \\
&&  \notag \\
&&\cdot \left[ q_{0}-ip^{0}+\left( q_{j}+ip^{j}\right) \overline{z}_{\infty
}^{\overline{j}}\right] \cdot  \notag \\
&&  \notag \\
&&\cdot \left[ (q_{i}-ip^{i})(1-\left| z_{\infty }\right|
^{2})+(q_{0}+ip^{0})\overline{z}_{\infty }^{\overline{i}}+(q_{m}-ip^{m})z_{%
\infty }^{m}\overline{z}_{\infty }^{\overline{i}}\right] \cdot  \notag \\
&&  \notag \\
&&\cdot \left[
\begin{array}{l}
\left( \delta ^{n\overline{p}}-z_{\infty }^{n}\overline{z}_{\infty }^{%
\overline{p}}\right) \cdot \\
\\
\cdot \left[ (q_{n}-ip^{n})(1-\left| z_{\infty }\right| ^{2})+(q_{0}+ip^{0})%
\overline{z}_{\infty }^{\overline{n}}+(q_{s}-ip^{s})z_{\infty }^{s}\overline{%
z}_{\infty }^{\overline{n}}\right] \cdot \\
\\
\cdot \left[ (q_{p}+ip^{p})(1-\left| z_{\infty }\right|
^{2})+(q_{0}-ip^{0})z_{\infty }^{p}+(q_{w}+ip^{w})\overline{z}_{\infty }^{%
\overline{w}}z_{\infty }^{p}\right]
\end{array}
\right] ^{-1/2}  \notag \\
&&  \label{euro2}
\end{eqnarray}
\begin{eqnarray}
R_{H,non-BPS}^{2} &=&\lim_{\tau \rightarrow 0^{-}}\left[
\begin{array}{l}
\mathcal{W}_{non-BPS}^{2}\left( z\left( \tau \right) ,\overline{z}\left(
\tau \right) ,p,q\right) + \\
\\
-4G^{i\overline{j}}\left( z\left( \tau \right) ,\overline{z}\left( \tau
\right) \right) \left( \partial _{i}\mathcal{W}_{non-BPS}\right) \left(
z\left( \tau \right) ,\overline{z}\left( \tau \right) ,p,q\right) \cdot \\
\\
\cdot \left( \overline{\partial }_{\overline{j}}\mathcal{W}_{non-BPS}\right)
\left( z\left( \tau \right) ,\overline{z}\left( \tau \right) ,p,q\right)
\end{array}
\right] =  \notag \\
&&  \notag \\
&&  \notag \\
&=&-\mathcal{I}_{2}\left( p,q\right) =V_{BH,non-BPS}=\frac{%
S_{BH,non-BPS}\left( p,q\right) }{\pi }.  \label{Lun-4}
\end{eqnarray}

Eq. (\ref{Lun-4}) proves Eq. (\ref{CERN-Thu-1}) for the non-BPS $Z=0$
attractor flow of the the $\mathcal{N}=2$, $d=4$ supergravity \textit{%
minimally coupled} to $n\equiv n_{V}$ Abelian vector multiplets. The
considerations made at the end of Subsect. \ref
{N=2-d=4-quadratic-BPS-Attractor-Flow} hold also for the considered
attractor flow.

It is worth noticing out that Eqs. (\ref{Lun-3}) and (\ref{Lun-4}) are
consistent, because, as pointed out above, the ($\frac{1}{2}$-)BPS- and
non-BPS ($Z=0$)- supporting BH charge configurations in the considered
theory are respectively defined by the \textit{quadratic} constraints $%
\mathcal{I}_{2}\left( p,q\right) >0$ and $\mathcal{I}_{2}\left( p,q\right)
<0 $.

As yielded by Eqs. (\ref{euro1}) and (\ref{euro2}) for both \textit{%
non-degenerate} attractor flows of the considered theory it holds the
following relation among \textit{scalar charges} and \textit{ADM mass}:
\begin{equation}
\Sigma _{i}=\frac{1}{M_{ADM}}\lim_{\tau \rightarrow 0^{-}}D_{i}\left( \left|
Z\right| ^{2}\right) .
\end{equation}

Furthermore, Eq. (\ref{Sat-aft-1}) yields that the non-BPS $Z=0$ attractor
flow of the $\mathcal{N}=2$, $d=4$ supergravity \textit{minimally coupled}
to $n\equiv n_{V}$ Abelian vector multiplets does \textit{not }saturate the
\textit{marginal stability bound} (see \cite{Marginal-Refs} and \cite{GLS1}).

As it will be proved in the next Sections, for \textit{all} non-degenerate
attractor flows of the considered $d=4$ supergravities the \textit{marginal
stability bound} is \textit{not} saturated. A more detailed discussion of
such an issue falls beyond the scope of the present investigation, and it
will be given elsewhere \cite{stu-Yeranyan}.

\section{\label{N=3,d=4}$\mathcal{N}=3$ Supergravity}

The (K\"{a}hler) scalar manifold is \cite{N=3-Ref}
\begin{equation}
\mathcal{M}_{\mathcal{N}=3,n}=\frac{G_{\mathcal{N}=3,n}}{H_{\mathcal{N}=3,n}}%
=\frac{SU\left( 3,n\right) }{SU\left( 3\right) \times SU\left( n\right)
\times U\left( 1\right) },~dim_{\mathbb{R}}=6n.
\end{equation}

The $3+n$ vector field strengths and their duals, as well as their
asymptotical fluxes, sit in the \textit{fundamental} $\mathbf{3+n}$ \
representation of the $U$-duality group $G_{\mathcal{N}=3,n}=SU\left(
3,n\right) $, in turn embedded in the symplectic group $Sp\left( 6+2n,%
\mathbb{R}\right) $.

$Z_{AB}=Z_{\left[ AB\right] }$ ($A,B=1,2,3=\mathcal{N}$) is the \textit{%
central charge matrix}, and $Z_{I}$ ($I=1,...,n$) are the \textit{matter
charges}, where $n\in \mathbb{N}$ is the number of matter (Abelian vector)
multiplets coupled to the gravity multiplet. By a suitable transformation of
the $\mathcal{R}$-symmetry $U\left( 3\right) $, $Z_{AB}$ can be \textit{%
skew-diagonalized} by putting it in the \textit{normal form} (see e.g. \cite
{ADOT-1} and Refs. therein):
\begin{equation}
Z_{AB}=\left(
\begin{array}{cc}
\mathcal{Z}_{1}\epsilon &  \\
& 0
\end{array}
\right) ,
\end{equation}
where $\epsilon $ is the $2\times 2$ symplectic metric, and $\mathcal{Z}%
_{1}\in \mathbb{R}_{0}^{+}$ is the unique $\mathcal{N}=3$ (moduli-dependent)
\textit{skew-eigenvalue}, which can be expressed in term of the unique $%
U\left( 3\right) $(and also $H_{\mathcal{N}=3,n}=U\left( 3\right) \times
U\left( n\right) $)-invariant as follows:
\begin{equation}
\mathcal{Z}_{1}=\sqrt{\frac{1}{2}Z_{AB}\overline{Z}^{AB}}.  \label{salsa-1}
\end{equation}
On the other hand, by a suitable rotation of $U\left( n\right) $, the vector
$Z_{I}$ of matter charges can be chosen real and pointing in a given
direction, \textit{e.g.}
\begin{equation}
Z_{I}=\rho \delta _{I1},
\end{equation}
where $\rho $ can be expressed the unique $U\left( n\right) $(and also $H_{%
\mathcal{N}=3,n}=U\left( 3\right) \times U\left( n\right) $)-invariant as:
\begin{equation}
\rho =\sqrt{Z_{I}\overline{Z}^{I}}.  \label{salsa-2}
\end{equation}

The simplest holomorphic parametrization of $\mathcal{M}_{\mathcal{N}=3,n}$
can be written in terms of the $\left( 3+n\right) \times \left( 3+n\right) $
coset representative \cite{Helgason,Gilmore}
\begin{equation}
L=\left(
\begin{array}{cc}
\sqrt{1+XX^{\dag }} & X \\
X^{\dag } & \sqrt{1+X^{\dag }X}
\end{array}
\right) \ ,
\end{equation}
where $X$ is a complex $n\times 3$ matrix in the bi-fundamental of $%
SU(3)\times SU\left( n\right) =H_{\mathcal{N}=3,n}\backslash U(1)$, whose
component are nothing but the $3n$ complex scalars $z^{i}$ ($i=1,...,3n$)
spanning $\mathcal{M}_{\mathcal{N}=3,n}$. The embedding of $G_{\mathcal{N}%
=3,n}$ into the symplectic group $Sp\left( 6+2n,\mathbb{R}\right) $:
\begin{equation}
SU(3,n)\rightarrow Sp\left( 6+2n,\mathbb{R}\right) \ ,~g\equiv
L(z)\rightarrow S(g)\equiv S(L(z))\ ,
\end{equation}
is determined by the $\left( 6+2n\right) \times \left( 6+2n\right) $ matrix
\begin{equation}
S(g)=\left(
\begin{array}{cc}
\phi _{0} & \overline{\phi _{1}} \\
\phi _{1} & \overline{\phi _{0}}
\end{array}
\right) \ \in SU(3,n)\subset Sp\left( 6+2n,\mathbb{R}\right) ,
\end{equation}
such that the $\left( 3+n\right) \times \left( 3+n\right) $ sub-blocks $\phi
_{0}$ and $\phi _{1}$ satisfy the relations
\begin{equation}
\phi _{0}^{\dag }\phi _{0}-\phi _{1}^{\dag }\phi _{1}=1\ ,~\phi _{0}^{\dag }%
\overline{\phi }_{1}-\phi _{1}^{\dag }\overline{\phi }_{0}=0\ .
\label{Fri-1}
\end{equation}
Let us here recall that in the \textit{Gaillard-Zumino formalism}\textbf{\ }
\cite{Gaillard-Zumino-1}, the vector kinetic matrix can be written as ($%
\Lambda =1,2,3,4,...,3+n$ throughout all the present Section)
\begin{equation}
\mathcal{N}_{\Lambda \Sigma }=(\phi _{0}^{\dag }+\phi _{1}^{\dag
})^{-1}(\phi _{0}^{\dag }-\phi _{1}^{\dag })\ .
\end{equation}
The embedding $SU(3,n)\rightarrow Sp\left( 6+2n,\mathbb{R}\right) $ is
determined once $S$ is written as a functions of $X\left( z\right) $, namely
\begin{equation}
S\left( X\right) =\left(
\begin{array}{cc|cc}
\sqrt{1+XX^{\dag }} & 0 & 0 & X \\
0 & \sqrt{1+X^{T}\overline{X}} & X^{T} & 0 \\ \hline
0 & \overline{X} & \sqrt{1+\overline{X}X^{T}} & 0 \\
X^{T} & 0 & 0 & \sqrt{1+X^{\dag }X}
\end{array}
\right) \ ,
\end{equation}
that is
\begin{eqnarray}
\phi _{1} &=&\left(
\begin{array}{cc}
0 & \overline{X} \\
X^{\dag } & 0
\end{array}
\right) \ ; \\
&&  \notag \\
\phi _{0} &=&\sqrt{1+\overline{\phi }_{1}\phi _{1}}=\left(
\begin{array}{cc}
\sqrt{1+XX^{\dag }} & 0 \\
0 & \sqrt{1+X^{T}\overline{X}}
\end{array}
\right) \ .
\end{eqnarray}

The vector kinetic matrix $\mathcal{N}_{\Lambda \Sigma }$ can be written in
terms of the $\left( 3+n\right) \times \left( 3+n\right) $ symplectic
sections (and their inverse) as follows (see \textit{e.g.} \cite{4}, and
Refs. therein):
\begin{equation}
\mathcal{N}_{\Lambda \Sigma }=h_{\Omega \Lambda }\left( f^{-1}\right)
_{\Sigma }^{\Omega }.  \label{Thu-aft-1}
\end{equation}
The explicit dependence of the symplectic sections on the sub-blocks of $%
S\left( X\right) $ is simply
\begin{equation}
h_{\Lambda \Sigma }=-\frac{i}{\sqrt{2}}\left( \phi _{0}-\phi _{1}\right)
,~f_{\Sigma }^{\Lambda }=\frac{1}{\sqrt{2}}\left( \phi _{0}+\phi _{1}\right)
,
\end{equation}
whereas in terms of the matrix $X\left( z\right) $ they read
\begin{eqnarray}
h_{\Lambda \Sigma } &=&-\frac{i}{\sqrt{2}}\left(
\begin{array}{cc}
\sqrt{1+XX^{\dag }} & -\overline{X} \\
-X^{\dag } & \sqrt{1+X^{T}\overline{X}}
\end{array}
\right) \equiv \left( h_{\Lambda \mid AB},\overline{h}_{\Lambda \mid
I}\right) \equiv \mathbf{h};  \label{h} \\
&&  \notag \\
f_{\Sigma }^{\Lambda } &=&\frac{1}{\sqrt{2}}\left(
\begin{array}{cc}
\sqrt{1+XX^{\dag }} & \overline{X} \\
X^{\dag } & \sqrt{1+X^{T}\overline{X}}
\end{array}
\right) \equiv \left( f_{AB}^{\Lambda },\overline{f}_{\overline{I}}^{\Lambda
}\right) \equiv \mathbf{f}.  \label{f}
\end{eqnarray}
By rewriting Eqs. (\ref{Fri-1}) in terms of the symplectic sections, one
finds \cite{ADF1,ADF2}
\begin{eqnarray}
i\left( \mathbf{f}^{\dag }\mathbf{h}-\mathbf{h}^{\dag }\mathbf{f}\right)
&=&1;  \label{Mon-5} \\
&&  \notag \\
\mathbf{h}^{T}\mathbf{f}-\mathbf{hf}^{T} &=&0.  \label{Mon-6}
\end{eqnarray}

The \textit{central charge matrix} $Z_{AB}$ and the \textit{matter charges} $%
Z_{I}$ are respectively defined as the integral over the $2$-sphere at
infinity $S_{\infty }^{2}$ of the\textit{\ dressed} graviphoton and matter
field strengths \cite{ADF1,ADF2,ADF-Duality-d=4}:
\begin{eqnarray}
Z_{AB} &\equiv &-\int_{S_{\infty }^{2}}T_{AB}=-\int_{S_{\infty
}^{2}}T_{AB}^{-}=f_{AB}^{\Lambda }q_{\Lambda }-h_{\Lambda \mid AB}p^{\Lambda
}\ ;  \label{Wed-8} \\
&&  \notag \\
Z_{I} &\equiv &-\int_{S_{\infty }^{2}}T_{I}=-\int_{S_{\infty
}^{2}}T_{I}^{-}=f_{I}^{\Lambda }q_{\Lambda }-h_{\Lambda \mid I}p^{\Lambda }\
.
\end{eqnarray}
Using the explicit expression for the symplectic sections given in Eqs. (\ref
{h}) and (\ref{f}), one obtains
\begin{eqnarray}
Z_{AB} &=&\frac{1}{\sqrt{2}}\left[ \left( \sqrt{1+XX^{\dag }}\right)
_{AB}^{C}(q_{C}+ip^{C})+\overline{X}_{AB}^{i}(q_{i}-ip^{i})\right] \ ;
\label{carMatN=3} \\
&&  \notag \\
Z_{I} &=&\frac{1}{\sqrt{2}}\left[ (X^{\dag })_{I}^{A}(q_{A}-ip^{A})+\left(
\sqrt{1+X^{T}\overline{X}}\right) _{I}^{\ i}(q_{i}+ip^{i})\right] \ .
\label{carMatN=3-2}
\end{eqnarray}

As recalled at the start of the next Subsection, only ($\frac{1}{3}$-)BPS
and non-BPS ($Z_{AB}=0$) attractor flows are \textit{non-degenerate} (%
\textit{i.e.} corresponding to \textit{large} BHs), and the corresponding
(square) \textit{first order fake superpotentials} are (\cite{ADOT-1};
recall Eq. (\ref{salsa-1}) and (\ref{salsa-2}), respectively)
\begin{eqnarray}
\mathcal{W}_{\left( \frac{1}{3}-\right) BPS}^{2} &=&\frac{1}{2}Z_{AB}%
\overline{Z}^{AB}=\mathcal{Z}_{1}^{2}=  \notag \\
&&  \notag \\
&=&\frac{1}{2}\left[
\begin{array}{l}
\left[ (q_{C}-ip^{C})\left( \sqrt{1+XX^{\dag }}\right) ^{\
C}+(q_{i}+ip^{i})\left( X^{T}\right) ^{i}\right] \cdot \\
\\
\cdot \left[ \left( \sqrt{1+XX^{\dag }}\right) ^{\ D}(q_{D}+ip^{D})+%
\overline{X}^{j}(q_{j}-ip^{j})\right]
\end{array}
\right] =  \notag \\
&&  \notag \\
&&  \notag \\
&=&\frac{1}{2}\left[
\begin{array}{l}
(1+XX^{\dag })^{AB}(q_{A}-ip^{A})(q_{B}+ip^{B})+ \\
\\
+(\sqrt{1+XX^{\dag }}\,X)^{A\,i}(q_{i}+ip^{i})(q_{A}+ip^{A})+ \\
\\
+(X^{\dag }\sqrt{1+XX^{\dag }})^{j\,B}(q_{A}-ip^{B})(q_{j}-ip^{j})+ \\
\\
+(X^{\dag }X)^{kl}(q_{l}+ip^{l})(q_{k}-ip^{k})\
\end{array}
\right] ;  \notag \\
&&  \label{sat-1}
\end{eqnarray}
\begin{eqnarray}
\mathcal{W}_{non-BPS(,Z_{AB}=0)}^{2} &=&Z_{I}\overline{Z}^{I}=\rho ^{2}=
\notag \\
&&  \notag \\
&=&\frac{1}{2}\left[
\begin{array}{l}
\left[ (q_{D}+ip^{D})X^{D}+(q_{l}-ip^{l})\left( \sqrt{1+X^{T}\overline{X}}%
\right) ^{\ l}\right] \cdot \\
\\
\cdot \left[ (X^{\dag })^{C}(q_{C}-ip^{C})+\left( \sqrt{1+X^{T}\overline{X}}%
\right) ^{\ i}(q_{i}+ip^{i})\right]
\end{array}
\right] =  \notag \\
&&  \notag \\
&&  \notag \\
&=&\frac{1}{2}\left[
\begin{array}{l}
(XX^{\dag })^{CD}(q_{C}+ip^{C})(q_{D}-ip^{D})+ \\
\\
+(\sqrt{1+X^{\dag }X}\,X^{\dag })^{l\,C}(q_{l}-ip^{l})(q_{C}-ip^{C})+ \\
\\
+(X\sqrt{1+X^{\dag }X})^{D\,i}(q_{D}+ip^{D})(q_{i}+ip^{i})+ \\
\\
+(1+X^{\dag }X)^{l\,i}(q_{l}-ip^{l})(q_{i}+ip^{i})\ .
\end{array}
\right] ,  \notag \\
&&  \label{sat-2}
\end{eqnarray}
where Eqs. (\ref{carMatN=3}) and (\ref{carMatN=3-2}) were used. Notice that,
since all the contractions of $SU(3)$ and $SU(n)$ indices of electric and
magnetic BH charges are uniquely defined with respect to the row or columns
of the matrix $X$, every transposition index has been suppressed in Eqs. (%
\ref{sat-1}) and (\ref{sat-2}).

By introducing the \textit{complexified graviphoton} and \textit{matter} BH
charges respectively as follows:
\begin{eqnarray}
Q_{C} &\equiv &q_{C}+ip^{C};  \label{tired-1} \\
Q_{i} &\equiv &q_{i}+ip^{i},  \label{tired-2}
\end{eqnarray}
Eqs. (\ref{sat-1}) and (\ref{sat-2}) can be rewritten as follows:
\begin{eqnarray}
\mathcal{W}_{BPS}^{2} &=&\frac{1}{2}\left[
\begin{array}{l}
(1+XX^{\dag })^{AB}\overline{Q}_{A}Q_{B}+(\sqrt{1+XX^{\dag }}\,X)^{\
Ai}Q_{i}Q_{A}+ \\
\\
+(X^{\dag }\sqrt{1+XX^{\dag }})^{j\,B}\overline{Q}_{B}\overline{Q}%
_{j}+(X^{\dag }X)^{k\,l}\overline{Q}_{k}Q_{l}\
\end{array}
\right] ;  \label{sat-3} \\
&&~  \notag \\
&&~  \notag
\end{eqnarray}
\begin{equation}
\mathcal{W}_{non-BPS}^{2}=\frac{1}{2}\left[
\begin{array}{l}
(XX^{\dag })^{AB}Q_{A}\overline{Q}_{B}+(\sqrt{1+X^{\dag }X}\,X^{\dag
})^{i\,A}\overline{Q}_{i}\overline{Q}_{A}+ \\
\\
+(X\sqrt{1+X^{\dag }X})^{B\,j}Q_{B}Q_{j}+(1+X^{\dag }X)^{k\,l}\overline{Q}%
_{k}Q_{l}
\end{array}
\right] .  \label{sat-4}
\end{equation}

\subsection{\label{N=3,d=4,AES}Attractor Equations and their Solutions}

The \textit{BH effective potential} can be written as
\begin{equation}
V_{BH}=\frac{1}{2}Z_{AB}\overline{Z}^{AB}+Z_{I}\overline{Z}^{I}=\mathcal{Z}%
_{1}^{2}+\rho ^{2}.  \label{VBH}
\end{equation}
The $\mathcal{N}=3$, $d=4$ \textit{Attractor Eqs.} are nothing but the
\textit{criticality conditions} for such an $H_{\mathcal{N}=3,n}$-invariant
(and K\"{a}hler-gauge-invariant) quantity. Such criticality conditions are
satisfied for two classes of critical points :

\begin{itemize}
\item  ($\frac{1}{3}$-)BPS:
\begin{eqnarray}
Z_{I} &=&0~\forall I=1,...,n\Leftrightarrow \rho =0;  \notag \\
Z_{AB} &\neq &0;  \label{1/3-BPS}
\end{eqnarray}
%\newline

\item  non-supersymmetric (non-BPS with $Z_{AB}=0$):
\begin{eqnarray}
Z_{I} &\neq &0~(\text{\textit{at least~}for~some~}I);  \notag \\
Z_{AB} &=&0\Leftrightarrow \mathcal{Z}_{1}=0.  \label{non-BPS-Z=0-N=3}
\end{eqnarray}
\end{itemize}

It is worth counting here the degrees of freedom related to Eqs. (\ref
{1/3-BPS}) and (\ref{non-BPS-Z=0-N=3}).

The $\frac{1}{3}$-BPS criticality conditions\ (\ref{1/3-BPS}) are $n$
\textit{complex} independent ones, thus a moduli space of $\frac{1}{3}$-BPS
attractors, spanned by the $2n$ \textit{complex} scalars unstabilized by Eq.
(\ref{1/3-BPS}) might - and actually does \cite{BFGM1} - exist.

Furthermore, there are only three \textit{complex} non-BPS $Z=0$ criticality
conditions\ (\ref{non-BPS-Z=0-N=3}). This fact paves the way to the
possibility to have a \textit{moduli space} of non-BPS ($Z_{AB}=0$)
attractors, spanned by the $3\left( n-1\right) $ \textit{complex} scalars
unstabilized by Eq. (\ref{non-BPS-Z=0-N=3}); this actually holds true \cite
{ferrara4}, as it will be explicitly found below (see Subsection \ref
{N=3,non-BPS,Z=0-Attractors}).

\subsubsection{\label{1/3-BPS-Attractors}$\frac{1}{3}$-BPS Attractors and
their Moduli Space}

By inserting the ($\frac{1}{3}$-)BPS criticality conditions (\ref{1/3-BPS})
into the expression (\ref{carMatN=3-2}) of the \textit{matter charges} $%
Z_{I} $ and recalling the definitions (\ref{tired-1}) and (\ref{tired-2}) of
the \textit{complexified} BH charges, one obtains
\begin{equation}
(X^{\dag })_{I}^{A}\overline{Q}_{A}=-\left( \sqrt{1+X^{T}\overline{X}}%
\right) _{I}^{\ i}Q_{i},~\forall I=1,...,n.  \label{tired-4}
\end{equation}
By plugging such an expression and its Hermitian conjugate into Eq. (\ref
{sat-3}), and using the identity (holding for any matrix $A$)
\begin{equation}
A^{\dag }\sqrt{1+AA^{\dag }}=\sqrt{1+A^{\dag }A}\ A^{\dag }\   \label{id1}
\end{equation}
and its Hermitian conjugate
\begin{equation}
\sqrt{1+AA^{\dag }}\ A=A\sqrt{1+A^{\dag }A}\ ,  \label{id2}
\end{equation}
one obtains that

\begin{eqnarray}
\left( Z_{AB}\overline{Z}^{AB}\right) _{BPS} &=&(1+XX^{\dag })^{AB}\overline{%
Q}_{A}Q_{B}+(X\sqrt{1+X^{\dag }X})^{\,Ai}Q_{i}Q_{A}+  \notag \\
&&  \notag \\
&&+(\sqrt{1+X^{\dag }X}X^{\dag })^{j\,B}\overline{Q}_{B}\overline{Q}%
_{j}+(X^{\dag }X)^{kl}\overline{Q}_{k}Q_{l}=  \notag \\
&&  \notag \\
&=&Q_{A}\overline{Q}_{A}+(XX^{\dag })^{AB}\overline{Q}_{A}Q_{B}-(XX^{\dag
})^{AB}Q_{A}\overline{Q}_{B}+  \notag \\
&&  \notag \\
&&-(1+X^{\dag }X)^{ij}\overline{Q}_{i}Q_{j}+(X^{\dag }X)^{ij}\overline{Q}%
_{i}Q_{j}=  \notag \\
&&  \notag \\
&=&Q_{A}\overline{Q}_{A}-Q_{i}\overline{Q}_{i}\ ,  \label{tired-3}
\end{eqnarray}
where in the last step we exploited the Hermiticity of $XX^{\dag }$,
yielding that ($\left\langle \cdot ,\cdot \right\rangle _{XX^{\dag }}$
denotes the $XX^{\dag }$-dependent square norm of \textit{complexified} BH
charges)
\begin{equation}
(XX^{\dag })^{AB}Q_{A}\overline{Q}_{B}\equiv \langle Q,Q\rangle _{XX^{\dag
}}=\langle \overline{Q},\overline{Q}\rangle _{XX^{\dag }}=(XX^{\dag })^{AB}%
\overline{Q}_{A}Q_{B}\ .  \label{tired-5}
\end{equation}
By recalling Eqs. (\ref{BH-entropy-I2}) and (\ref{VBH}), and using Eq. (\ref
{tired-3}), one achieves the following result:
\begin{eqnarray}
\frac{S_{BH,BPS}}{\pi } &=&\frac{A_{H,BPS}}{4}=\left. V_{BH}\right| _{BPS}=%
\mathcal{Z}_{1,BPS}^{2}=\frac{1}{2}\left( Z_{AB}\overline{Z}^{AB}\right)
_{BPS}=  \notag \\
&=&\frac{1}{2}\left( Q_{A}\overline{Q}_{A}-Q_{i}\overline{Q}_{i}\right) =%
\mathcal{I}_{2}>0.
\end{eqnarray}

Here $\mathcal{I}_{2}$ denotes the (unique) invariant of the \textit{%
fundamental/anti-fundamental} $\left( \mathbf{3+n},\overline{\mathbf{3+n}}%
\right) $ \ representation of the $U$-duality group $G_{\mathcal{N}=3,n}$
(not \textit{irreducible} with respect to $G_{\mathcal{N}=3,n}$ itself),
\textit{quadratic} in BH charges (see Eq. (\ref{ven2}) below):
\begin{equation}
\mathcal{I}_{2}=\frac{1}{2}\left[ q_{A}^{2}-q_{i}^{2}+\left( p^{A}\right)
^{2}-\left( p^{i}\right) ^{2}\right] =\frac{1}{2}\left( q^{2}+p^{2}\right) ,
\label{tired-7}
\end{equation}
where $q^{2}\equiv \eta ^{\Lambda \Sigma }q_{\Lambda }q_{\Sigma }$ and $%
p^{2}\equiv \eta _{\Lambda \Sigma }p^{\Lambda }p^{\Sigma }$, $\eta ^{\Lambda
\Sigma }=\eta _{\Lambda \Sigma }$ being the $\left( 3+n\right) $-dim.
Lorentzian metric with signature $\left( 3,n\right) $ (see the discussion in
Sect. \ref{Invariance-Props}). In terms of the \textit{dressed} charges $%
Z_{AB}$ and $Z_{I}$, the (\textit{only apparently} moduli-dependent)
expression of $\mathcal{I}_{2}$ reads (see \textit{e.g.} \cite
{ADF-Duality-d=4,ADF1,ADF2}):
\begin{equation}
\mathcal{I}_{2}=\frac{1}{2}Z_{AB}\overline{Z}^{AB}-Z_{I}\overline{Z}^{I}=%
\mathcal{Z}_{1}^{2}-\rho ^{2}  \label{tired-8}
\end{equation}

As mentioned above, the $\frac{1}{3}$-BPS criticality conditions (\ref
{1/3-BPS}) or (\ref{tired-4}) are a set of $n$ \textit{complex} equations,
which thus does not stabilize \textit{all} the $3n$ \textit{complex} scalar
fields $z^{i}$ in terms of the BH electric and magnetic charges. In \cite
{bellucci2} the residual $2n$ unstabilized scalars have been shown to span
the $\frac{1}{3}$-BPS moduli space
\begin{equation}
\mathcal{M}_{\mathcal{N}=3,n,BPS}=\frac{SU(2,n)}{SU(2)\times SU(n)\times U(1)%
}\ ,~dim_{\mathbb{R}}=4n.
\end{equation}

\subsubsection{\label{N=3,non-BPS,Z=0-Attractors}Non-BPS ($Z_{AB}=0$)
Attractors and their Moduli Space}

By inserting the non-BPS ($Z_{AB}=0$) criticality conditions (\ref
{non-BPS-Z=0-N=3}) into the expression (\ref{carMatN=3}) of the \textit{%
central charge matrix} $Z_{AB}$ and recalling the definitions (\ref{tired-1}%
) and (\ref{tired-2}) of the \textit{complexified} BH charges, one obtains
\begin{equation}
\left( \sqrt{1+XX^{\dag }}\right) _{AB}^{C}Q_{C}=-(\overline{X})_{AB}^{i}%
\overline{Q}_{i},~\forall A,B=1,2,3.  \label{tired-9}
\end{equation}
By plugging such an expression and its Hermitian conjugate into Eq. (\ref
{sat-4}), and using the identities (\ref{id1}) and (\ref{id2}), one obtains
that

\begin{eqnarray}
2\left( Z_{I}\overline{Z}^{I}\right) _{non-BPS} &=&(XX^{\dag })^{AB}Q_{A}%
\overline{Q}_{B}+(X^{\dag }\sqrt{1+XX^{\dag }})^{i\,A}\overline{Q}_{i}%
\overline{Q}_{A}+  \notag \\
&&  \notag \\
&&+(\sqrt{1+XX^{\dag }}X)^{B\,j}Q_{B}Q_{j}+(1+X^{\dag }X)^{kl}\overline{Q}%
_{k}Q_{l}=  \notag \\
&&  \notag \\
&=&(XX^{\dag })^{AB}Q_{A}\overline{Q}_{B}-(1+XX^{\dag })^{AB}Q_{A}\overline{Q%
}_{B}+  \notag \\
&&  \notag \\
&&-(X^{\dag }X)^{ij}\overline{Q}_{i}Q_{j}+(1+X^{\dag }X)^{ij}\overline{Q}%
_{i}Q_{j}=  \notag \\
&&  \notag \\
&=&Q_{i}\overline{Q}_{i}-Q_{A}\overline{Q}_{A}\ .  \label{tired-6}
\end{eqnarray}
where once again Eq. (\ref{tired-5}) was used.

By recalling Eqs. (\ref{BH-entropy-I2}) and (\ref{VBH}), and using Eq. (\ref
{tired-6}), one achieves the following result:
\begin{eqnarray}
\frac{S_{BH,non-BPS}}{\pi } &=&\frac{A_{H,non-BPS}}{4}=\left. V_{BH}\right|
_{non-BPS}=\rho _{non-BPS}^{2}=\left( Z_{I}\overline{Z}^{I}\right)
_{non-BPS}=  \notag \\
&=&-\frac{1}{2}\left( Q_{A}\overline{Q}_{A}-Q_{i}\overline{Q}_{i}\right) =-%
\mathcal{I}_{2}>0,
\end{eqnarray}
where $\mathcal{I}_{2}$ is the \textit{quadratic} $G_{\mathcal{N}=3,n}$%
-invariant, given by Eqs. (\ref{tired-7}) and (\ref{tired-8}).

As mentioned above, the non-BPS criticality conditions (\ref{non-BPS-Z=0-N=3}%
) or (\ref{tired-9}) are a set of $3$ \textit{complex} equations, which thus
does not stabilize \textit{all} the $3n$ \textit{complex} scalar fields $%
z^{i}$ in terms of the BH electric and magnetic charges. In \cite{bellucci2}
the residual $3\left( n-1\right) $ unstabilized scalars have been shown to
span the non-BPS ($Z_{AB}=0$) moduli space
\begin{equation}
\mathcal{M}_{\mathcal{N}=3,n,non-BPS}=\frac{SU(3,n-1)}{SU(3)\times
SU(n-1)\times U(1)}\ =\mathcal{M}_{\mathcal{N}=3,n-1},\text{ }dim_{\mathbb{R}%
}=6\left( n-1\right) .
\end{equation}

Thus, as it holds in symmetric $\mathcal{N}=2$, $d=4$ supergravity \textit{%
minimally coupled} to $n_{V}=n$ Abelian vector multiplets, also in the
considered $\mathcal{N}=3$, $d=4$ supergravity the BH charges supporting
\textit{non-degenerate} critical points of $V_{BH}$ are split in two
branches: the ($\frac{1}{3}$-)BPS one, defined by $\mathcal{I}_{2}>0$, and
the non-BPS ($Z_{AB}=0$) one, corresponding to $\mathcal{I}_{2}<0$.

\subsection{\label{N=3,d=4-BPS}Black Hole Parameters for $\frac{1}{3}$-BPS
Flow}

By using the \textit{Maurer-Cartan Eqs.} of $\mathcal{N}=3$, $d=4$
supergravity (see \textit{e.g.} \cite{ADF-Duality-d=4,ADF1,ADF2}), one gets
\cite{ADOT-1}
\begin{equation}
\partial _{i}\mathcal{Z}_{1}=\partial _{i}\mathcal{W}_{BPS}=\frac{1}{2\sqrt{2%
}}\frac{P_{IAB,i}\overline{Z}^{I}\overline{Z}^{AB}}{\sqrt{Z_{CD}\overline{Z}%
^{CD}}}=\frac{1}{4\mathcal{Z}_{1}}P_{IAB,i}\overline{Z}^{I}\overline{Z}^{AB},
\label{CERN-2}
\end{equation}
where $P_{IAB}\equiv P_{IAB,i}dz^{i}$ is the holomorphic Vielbein of $%
\mathcal{M}_{\mathcal{N}=3,n}$. Here, $\nabla $ denotes the $U\left(
1\right) $-K\"{a}hler and $H_{\mathcal{N}=3,n}$-covariant differential
operator.

Thus, by using the explicit expressions of $\mathcal{W}_{BPS}^{2}$ given by
Eq. (\ref{sat-3}), using the \textit{Maurer-Cartan Eqs.} of $\mathcal{N}=3$,
$d=4$ supergravity (see \textit{e.g.} \cite{ADF-Duality-d=4,ADF1,ADF2}), and
exploiting the \textit{first order (fake supergravity) formalism} discussed
in Sect. \ref{First-Order}, one respectively obtains the following
expressions of the \textit{(square) ADM mass}, \textit{covariant scalar
charges} and \textit{(square) effective horizon radius} for the $\frac{1}{3}$%
-BPS attractor flow:
\begin{eqnarray}
r_{H,BPS}^{2}\left( z_{\infty },\overline{z}_{\infty },p,q\right)
&=&M_{ADM,BPS}^{2}\left( z_{\infty },\overline{z}_{\infty },p,q\right) =%
\mathcal{W}_{BPS}^{2}\left( z_{\infty },\overline{z}_{\infty },p,q\right) =
\notag \\
&&  \notag \\
&=&\frac{1}{2}\lim_{\tau \rightarrow 0^{-}}\left( Z_{AB}\overline{Z}%
^{AB}\right) \left( z\left( \tau \right) ,\overline{z}\left( \tau \right)
,p,q\right) =  \notag \\
&&  \notag \\
&=&\frac{1}{2}\left[
\begin{array}{l}
(1+X_{\infty }X_{\infty }^{\dag })^{AB}\overline{Q}_{A}Q_{B}+(\sqrt{%
1+X_{\infty }X_{\infty }^{\dag }}\,X_{\infty })^{\ Ai}Q_{i}Q_{A}+ \\
\\
+(X_{\infty }^{\dag }\sqrt{1+X_{\infty }X_{\infty }^{\dag }})^{j\,B}%
\overline{Q}_{B}\overline{Q}_{j}+(X_{\infty }^{\dag }X_{\infty })^{k\,l}%
\overline{Q}_{k}Q_{l}\
\end{array}
\right] ;  \notag \\
&&  \label{sat-night-2}
\end{eqnarray}
\begin{eqnarray}
\Sigma _{i,BPS}\left( z_{\infty },\overline{z}_{\infty },p,q\right)
&=&2\lim_{\tau \rightarrow 0^{-}}\left( \partial _{i}\mathcal{W}%
_{BPS}\right) \left( z\left( \tau \right) ,\overline{z}\left( \tau \right)
,p,q\right) =  \notag \\
&&  \notag \\
&=&\frac{1}{\sqrt{2}}\left[ \frac{P_{IAB,i}\overline{Z}^{I}\overline{Z}^{AB}%
}{\sqrt{Z_{CD}\overline{Z}^{CD}}}\right] _{\infty }=\frac{1}{2}\left[ \frac{1%
}{\mathcal{Z}_{1}}P_{IAB,i}\overline{Z}^{I}\overline{Z}^{AB}\right] _{\infty
}=  \notag \\
&&  \notag \\
&=&\frac{1}{2M_{ADM,BPS}\left( z_{\infty },\overline{z}_{\infty },p,q\right)
}\left( P_{IAB,i}\overline{Z}^{I}\overline{Z}^{AB}\right) _{\infty };
\label{euro5}
\end{eqnarray}
\begin{eqnarray}
R_{H,BPS}^{2} &=&\lim_{\tau \rightarrow 0^{-}}\left[
\begin{array}{l}
\mathcal{W}_{BPS}^{2}\left( z\left( \tau \right) ,\overline{z}\left( \tau
\right) ,p,q\right) + \\
\\
-4G^{i\overline{j}}\left( z\left( \tau \right) ,\overline{z}\left( \tau
\right) \right) \left( \partial _{i}\mathcal{W}_{BPS}\right) \left( z\left(
\tau \right) ,\overline{z}\left( \tau \right) ,p,q\right) \cdot \\
\\
\cdot \left( \overline{\partial }_{\overline{j}}\mathcal{W}_{BPS}\right)
\left( z\left( \tau \right) ,\overline{z}\left( \tau \right) ,p,q\right)
\end{array}
\right] =  \notag \\
&&  \notag \\
&&  \notag \\
&=&\mathcal{I}_{2}\left( p,q\right) =V_{BH,BPS}=\frac{S_{BH,BPS}\left(
p,q\right) }{\pi },  \label{euro3}
\end{eqnarray}
where
\begin{equation}
X_{\infty }\equiv \lim_{\tau \rightarrow 0^{-}}X\left( \tau \right) .
\end{equation}
Throughout all the treatment, the subscript ``$\infty $'' indicates the
evaluation at the scalars at \textit{radial infinity }$z_{\infty }^{i}$.

Eq. (\ref{euro3}) proves Eq. (\ref{CERN-Thu-1}) for the $\frac{1}{3}$-BPS
attractor flow of the considered $\mathcal{N}=3$, $d=4$ supergravity. Such a
result was obtained by using Eq. (\ref{CERN-2}) and computing that
\begin{eqnarray}
4G^{i\overline{j}}\left( \partial _{i}\mathcal{W}_{BPS}\right) \overline{%
\partial }_{\overline{j}}\mathcal{W}_{BPS} &=&4G^{i\overline{j}}\left(
\partial _{i}\mathcal{Z}_{1}\right) \overline{\partial }_{\overline{j}}%
\mathcal{Z}_{1}=  \notag \\
&=&\frac{G^{i\overline{j}}P_{IAB,i}\overline{P}_{JEF,\overline{j}}\overline{Z%
}^{I}Z^{J}\overline{Z}^{AB}Z^{EF}}{2Z_{CD}\overline{Z}^{CD}}=Z_{I}\overline{Z%
}^{I}=\rho ^{2},  \label{CERN-3}
\end{eqnarray}
where the relation
\begin{equation}
G^{i\overline{j}}P_{IAB,i}\overline{P}_{JEF,\overline{j}}=\delta _{IJ}\left(
\delta _{AE}\delta _{BF}-\delta _{AF}\delta _{BE}\right)
\end{equation}
was exploited.

The considerations made at the end of Subsect. \ref
{N=2-d=4-quadratic-BPS-Attractor-Flow} hold also for the considered
attractor flow.

As pointed out above, the same also holds for ($\frac{1}{2}$-BPS attractor
flow of) $\mathcal{N}=2$, $d=4$ supergravity \textit{minimally coupled} to
Abelian vector multiplets (see Eq. (150) of \cite{FHM}), in which the (%
\textit{unique}) invariant of the $U$-duality group $SU\left( 1,n\right) $
is \textit{quadratic} in BH electric and magnetic charges. Such a similarity
is ultimately due to the fact that $SU\left( m,n\right) $ is endowed with a
pseudo-Hermitian \textit{quadratic} form built out of the \textit{fundamental%
} $\mathbf{m+n}$ and \textit{antifundamental} $\overline{\mathbf{m+n}}$
representations.

Furthermore, Eq. (\ref{sat-night-2}) yields that the $\frac{1}{3}$-BPS
attractor flow of the $\mathcal{N}=3$, $d=4$ supergravity does \textit{not }%
saturate the \textit{marginal stability bound} (see \cite{Marginal-Refs} and
\cite{GLS1}; see also the discussion at the end of Subsect. \ref
{N=2-d=4-quadratic-non-BPS-Z=0-Attractor-Flow}).

\subsection{\label{N=3-d=4-non-BPS}Black Hole Parameters for Non-BPS ($%
Z_{AB}=0$) Flow}

By using the \textit{Maurer-Cartan Eqs.} of $\mathcal{N}=3$, $d=4$
supergravity (see \textit{e.g.} \cite{ADF-Duality-d=4,ADF1,ADF2}), one gets
\cite{ADOT-1}
\begin{equation}
\partial _{i}\rho =\partial _{i}\mathcal{W}_{non-BPS}=\frac{1}{4}\frac{%
P_{IAB,i}\overline{Z}^{I}\overline{Z}^{AB}}{\sqrt{Z_{J}\overline{Z}^{J}}}=%
\frac{P_{IAB,i}\overline{Z}^{I}\overline{Z}^{AB}}{4\rho }.  \label{CERN-5}
\end{equation}
Thus, by using the explicit expressions of $\mathcal{W}_{non-BPS}^{2}$ given
by Eq. (\ref{sat-4}), using the \textit{Maurer-Cartan Eqs.} of $\mathcal{N}%
=3 $, $d=4$ supergravity (see \textit{e.g.} \cite{ADF-Duality-d=4,ADF1,ADF2}%
), and exploiting the \textit{first order (fake supergravity) formalism}
discussed in Sect. \ref{First-Order}, one respectively obtains the following
expressions of the \textit{(square) ADM mass}, \textit{covariant scalar
charges} and \textit{(square) effective horizon radius} for the non-BPS
attractor flow:
\begin{eqnarray}
r_{H,non-BPS}^{2}\left( z_{\infty },\overline{z}_{\infty },p,q\right)
&=&M_{ADM,non-BPS}^{2}\left( z_{\infty },\overline{z}_{\infty },p,q\right) =%
\mathcal{W}_{non-BPS}^{2}\left( z_{\infty },\overline{z}_{\infty
},p,q\right) =  \notag \\
&&  \notag \\
&=&\lim_{\tau \rightarrow 0^{-}}\left( Z_{I}\overline{Z}^{I}\right) \left(
z\left( \tau \right) ,\overline{z}\left( \tau \right) ,p,q\right) =  \notag
\\
&&  \notag \\
&=&\frac{1}{2}\left[
\begin{array}{l}
(X_{\infty }X_{\infty }^{\dag })^{AB}Q_{A}\overline{Q}_{B}+(\sqrt{%
1+X_{\infty }^{\dag }X_{\infty }}\,X_{\infty }^{\dag })^{i\,A}\overline{Q}%
_{i}\overline{Q}_{A}+ \\
\\
+(X_{\infty }\sqrt{1+X_{\infty }^{\dag }X_{\infty }})^{B%
\,j}Q_{B}Q_{j}+(1+X_{\infty }^{\dag }X_{\infty })^{k\,l}\overline{Q}_{k}Q_{l}
\end{array}
\right] ;  \notag \\
&&  \label{sat-night-3}
\end{eqnarray}
\begin{eqnarray}
\Sigma _{i,non-BPS}\left( z_{\infty },\overline{z}_{\infty },p,q\right)
&=&2\lim_{\tau \rightarrow 0^{-}}\left( \partial _{i}\mathcal{W}%
_{non-BPS}\right) \left( z\left( \tau \right) ,\overline{z}\left( \tau
\right) ,p,q\right) =  \notag \\
&&  \notag \\
&=&\frac{1}{2}\left[ \frac{P_{IAB,i}\overline{Z}^{I}\overline{Z}^{AB}}{\sqrt{%
Z_{J}\overline{Z}^{J}}}\right] _{\infty }=\frac{1}{2}\left[ \frac{P_{IAB,i}%
\overline{Z}^{I}\overline{Z}^{AB}}{\rho }\right] _{\infty }=  \notag \\
&&  \notag \\
&=&\frac{1}{2M_{ADM,non-BPS}\left( z_{\infty },\overline{z}_{\infty
},p,q\right) }\left( P_{IAB,i}\overline{Z}^{I}\overline{Z}^{AB}\right)
_{\infty };  \notag \\
&&  \label{euro6}
\end{eqnarray}
\begin{eqnarray}
R_{H,non-BPS}^{2} &=&\lim_{\tau \rightarrow 0^{-}}\left[
\begin{array}{l}
\mathcal{W}_{non-BPS}^{2}\left( z\left( \tau \right) ,\overline{z}\left(
\tau \right) ,p,q\right) + \\
\\
-4G^{i\overline{j}}\left( z\left( \tau \right) ,\overline{z}\left( \tau
\right) \right) \left( \partial _{i}\mathcal{W}_{non-BPS}\right) \left(
z\left( \tau \right) ,\overline{z}\left( \tau \right) ,p,q\right) \cdot \\
\\
\cdot \left( \overline{\partial }_{\overline{j}}\mathcal{W}_{non-BPS}\right)
\left( z\left( \tau \right) ,\overline{z}\left( \tau \right) ,p,q\right)
\end{array}
\right] =  \notag \\
&&  \notag \\
&=&-\mathcal{I}_{2}\left( p,q\right) =V_{BH,non-BPS}=\frac{%
S_{BH,non-BPS}\left( p,q\right) }{\pi }.  \label{euro4}
\end{eqnarray}
Eq. (\ref{euro3}) proves Eq. (\ref{CERN-Thu-1}) for the non-BPS ($Z_{AB}=0$)
attractor flow of the considered $\mathcal{N}=3$, $d=4$ supergravity. Such a
result was obtained by using Eq. (\ref{CERN-5}) and computing that
\begin{eqnarray}
4G^{i\overline{j}}\left( \partial _{i}\mathcal{W}_{non-BPS}\right) \overline{%
\partial }_{\overline{j}}\mathcal{W}_{non-BPS} &=&4G^{i\overline{j}}\left(
\partial _{i}\rho \right) \overline{\partial }_{\overline{j}}\rho =  \notag
\\
&=&\frac{1}{4}\delta _{IK}\left( \delta _{AC}\delta _{BD}-\delta _{AD}\delta
_{BC}\right) \frac{\overline{Z}^{AB}Z^{CD}\overline{Z}^{I}Z^{K}}{Z_{J}%
\overline{Z}^{J}}=  \notag \\
&=&\frac{1}{2}Z_{AB}\overline{Z}^{AB}=\mathcal{Z}_{1}^{2}.  \label{CERN-7}
\end{eqnarray}

The considerations made at the end of Subsect. \ref
{N=2-d=4-quadratic-BPS-Attractor-Flow} hold also for the considered
attractor flow.

It is worth noticing out that Eqs. (\ref{euro3}) and (\ref{euro4}) are
consistent, because, as pointed out above, the ($\frac{1}{3}$-BPS)- and
non-BPS ($Z_{AB}=0$)- supporting BH charge configurations in the considered
theory are respectively defined by the \textit{quadratic} constraints $%
\mathcal{I}_{2}\left( p,q\right) >0$ and $\mathcal{I}_{2}\left( p,q\right)
<0 $.

As yielded by Eqs. (\ref{euro5}) and (\ref{euro6}) for both \textit{%
non-degenerate} attractor flows of the considered theory it holds the
following relation among \textit{scalar charges} and \textit{ADM mass}:
\begin{equation}
\Sigma _{i}=\frac{1}{2M_{ADM}}\lim_{\tau \rightarrow 0^{-}}P_{IAB,i}%
\overline{Z}^{I}\overline{Z}^{AB}.
\end{equation}

Furthermore, Eq. (\ref{sat-night-3}) yields that the non-BPS $Z_{AB}=0$
attractor flow of the $\mathcal{N}=3$, $d=4$ supergravity does \textit{not }%
saturate the \textit{marginal stability bound} (see \cite{Marginal-Refs} and
\cite{GLS1}; see also the discussion at the end of Subsect. \ref
{N=2-d=4-quadratic-non-BPS-Z=0-Attractor-Flow}).

\section{\label{Invariance-Props}Black Hole Entropy\newline
in \textit{Minimally Coupled }$\mathcal{N}=2$ and $\mathcal{N}=3$
Supergravity}

It is here worth remarking that the classical Bekenstein-Hawking \cite
{hawking2} $d=4$ BH entropy $S_{BH}$ for \textit{minimally coupled} $%
\mathcal{N}=2$ and $\mathcal{N}=3$ supergravity is given by the following $%
SU\left( m,n\right) $-invariant expression:
\begin{equation}
\frac{S_{BH}}{\pi }=\frac{1}{2}\left| q^{2}+p^{2}\right| ,  \label{ven2}
\end{equation}
where $q^{2}\equiv \eta ^{\Lambda \Sigma }q_{\Lambda }q_{\Sigma }$ and $%
p^{2}\equiv \eta _{\Lambda \Sigma }p^{\Lambda }p^{\Sigma }$, $\eta ^{\Lambda
\Sigma }=\eta _{\Lambda \Sigma }$ being the Lorentzian metric with signature
$\left( m,n\right) $. As said above, $\mathcal{N}=2$ is obtained by putting $%
m=1$, whereas $\mathcal{N}=3$ is given by $m=3$ (see Eqs. (\ref
{I2-N=2-quadr.-BH-charges}) and (\ref{tired-7}) above, respectively). Thus,
in Eq. (\ref{ven2}) the positive signature pertains to the \textit{%
graviphoton charges}, while the negative signature corresponds to the
charges given by the fluxes of the vector field strengths from the matter
multiplets.

The supersymmetry-preserving features of the attractor solution depend on
the sign of $q^{2}+p^{2}$. The limit case $q^{2}+p^{2}=0$ corresponds to the
so-called \textit{small} BHs (which however, in the case $\mathcal{N}=3$, do
\textit{not} enjoy an enhancement of supersymmetry, contrarily to what
usually happens in $\mathcal{N}\geqslant 4$, $d=4$ supergravities; see
\textit{e.g.} the treatment in \cite{ADFT}).

By setting $n=0$ in $\mathcal{N}=3$, $d=4$ supergravity (with resulting $U$%
-duality $U\left( 3\right) $ which, due to the absence of scalars, coincides
with the $\mathcal{N}=3$ $\mathcal{R}$-symmetry $U\left( 3\right) $ \cite
{FSZ}), one gets
\begin{equation}
\frac{S_{BH}}{\pi }=\frac{1}{2}\left[ q_{1}^{2}+q_{2}^{2}+q_{3}^{2}+\left(
p^{1}\right) ^{2}+\left( p^{2}\right) ^{2}+\left( p^{3}\right) ^{2}\right] ,
\label{ven1}
\end{equation}
which is nothing but the sum of the entropies of three \textit{extremal}
(and thus BPS; see \textit{e.g.} the discussion in \cite{FHM})
Reissner-N\"{o}rdstrom BHs, \textit{without any interference terms}. Such a
result can be simply understood by recalling that the generalization of the
Maxwell electric-magnetic duality $U\left( 1\right) $ to the case of $n$
Abelian gauge fields is $U\left( n\right) $ \cite{Gaillard-Zumino-1}, and
that the expression in the right-hand side of Eq. (\ref{ven1}) is the unique
possible $U\left( 3\right) $-invariant combination of charges.

Moreover, it is here worth noticing that $\mathcal{N}=3$, $d=4$ supergravity
is the only $\mathcal{N}>2$ supergravity in which the gravity multiplet does
\textit{not} contain any scalar field at all, analogously to what happens in
the case $\mathcal{N}=2$. Thus, in \textit{minimally coupled} $\mathcal{N}=2$%
\footnote{%
Let us notice also that $\mathcal{N}=2$ \textit{minimally coupled} theory is
the only (symmetric) $\mathcal{N}=2$, $d=4$ supergravity which yields the
\textit{pure }$\mathcal{N}=2$ supergravity simply by setting $n=0$.} and $%
\mathcal{N}=3$, $d=4$ supergravity the \textit{pure }supergravity theory,
obtained by setting $n=0$, is \textit{scalarless}, with the $U$-duality
coinciding with the $\mathcal{R}$-symmetry \cite{FSZ}.

This does \textit{not} happen for all other $\mathcal{N}>2$ theories. For
instance, the $\mathcal{N}=4$, $d=4$ gravity multiplet does contain one
complex scalar field (usually named \textit{axion-dilaton}) and six Abelian
vectors; thus, the \textit{pure }$\mathcal{N}=4$ theory, obtained by setting
$n=0$ (see Eqs. (\ref{Sun-3}) and (\ref{Sun-2})), \textit{is not scalarless}%
. By further truncating four vectors out (\textit{i.e.} by performing a $%
\left( U\left( 1\right) \right) ^{6}\rightarrow \left( U\left( 1\right)
\right) ^{2}$ gauge truncation) and analyzing the bosonic field content, one
gets the bosonic sector of $\mathcal{N}=2$, $d=4$ supergravity \textit{%
minimally coupled} to one vector multiplet, the so-called \textit{%
axion-dilaton supergravity} (for a discussion of the invariance properties
of the classical BH entropy in such cases, see \textit{e.g.} \cite{FHM} and
Refs. therein). \setcounter{equation}0

\section{\label{N=5,d=4}$\mathcal{N}=5$ Supergravity}

The (special K\"{a}hler) \textit{scalar manifold} is \cite{N=5-Ref}
\begin{equation}
\mathcal{M}_{\mathcal{N}=5}=\frac{G_{\mathcal{N}=5}}{H_{\mathcal{N}=5}}=%
\frac{SU\left( 1,5\right) }{SU\left( 5\right) \times U\left( 1\right) }%
,~dim_{\mathbb{R}}=10.
\end{equation}
As previously mentioned, \textit{no} matter coupling is allowed (\textit{pure%
} supergravity).

The $10$ vector field strengths and their duals, as well as their
asymptotical fluxes, sit in the \textit{three-fold antisymmetric} irrepr. $%
\mathbf{20}$ of the $U$-duality group $G_{\mathcal{N}=5}=SU\left( 1,5\right)
$ (or equivalently of the compact form $SU\left( 6\right) _{\mathbb{C}}$),
and \textit{not} in its \textit{fundamental} repr. $\mathbf{6}$.

$Z_{AB}=Z_{\left[ AB\right] }$, $A,B=1,2,3,4,5=\mathcal{N}$ is the \textit{%
central charge matrix}. By means of a suitable transformation of the $%
\mathcal{R}$-symmetry $H_{\mathcal{N}=5}=U\left( 5\right) $, $Z_{AB}$ can be
\textit{skew-diagonalized} by putting it in the \textit{normal form} (see
e.g. \cite{ADOT-1} and Refs. therein):
\begin{equation}
Z_{AB}=\left(
\begin{array}{ccc}
\mathcal{Z}_{1}\epsilon &  &  \\
& \mathcal{Z}_{2}\epsilon &  \\
&  & 0
\end{array}
\right) ,
\end{equation}
where $\mathcal{Z}_{1},\mathcal{Z}_{2}\in \mathbb{R}_{0}^{+}$ are the $%
\mathcal{N}=5$ (moduli-dependent) \textit{skew-eigenvalues}, which can be
ordered as $\mathcal{Z}_{1}\geqslant \mathcal{Z}_{2}$ without any loss of
generality (up to renamings; see \textit{e.g.} \cite{ADOT-1}), and can be
expressed as follows (see also the treatment of \cite{DFL}):
\begin{equation}
\left\{
\begin{array}{l}
\mathcal{Z}_{1}=\frac{1}{\sqrt{2}}\sqrt{I_{1}+\sqrt{2I_{2}-I_{1}^{2}}}; \\
\\
\mathcal{Z}_{2}=\frac{1}{\sqrt{2}}\sqrt{I_{1}-\sqrt{2I_{2}-I_{1}^{2}}};
\end{array}
\right. \Longleftrightarrow \left\{
\begin{array}{l}
I_{1}=\mathcal{Z}_{1}^{2}+\mathcal{Z}_{2}^{2}; \\
\\
I_{2}=\mathcal{Z}_{1}^{4}+\mathcal{Z}_{2}^{4},
\end{array}
\right.  \label{cuba-3}
\end{equation}
where
\begin{eqnarray}
I_{1} &\equiv &\frac{1}{2}Z_{AB}\overline{Z}^{AB};  \label{cuba-1} \\
I_{2} &\equiv &\frac{1}{2}Z_{AB}\overline{Z}^{BC}Z_{CD}\overline{Z}^{DA}
\label{cuba-2}
\end{eqnarray}
are the two \textit{unique} (moduli-dependent) $H_{\mathcal{N}=5}$%
-invariants.

From the Lagrangian density of $\mathcal{N}=5$, $d=4$ supergravity \cite
{N=5-Ref}, the vector kinetic terms read as follows:
\begin{equation}
\mathcal{L}_{kin}^{vec}=-\frac{1}{4}\sqrt{-g}(\mathcal{S}^{ij\mid kl}-\frac{1%
}{2}\delta ^{ik}\delta ^{jl})F_{\mu \nu \mid \,ij}^{+}F_{\phantom{+\mu\nu}%
kl}^{+\,\mu \nu }+h.c.\ ;  \label{Mon-1}
\end{equation}
here and below $i=1,...,5$ and antisymmetrization of couple of $i$-indices
is understood. In the Gaillard-Zumino formalism, it also holds (see Eq. (36)
of \cite{ADFT})
\begin{equation}
\mathcal{L}_{kin}^{vec}=\sqrt{-g}\,\frac{i}{4}\mathcal{N}_{ij\mid kl}F_{\mu
\nu \mid \,ij}^{+}F_{\phantom{+\mu\nu}kl}^{+\,\mu \nu }+h.c.\ .
\label{Mon-2}
\end{equation}
By comparing Eqs. (\ref{Mon-1}) and (\ref{Mon-2}), can write the $10\times
10 $ vector kinetic matrix $\mathcal{N}^{ij,kl}$ as follows:
\begin{equation}
\mathcal{N}^{ij\mid kl}=i(\mathcal{S}^{ij\mid kl}-\frac{1}{2}\delta
^{ik}\delta ^{jl})\ .
\end{equation}
The matrix $\mathcal{S}$ satisfies the relation
\begin{equation}
(\delta _{kl}^{ij}-\overline{S}^{ij\mid kl})\mathcal{S}^{kl\mid mn}=\delta
_{mn}^{ij}\ ,
\end{equation}
and, in a suitable choice of the scalar fields, it holds that
\begin{equation}
\overline{S}^{ij\mid kl}=-\frac{1}{2}\epsilon ^{ijklm}z_{m}\ .
\end{equation}
Thus, one finds (square brackets denote antisymmetrization of enclosed
indices throughout)
\begin{equation}
\mathcal{S}^{ij\mid kl}=\frac{1}{1-(z_{n})^{2}}\left[ \delta _{kl}^{ij}-%
\frac{1}{2}\epsilon ^{ijklm}z_{m}-2\delta _{\lbrack i\,[k}z_{l]}z_{j]}\right]
\ ,
\end{equation}
where the last term is normalized as
\begin{equation}
\delta _{\lbrack k}^{[i}z^{j]}z_{l]}=\frac{1}{4}(\delta
_{k}^{i}z^{j}z_{l}\pm {permutations}...).
\end{equation}
Consequently, one achieves the following expression of the vector kinetic
matrix:
\begin{equation}
\mathcal{N}_{ij\mid kl}=\frac{\alpha }{1-(z_{m})^{2}}\left( \frac{1}{2}\left[
1+(z_{n})^{2}\right] \delta _{kl}^{ij}-\frac{1}{2}\epsilon
^{ijklp}z_{p}-2\delta _{\lbrack i\,[k}z_{l]}z_{j]}\right) \ ,  \label{Wed-6}
\end{equation}
where $\alpha $ is a factor to be determined by the relations satisfied by
the symplectic sections $\mathbf{h}$ and $\mathbf{f}$ defined in the last
steps of Eqs. (\ref{h}) and (\ref{f}) above.

The supersymmetry transformation of the vector field can be written as \cite
{DF-Fermion}
\begin{equation}
\delta A_{\mu }^{ij}=2f^{ij\,\mid AB}\bar{\psi}_{A\mu }+2f_{AB}^{ij}\bar{\psi%
}_{\mu }^{A}\epsilon ^{B}\ .  \label{Mon-3}
\end{equation}
Such a formula is equivalent to the following expression \cite{N=5-Ref} (see
also \cite{dW-N=8}):
\begin{equation}
\delta A_{\mu }^{ij}=(\mathcal{S}^{ij\mid kl}-\delta ^{ij\mid kl})(\mathcal{C%
}^{-1})_{kl}^{\phantom{-1}AB}(\bar{\epsilon}^{C}\gamma _{\mu }\chi _{ABC}+2%
\sqrt{2}\bar{\epsilon}_{A}\psi _{B})\ ,  \label{Mon-4}
\end{equation}
where
\begin{eqnarray}
\mathcal{C}_{ij}^{\phantom{ij}kl} &\equiv &\frac{1}{e_{1}}\delta _{kl}^{ij}-2%
\frac{e_{2}}{e_{1}}\delta _{\lbrack k}^{[i}z^{j]}z_{l]}\ ; \\
&&  \notag \\
(\mathcal{C}^{-1})_{ij}^{\phantom{ij}kl} &\equiv &\left( e_{1}\delta
_{ij}^{kl}+2e_{2}\delta _{\lbrack k}^{[i}z^{j]}z_{l]}\right) \ ,
\end{eqnarray}
and
\begin{eqnarray}
e_{1}^{2} &\equiv &\frac{1}{1-|z|^{2}}\ ;  \label{Wed-4} \\
e_{2} &\equiv &\frac{1}{|z|^{2}}(1-e_{1})\ .  \label{Wed-5}
\end{eqnarray}

By comparing Eqs. (\ref{Mon-3}) and (\ref{Mon-4}), one obtains that
\begin{equation}
f_{\phantom{ij}AB}^{ij}=\left( e_{1}\delta _{AB}^{ij}+\frac{e_{1}}{2}%
\epsilon ^{ijABm}z_{m}+2e_{2}\delta _{\lbrack i}^{[A}z^{B]}z_{j]}\right) \ .
\label{Wed-9}
\end{equation}
The symplectic section $\mathbf{h}$ is
\begin{equation}
h_{ij\mid AB}=\mathcal{N}_{ij\mid mn}f_{AB}^{mn},  \label{Wed-10}
\end{equation}
and thus it explicitly reads
\begin{eqnarray}
h_{ij\mid AB} &=&\frac{\alpha }{1-\left( z_{p}\right) ^{2}}\left\{
\begin{array}{l}
\left[ \frac{1}{2}\left( 1-\left( z_{q}\right) ^{2}\right) \delta _{kl}^{ij}-%
\frac{1}{2}\epsilon ^{ijklm}z_{m}-2\delta _{\lbrack i[k}z_{l]}z_{j]}\right]
\cdot \\
\\
\cdot \left[ e_{1}\delta _{AB}^{kl}+\frac{e_{1}}{2}\epsilon
^{klABn}z_{n}+2e_{2}\delta _{\lbrack k}^{[A}z^{B]}z_{l]}\right]
\end{array}
\right\} =  \notag \\
&&~  \notag \\
&&~  \notag \\
&=&\alpha \left[ \frac{e_{1}}{2}\delta _{AB}^{ij}-\frac{e_{1}}{4}\epsilon
^{ijABk}z_{k}+e_{2}\delta _{\lbrack i}^{[A}z^{B]}z_{j]}\right] .
\label{Wed-11}
\end{eqnarray}

Next, we check the above results and determine the overall numerical factor $%
\alpha $ of the matrix $\mathcal{N}$, by expliciting writing down the
identities (\ref{Mon-5}) and (\ref{Mon-6}) in the considered framework.

By recalling that
\begin{eqnarray}
\frac{1}{4}\epsilon ^{ijABk}\epsilon ^{ijCDl}z_{k}z_{l} &=&\left(
z_{m}\right) ^{2}\delta _{CD}^{AB}-2\delta _{\lbrack A[C}z_{D]}z_{B]}; \\
&&  \notag \\
\delta _{\lbrack i}^{[A}z^{B]}z_{j]}\delta _{\lbrack i}^{[C}z^{D]}z_{j]} &=&%
\frac{1}{2}\left( z_{m}\right) ^{2}\delta ^{\lbrack A[C}z^{D]}z^{B]},
\end{eqnarray}
the identity (\ref{Mon-6}) is easily verified, because
\begin{eqnarray}
\left( \mathbf{f}^{T}\mathbf{h}\right) _{CD}^{AB} &=&\left( \mathbf{h}^{T}%
\mathbf{f}\right) _{CD}^{AB}=  \notag \\
&&  \notag \\
&=&\alpha \left\{
\begin{array}{l}
\left[ \frac{e_{1}}{2}\delta _{AB}^{ij}-\frac{e_{1}}{4}\epsilon
^{ijABk}z_{k}+e_{2}\delta _{\lbrack i}^{[A}z^{B]}z_{j]}\right] \cdot \\
\\
\cdot \left[ e_{1}\delta _{CD}^{ij}+\frac{e_{1}}{2}\epsilon
^{ijCDl}z_{l}+2e_{2}\delta _{\lbrack i}^{[C}z^{D]}z_{j]}\right]
\end{array}
\right\} =  \notag \\
&&  \notag \\
&&  \notag \\
&=&\alpha \left\{
\begin{array}{l}
\frac{e_{1}^{2}}{2}\delta _{CD}^{AB}\left( 1-\left( z_{i}\right) ^{2}\right)
+e_{1}^{2}\delta _{\lbrack A[C}z_{D]}z_{B]}+e_{2}^{2}\left( z_{i}\right)
^{2}\delta ^{\lbrack A[C}z^{D]}z^{B]}+ \\
\\
+2e_{1}e_{2}Re\left( \delta _{\lbrack A}^{[C}z^{D]}z_{B]}\right)
\end{array}
\right\} .  \notag \\
&&
\end{eqnarray}

In order to check the identity (\ref{Mon-5}) and determine $\alpha $, we
compute
\begin{eqnarray}
\mathbf{f}^{\dag }\mathbf{h} &=&f_{ij}^{~~AB}h_{ij\mid CD}=  \notag \\
&&  \notag \\
&=&\alpha \left\{
\begin{array}{l}
\left[ e_{1}\delta _{ij}^{AB}+\frac{e_{1}}{2}\epsilon
_{ijABk}z^{k}+2e_{2}\delta _{\lbrack A}^{[i}z^{j]}z_{B]}\right] \cdot \\
\\
\cdot \left[ \frac{e_{1}}{2}\delta _{CD}^{ij}-\frac{e_{1}}{4}\epsilon
^{ijCDl}z_{l}+e_{2}\delta _{\lbrack i}^{[C}z^{D]}z_{j]}\right]
\end{array}
\right\} =  \notag \\
&&  \notag \\
&&  \notag \\
&=&\alpha \left[ \frac{1}{2}\delta _{CD}^{AB}+\frac{e_{1}^{2}}{4}\left(
\epsilon _{ijABk}z^{k}-\epsilon ^{ijABk}z_{k}\right) -\left(
e_{1}^{2}+e_{1}e_{2}+e_{2}\right) \delta _{\lbrack A}^{[C}z^{D]}z_{B]}\right]
=  \notag \\
&&  \notag \\
&=&\alpha \left[ \frac{1}{2}\delta _{CD}^{AB}+i\frac{e_{1}^{2}}{2}Im\left(
\epsilon _{ijABk}z^{k}\right) \right] ,  \notag \\
&&
\end{eqnarray}
where we used the facts that $\left| z\right| ^{2}e_{2}=1-e_{1}$, and $%
e_{1}^{2}+e_{1}e_{2}+e_{2}=0$, directly following from the definitions (\ref
{Wed-4}) and (\ref{Wed-5}).

By an analogous calculation, one achieves the following result:
\begin{equation}
\mathbf{h}^{\dag }\mathbf{f}=\left( h_{ij\mid CD}\right) ^{\dag
}f_{ij}^{~~AB}=\overline{\alpha }\left[ \frac{1}{2}\delta _{CD}^{AB}+i\frac{%
e_{1}^{2}}{2}Im\left( \epsilon ^{ijABk}z_{k}\right) \right] .
\end{equation}
Consequently, the identity (\ref{Mon-5}) is satisfied iff $\alpha =-i$. By
substituting such a value into Eq. (\ref{Wed-6}), one obtains the following
expression of the vector kinetic matrix of $\mathcal{N}=5$, $d=4$ \textit{%
ungauged} supergravity:
\begin{equation}
\mathcal{N}_{ij\mid kl}=-\frac{i}{1-(z_{m})^{2}}\left( \frac{1}{2}\left[
1+(z_{n})^{2}\right] \delta _{kl}^{ij}-\frac{1}{2}\epsilon
^{ijklp}z_{p}-2\delta _{\lbrack i\,[k}z_{l]}z_{j]}\right) \ .  \label{Wed-7}
\end{equation}

From the general definition\footnote{%
We rescale the symplectic section $\mathbf{f}$ by a factor $\frac{1}{2}$,
and the kinetic matrix $\mathcal{N}$ correspondingly by a factor $2$. The
definition of the symplectic section $\mathbf{h}$ through Eq. (\ref{Wed-10})
and the identities (\ref{Mon-5}) and (\ref{Mon-6}) are left unchanged. Such
a redefinition is performed in order to avoid an unsuitable rescaling of the
magnetic charges, and thus to define the \textit{complexified} BH charges as
in Eq. (\ref{Wed-aft-1}).} (\ref{Wed-8}), the \textit{central charge matrix}
$Z_{AB}$ is defined as the integral over the $2$-sphere at infinity $%
S_{\infty }^{2}$ of the\textit{\ dressed} graviphoton field strength \cite
{ADF1,ADF2,ADF-Duality-d=4}:
\begin{eqnarray}
Z_{AB} &\equiv &-\int_{S_{\infty }^{2}}T_{AB}=-\int_{S_{\infty
}^{2}}T_{AB}^{-}=f_{AB}^{\Lambda }q_{\Lambda }-h_{\Lambda \mid AB}p^{\Lambda
}\ =  \notag \\
&=&f_{AB}^{ij}q_{ij}-h_{ij\mid AB}p^{ij}\ .
\end{eqnarray}
By recalling Eqs. (\ref{Wed-9}) and (\ref{Wed-11}) with $\alpha =-i$, and,
analogously to definitions (\ref{tired-1}) and (\ref{tired-2}), by
introducing the \textit{complexified} BH charges as
\begin{equation}
Q^{ij}\equiv \frac{1}{2}\left( q_{ij}+ip^{ij}\right) ,  \label{Wed-aft-1}
\end{equation}
one gets
\begin{equation}
Z_{AB}\left( z,\overline{z},q,p\right) =e_{1}Q^{AB}+\frac{e_{1}}{2}\epsilon
^{ABijk}\overline{Q}^{ij}z_{k}-2e_{2}z^{[A}Q^{B]C}z_{C}.  \label{Wed-aft-2}
\end{equation}

As recalled at the start of the next Subsection, only ($\frac{1}{5}$-)BPS
attractor flow is \textit{non-degenerate} (\textit{i.e.} corresponding to
\textit{large} BHs; see \textit{e.g.} the discussion in \cite{ADFT}), and
the corresponding (squared) \textit{first order fake superpotential} is (
\cite{ADOT-1}; recall Eqs. (\ref{cuba-3})-(\ref{cuba-2}))
\begin{eqnarray}
\mathcal{W}_{\left( \frac{1}{5}-\right) BPS}^{2} &=&\frac{1}{2}\left[ \frac{1%
}{2}Z_{AB}\overline{Z}^{AB}+\sqrt{Z_{AB}\overline{Z}^{BC}Z_{CD}\overline{Z}%
^{DA}-\frac{1}{4}\left( Z_{AB}\overline{Z}^{AB}\right) ^{2}}\right] =  \notag
\\
&=&\frac{1}{2}\left[ I_{1}+\sqrt{2I_{2}-I_{1}^{2}}\right] =\mathcal{Z}%
_{1}^{2}.  \label{W-BPS}
\end{eqnarray}

\subsection{\label{N=5-d=4-AEs}Attractor Equations and their Solutions}

Due to the absence of \textit{matter charges}, the \textit{BH effective
potential} can be written as
\begin{equation}
V_{BH}=\frac{1}{2}Z_{AB}\overline{Z}^{AB}=\mathcal{Z}_{1}^{2}+\mathcal{Z}%
_{2}^{2}=I_{1}.  \label{VBH-N=5}
\end{equation}
The $\mathcal{N}=5$, $d=4$ \textit{Attractor Eqs.} are nothing but the
\textit{criticality conditions} for the $H_{\mathcal{N}=5}$-invariant $I_{1}$%
. By using the \textit{Maurer-Cartan Eqs.} of $\mathcal{N}=5$, $d=4$
supergravity (see \textit{e.g.} \cite{ADF-Duality-d=4,ADF1,ADF2}), such
\textit{Attractor Eqs.} can be written as follows:
\begin{eqnarray}
\epsilon ^{ABCDE}Z_{AB}Z_{CD} &=&0;  \label{Wed-aft-3} \\
\epsilon _{ABCDE}\overline{Z}^{AB}\overline{Z}^{CD} &=&0,  \label{Wed-aft-4}
\end{eqnarray}
where the two lines are reciprocally complex conjugated. They can also be
written more explicitly, by using the expression of central charge matrix $%
Z_{AB}$ given by Eq. (\ref{Wed-aft-2}). For instance, Eqs. (\ref{Wed-aft-3})
can be rewritten in the following way:
\begin{eqnarray}
0 &=&\epsilon ^{ABCDE}Z_{AB}Z_{CD}=  \notag \\
&&  \notag \\
&=&e_{1}^{2}\epsilon ^{ABCDE}Q^{AB}Q^{CD}+e_{1}^{2}(\epsilon ^{ABCDi}%
\overline{Q}^{AB}\overline{Q}^{CD}z_{i})z_{E}+8e_{1}^{2}(Q^{AB}Q^{AB})z_{E}+
\notag \\
&&  \notag \\
&&+16e_{1}^{2}z_{i}Q^{iA}\overline{Q}^{AE}-4e_{1}e_{2}\epsilon
^{ABCDE}Q^{AB}z^{C}Q^{Di}z_{i}+  \notag \\
&&  \notag \\
&&-8e_{1}e_{2}(z_{i}Q^{Di}\overline{Q}%
^{CD}z^{C})z_{E}-16e_{1}e_{2}|z|^{2}z_{j}Q^{ij}\overline{Q}^{iE}\ .
\label{Wed-aft-5}
\end{eqnarray}

The criticality conditions (\ref{Wed-aft-4}) and (\ref{Wed-aft-5}) are
satisfied for a unique class of critical points (for further elucidation,
see \textit{e.g.} the treatment in \cite{ADFT}):

\begin{itemize}
\item  ($\frac{1}{5}$-)BPS:
\begin{equation}
\mathcal{Z}_{2}=0,~\mathcal{Z}_{1}>0.  \label{1/5-BPS}
\end{equation}
%\newline
\end{itemize}

It is worth counting here the degrees of freedom related to Eqs. (\ref
{Wed-aft-3}) and (\ref{Wed-aft-4}), or equivalently to the unique $\frac{1}{5%
}$-BPS solution given by Eq. (\ref{1/5-BPS}). Eqs. (\ref{Wed-aft-3}) and (%
\ref{Wed-aft-4}) are $10$ \textit{real} equations, but actually only $6$
\textit{real} among them are independent. Thus a moduli space of $\frac{1}{5}
$-BPS attractors, spanned by the $2$ \textit{complex} scalars unstabilized
by Eq. (\ref{1/5-BPS}), might - and actually does \cite{bellucci2} - exist.
Such a counting of ``flat'' directions of $V_{BH}$ at its $\frac{1}{5}$-BPS
critical points was given in terms of the left-over would-be $\mathcal{N}=2$
hyperscalars' degrees of freedom in the $\mathcal{N}=5\rightarrow \mathcal{N}%
=2$ supersymmetry reduction in \cite{ADF-Duality-d=4}.

\subsubsection{\label{1/5-BPS-Attractors}Entropy of $\frac{1}{5}$-BPS
Attractors and their Moduli Space}

By recalling Eqs. (\ref{BH-entropy-I4}) and (\ref{VBH-N=5}), and using Eq. (%
\ref{1/5-BPS}), one achieves the following result:
\begin{equation}
\frac{S_{BH,BPS}}{\pi }=\frac{A_{H,BPS}}{4}=\left. V_{BH}\right| _{BPS}=%
\mathcal{Z}_{1,BPS}^{2}=\sqrt{\mathcal{I}_{4}}\ .  \label{Tue-1}
\end{equation}
$\mathcal{I}_{4}$ here denotes the (unique) invariant of the \textit{%
three-fold antisymmetric} $\mathbf{20}$ representation of the $U$-duality
group $G_{\mathcal{N}=5}$. Such a representation is \textit{symplectic},
containing the singlet $\mathbf{1}_{a}$ in the tensor product $\mathbf{20}%
\times \mathbf{20}$ \cite{Slansky}, and it is thus \textit{irreducible} with
respect to both $G_{\mathcal{N}=5}$ and $Sp\left( 20,\mathbb{R}\right) $). $%
\mathcal{I}_{4}$ is \textit{quartic} in BH charges.

In terms of the \textit{dressed} charges, \textit{i.e.} of the central
charge matrix $Z_{AB}$, $\mathcal{I}_{4}$ can be written as follows:
\begin{eqnarray}
\mathcal{I}_{4} &=&Z_{AB}\overline{Z}^{BC}Z_{CD}\overline{Z}^{DA}-\frac{1}{4}%
\left( Z_{AB}\overline{Z}^{AB}\right) ^{2}=2I_{2}-I_{1}^{2}=  \notag \\
&=&Tr(A^{2})-\frac{1}{4}\left( Tr\,\left( A\right) \right) ^{2}.
\label{cocco-1}
\end{eqnarray}
where the (moduli-dependent) matrix $A_{A}^{~~B}\equiv Z_{AC}\overline{Z}%
^{BC}$ and the related quantities \cite{ADFT}
\begin{eqnarray}
Tr\,\left( A\right) &=&Z_{AB}\overline{Z}^{AB}\ =2I_{1}=2V_{BH};
\label{Tr-A2} \\
Tr(A^{2}) &=&Z_{AB}\overline{Z}^{BC}Z_{CD}\overline{Z}^{DA}\ =2I_{2},
\label{Tr-A3}
\end{eqnarray}
were introduced, $I_{1}$ and $I_{2}$ being the two \textit{unique}
(moduli-dependent) $H_{\mathcal{N}=5}$-invariants, respectively defined by
Eqs. (\ref{cuba-1}) and (\ref{cuba-2}).

The $U$-invariant $\mathcal{I}_{4}$, introduced in Eq. (\ref{Tue-1}) and
expressed in terms of \textit{dressed} charges by Eq. (\ref{cocco-1}), is
the \textit{unique} (moduli-independent) independent $G_{\mathcal{N}=5}=$%
-invariant combination of (moduli-dependent) $H_{\mathcal{N}=5}$-invariant
quantities (see \textit{e.g.} \cite{ADF-Duality-d=4}, and \cite{ADFT}, as
well as Refs. therein). Its \textit{quarticity} in BH charges is ultimately
due to the \textit{symplectic} nature of the $\mathbf{20}$ of $SU\left(
1,5\right) $ (\textit{irreducible} with respect to \textit{both} $SU\left(
1,5\right) $ and $Sp\left( 20,\mathbb{R}\right) $), \textit{i.e.} to the
fact that the tensor product $\mathbf{20}\times \mathbf{20}$\textbf{\ }
contains an \textit{antisymmetric} singlet $\mathbf{1}_{a}$ \cite{Slansky},
thus yielding a vanishing \textit{quadratic} invariant of $SU\left(
1,5\right) $.

Thus, the expression (\ref{cocco-1}) is moduli-dependent \textit{only
apparently}. Being moduli-independent, $\mathcal{I}_{4}$ can actually be
written only in terms of the BH charges. In order to determine such an
expression, one can use the fact that, in the considered coordinate
parametrization, the origin $O$ of $\mathcal{M}_{\mathcal{N}=5}$ (determined
by $z^{i}=0$ $\forall i=1,...,5$) is the invariant point under the action of
$H_{\mathcal{N}=5}=U(5)$ (see \textit{e.g.} Eq. (2.17) of \cite{N=5-Ref}).
Thus, the explicit dependence of $\mathcal{I}_{4}$ on BH charges is obtained
simply by computing $\left. Z_{AB}\right| _{O}$ and using Eq. (\ref{cocco-1}%
). By recalling Eq. (\ref{Wed-aft-2}) and Eqs. (\ref{Tr-A2}) and (\ref{Tr-A3}%
), one obtains:
\begin{eqnarray}
\mathcal{I}_{4} &=&Q_{AB}Q^{BC}Q_{CD}Q^{DA}-\frac{1}{4}(Q_{AB}\,Q^{AB})^{2}\
=  \notag \\
&&  \notag \\
&=&\frac{1}{2^{4}}\,\left( q_{AB}-ip^{AB}\right) \left(
q_{BC}+ip^{BC}\right) \left( q_{CD}-ip^{CD}\right) \left(
q_{DA}+ip^{DA}\right) +  \notag \\
&&  \notag \\
&&-\frac{1}{2^{6}}\left[ \left( q_{AB}-ip^{AB}\right) \,\left(
q_{AB}+ip^{AB}\right) \right] ^{2},  \label{Wed-aft-7}
\end{eqnarray}
where the \textit{complexified} BH charges $Q^{ij}$ defined by Eq. (\ref
{Wed-aft-1}) have been used, with $Q_{AB}\equiv \overline{Q^{AB}}$. Eq. (\ref
{Wed-aft-7}) is manifestly $H_{N=5}$-invariant, the $10+10$ \textit{real }BH
charges being arranged in the (reciprocally conjugated) \textit{complex rank-%
}$2$ tensors $Q^{ij}$ and $Q_{ij}$ in the \textit{two-fold antisymmetric} $%
\mathbf{10}$ and $\mathbf{10}^{\prime }$ of $SU\left( 5\right) $, or
equivalently in the \textit{real} \textit{rank-}$2$ tensors $q_{ij}$ and $%
p^{ij}$ in a pair of \textit{two-fold antisymmetric} $\mathbf{10}$ of $%
SO\left( 5\right) $.

Clearly, the BH electric and magnetic BH charges, being the asymptotical
fluxes of the vector field strengths and of their duals, can actually be
arranged to sit in the \textit{three-fold antisymmetric}, \textit{symplectic}
representation $\mathbf{20}$ of the $U$-duality group $G_{\mathcal{N}%
=5}=SU\left( 1,5\right) $. The embedding of the \textit{two-fold
antisymmetric} $\mathbf{10}^{\left( \prime \right) }$ of $SU\left( 5\right) $
into the \textit{three-fold antisymmetric} $\mathbf{20}$ of $SU\left(
1,5\right) $ is given by the formula ($a,b,c,d,e=1,...,5$)
\begin{equation}
\mathbf{~}t^{abc}\equiv \frac{1}{3!}\epsilon ^{abcde6}t_{de}\ ,
\label{Thu-12}
\end{equation}
or, more precisely (recall $A,B,C=1,...,5$, and moreover $\widehat{A},%
\widehat{B},\widehat{C}=1,...,6$, here and below):
\begin{eqnarray}
Q_{\widehat{A}\widehat{B}\widehat{C}} &\equiv &q_{\widehat{A}\widehat{B}%
\widehat{C}}+ip^{\widehat{A}\widehat{B}\widehat{C}}=\left(
Q_{ABC},Q_{AB6}\right) ;  \label{Thu-13} \\
Q_{ABC} &\equiv &\frac{1}{3!}\epsilon _{ABCDE6}Q^{DE}\equiv \frac{1}{2}%
\left( q_{ABC}+ip^{ABC}\right) ;  \label{Thu-14} \\
Q_{6AB} &\equiv &Q_{AB}\equiv \frac{1}{2}\left( q_{6AB}+ip^{6AB}\right) .
\label{Thu-15}
\end{eqnarray}
Eq. (\ref{Thu-13}) describes the splitting of the $\mathbf{20}$ of $SU\left(
1,5\right) $ into the $\mathbf{10}$ and $\mathbf{10}^{\prime }$ of $SU\left(
5\right) $, whose embedding is determined by Eqs. (\ref{Thu-14}) and (\ref
{Thu-15}), respectively. As given by Eq. (\ref{Thu-15}), by identifying $%
t_{de}\equiv t_{de6}$, Eq. (\ref{Thu-12}) can be seen as part of a
self-reality condition (admitting solutions in $SU\left( 1,5\right) $, but
not in $SU\left( 6\right) $).

Thus, it can be easily shown that Eq. (\ref{Wed-aft-7}) can be recast in a
manifestly $U=G_{\mathcal{N}=5}$-invariant way, with BH charges in the
\textit{three-fold antisymmetric} $\mathbf{20}$ of $SU\left( 1,5\right) $,
as follows:
\begin{eqnarray}
\mathcal{I}_{4} &=&\frac{1}{3\cdot 2^{5}}\epsilon ^{\widehat{A}\widehat{B}%
\widehat{C}\widehat{A}^{\prime }\widehat{B}^{\prime }\widehat{C}^{\prime
\prime \prime }}\epsilon ^{\widehat{A}^{\prime \prime }\widehat{B}^{\prime
\prime }\widehat{C}^{\prime \prime }\widehat{A}^{\prime \prime \prime }%
\widehat{B}^{\prime \prime \prime }\widehat{C}^{\prime }}Q_{\widehat{A}%
\widehat{B}\widehat{C}}Q_{\widehat{A}^{\prime }\widehat{B}^{\prime }\widehat{%
C}^{\prime }}Q_{\widehat{A}^{\prime \prime }\widehat{B}^{\prime \prime }%
\widehat{C}^{\prime \prime }}Q_{\widehat{A}^{\prime \prime \prime }\widehat{B%
}^{\prime \prime \prime }\widehat{C}^{\prime \prime \prime }}\ =
\label{Thu-mid-1} \\
&&  \notag \\
&=&\frac{1}{3\cdot 2^{5}}\epsilon ^{\widehat{A}\widehat{B}\widehat{C}%
\widehat{A}^{\prime }\widehat{B}^{\prime }\widehat{C}^{\prime \prime \prime
}}\epsilon ^{\widehat{A}^{\prime \prime }\widehat{B}^{\prime \prime }%
\widehat{C}^{\prime \prime }\widehat{A}^{\prime \prime \prime }\widehat{B}%
^{\prime \prime \prime }\widehat{C}^{\prime }}\left( q_{\widehat{A}\widehat{B%
}\widehat{C}}+ip^{\widehat{A}\widehat{B}\widehat{C}}\right) \left( q_{%
\widehat{A}^{\prime }\widehat{B}^{\prime }\widehat{C}^{\prime }}+ip^{%
\widehat{A}^{\prime }\widehat{B}^{\prime }\widehat{C}^{\prime }}\right) \cdot
\notag \\
&&  \notag \\
&&\cdot \left( q_{\widehat{A}^{\prime \prime }\widehat{B}^{\prime \prime }%
\widehat{C}^{\prime \prime }}+ip^{\widehat{A}^{\prime \prime }\widehat{B}%
^{\prime \prime }\widehat{C}^{\prime \prime }}\right) \left( q_{\widehat{A}%
^{\prime \prime \prime }\widehat{B}^{\prime \prime \prime }\widehat{C}%
^{\prime \prime \prime }}+ip^{\widehat{A}^{\prime \prime \prime }\widehat{B}%
^{\prime \prime \prime }\widehat{C}^{\prime \prime \prime }}\right) .
\label{Thu-mid-1-bis}
\end{eqnarray}
In order to prove such a formula, let us explicit the entries ``$6$'' in Eq.
(\ref{Thu-mid-1}), obtaining
\begin{equation}
\mathcal{I}_{4}=\frac{1}{3\cdot 2^{5}}\left[
\begin{array}{l}
9\,\epsilon ^{6BCA^{\prime }B^{\prime }C^{\prime \prime \prime }}\epsilon
^{6B^{\prime \prime }C^{\prime \prime }A^{\prime \prime \prime }B^{\prime
\prime \prime }C^{\prime }}Q_{6BC}Q_{A^{\prime }B^{\prime }C^{\prime
}}Q_{6B^{\prime \prime }C^{\prime \prime }}Q_{A^{\prime \prime \prime
}B^{\prime \prime \prime }C^{\prime \prime \prime }}+ \\
\\
+12\,\epsilon ^{ABC6B^{\prime }C^{\prime \prime \prime }}\epsilon
^{6B^{\prime \prime }C^{\prime \prime }A^{\prime \prime \prime }B^{\prime
\prime \prime }C^{\prime }}Q_{ABC}Q_{6B^{\prime }C^{\prime }}Q_{6B^{\prime
\prime }C^{\prime \prime }}Q_{A^{\prime \prime \prime }B^{\prime \prime
\prime }C^{\prime \prime \prime }}+ \\
\\
+6\,\epsilon ^{ABCA^{\prime }B^{\prime }6}\epsilon ^{6B^{\prime \prime
}C^{\prime \prime }A^{\prime \prime \prime }B^{\prime \prime \prime
}C^{\prime }}Q_{ABC}Q_{6B^{\prime }C^{\prime }}Q_{A^{\prime \prime
}B^{\prime \prime }C^{\prime \prime }}Q_{A^{\prime \prime \prime }B^{\prime
\prime \prime }6}+ \\
\\
+4\,\epsilon ^{ABC6B^{\prime }C^{\prime \prime \prime }}\epsilon ^{A^{\prime
\prime }B^{\prime \prime }C^{\prime \prime }6B^{\prime \prime \prime
}C^{\prime }}Q_{ABC}Q_{6B^{\prime }C^{\prime }}Q_{A^{\prime \prime
}B^{\prime \prime }C^{\prime \prime }}Q_{6B^{\prime \prime \prime }C^{\prime
\prime \prime }}+ \\
\\
+\epsilon ^{ABCA^{\prime }B^{\prime }6}\epsilon ^{A^{\prime \prime
}B^{\prime \prime }C^{\prime \prime }A^{\prime \prime \prime }B^{\prime
\prime \prime }6}Q_{ABC}Q_{A^{\prime }B^{\prime }6}Q_{A^{\prime \prime
}B^{\prime \prime }C^{\prime \prime }}Q_{A^{\prime \prime \prime }B^{\prime
\prime \prime }6}
\end{array}
\right] .  \label{Thu-mid-2}
\end{equation}
By using the embedding Eqs. (\ref{Thu-14}) and (\ref{Thu-15}), it is
immediate to check that Eq. (\ref{Thu-mid-2}) yields Eq. (\ref{Wed-aft-7}%
).\smallskip

As mentioned above, only $6$ out of the $10$ \textit{real} $\frac{1}{5}$-BPS
criticality conditions (\ref{Wed-aft-3}) and (\ref{Wed-aft-4}) are actually
independent. Thus, they do not stabilize \textit{all} the $5$ \textit{complex%
} scalar fields $z^{i}$ in terms of the BH electric and magnetic charges,
but only $3$ of them. In \cite{bellucci2} the residual $2$ unstabilized
\textit{complex} scalar degrees of freedom have been shown to span the $%
\frac{1}{5}$-BPS moduli space
\begin{equation}
\mathcal{M}_{\mathcal{N}=5,BPS}=\frac{SU(2,1)}{SU(2)\times SU(1)}\ =\mathcal{%
M}_{\mathcal{N}=3,n=1,BPS},~dim_{\mathbb{R}}=4.
\end{equation}

\subsection{\label{N=5,d=4,BPS}Black Hole Parameters for $\frac{1}{5}$-BPS
Flow}

By using the \textit{Maurer-Cartan Eqs.} of $\mathcal{N}=5$, $d=4$
supergravity (see \textit{e.g.} \cite{ADF-Duality-d=4,ADF1,ADF2}), one gets
\cite{ADOT-1}
\begin{eqnarray}
\partial _{i}\mathcal{Z}_{1} &=&\partial _{i}\mathcal{W}_{BPS}=  \notag \\
&=&\frac{P_{,i}}{\sqrt{2}}\sqrt{\frac{1}{2}Z_{AB}\overline{Z}^{AB}-\sqrt{%
Z_{AB}\overline{Z}^{BC}Z_{CD}\overline{Z}^{DA}-\frac{1}{4}\left( Z_{AB}%
\overline{Z}^{AB}\right) ^{2}}}=P_{,i}\mathcal{Z}_{2},  \notag \\
&&  \label{CERN-al-1}
\end{eqnarray}
where $P\equiv P_{1234}$, $P_{ABCD}\equiv P_{ABCD,i}dz^{i}=\epsilon
_{ABCDE}P^{E}$ being the holomorphic Vielbein of $\mathcal{M}_{\mathcal{N}%
=5} $. Here, $\nabla $ denotes the $U\left( 1\right) $-K\"{a}hler and $H_{%
\mathcal{N}=5}$-covariant differential operator.

Thus, by using the explicit expressions of $\mathcal{W}_{BPS}^{2}$ given by
Eq. (\ref{W-BPS}), using the \textit{Maurer-Cartan Eqs.} of $\mathcal{N}=5$,
$d=4$ supergravity (see \textit{e.g.} \cite{ADF-Duality-d=4,ADF1,ADF2}), and
exploiting the \textit{first order (fake supergravity) formalism} discussed
in Sect. \ref{First-Order}, one respectively obtains the following
expressions of the \textit{(square) ADM mass}, \textit{covariant scalar
charges} and \textit{(square) effective horizon radius} for the $\frac{1}{5}$%
-BPS attractor flow:
\begin{eqnarray}
r_{H,BPS}^{2}\left( z_{\infty },\overline{z}_{\infty },p,q\right)
&=&M_{ADM,BPS}^{2}\left( z_{\infty },\overline{z}_{\infty },p,q\right) =%
\mathcal{W}_{BPS}^{2}\left( z_{\infty },\overline{z}_{\infty },p,q\right) =
\notag \\
&=&\frac{1}{2}\lim_{\tau \rightarrow 0^{-}}\left[ \frac{1}{2}Z_{AB}\overline{%
Z}^{AB}+\sqrt{Z_{AB}\overline{Z}^{BC}Z_{CD}\overline{Z}^{DA}-\frac{1}{4}%
\left( Z_{AB}\overline{Z}^{AB}\right) ^{2}}\right] =  \notag \\
&=&\left. \mathcal{Z}_{1}^{2}\right| _{\infty };  \notag \\
&&  \label{Sun-1}
\end{eqnarray}
\begin{eqnarray}
\Sigma _{i,BPS}\left( z_{\infty },\overline{z}_{\infty },p,q\right) &\equiv
&2\lim_{\tau \rightarrow 0^{-}}\left( \partial _{i}\mathcal{W}_{BPS}\right)
\left( z\left( \tau \right) ,\overline{z}\left( \tau \right) ,p,q\right) =
\notag \\
&=&\sqrt{2}\lim_{\tau \rightarrow 0^{-}}\left[ P_{,i}\sqrt{\frac{1}{2}Z_{AB}%
\overline{Z}^{AB}-\sqrt{Z_{AB}\overline{Z}^{BC}Z_{CD}\overline{Z}^{DA}-\frac{%
1}{4}\left( Z_{AB}\overline{Z}^{AB}\right) ^{2}}}\right] =  \notag \\
&=&2\left( P_{,i}\mathcal{Z}_{2}\right) _{\infty };  \notag \\
&&
\end{eqnarray}
\begin{equation}
\begin{array}{l}
R_{H,BPS}^{2}=\mathcal{W}_{BPS}^{2}\left( z_{\infty },\overline{z}_{\infty
},p,q\right) + \\
\\
-4G^{i\overline{j}}\left( z_{\infty },\overline{z}_{\infty }\right) \left(
\partial _{i}\mathcal{W}_{BPS}\right) \left( z_{\infty },\overline{z}%
_{\infty },p,q\right) \left( \overline{\partial }_{\overline{j}}\mathcal{W}%
_{BPS}\right) \left( z_{\infty },\overline{z}_{\infty },p,q\right) = \\
\\
=\sqrt{Z_{AB}\overline{Z}^{BC}Z_{CD}\overline{Z}^{DA}-\frac{1}{4}\left(
Z_{AB}\overline{Z}^{AB}\right) ^{2}}= \\
\\
=\sqrt{2\mathcal{Z}_{1}^{4}+2\mathcal{Z}_{2}^{4}-\left( \mathcal{Z}_{1}^{2}+%
\mathcal{Z}_{2}^{2}\right) ^{2}}= \\
\\
=\mathcal{Z}_{1}^{2}-\mathcal{Z}_{2}^{2}=\sqrt{\mathcal{I}_{4}\left(
p,q\right) }>0.
\end{array}
\label{euro3!}
\end{equation}

Eq. (\ref{euro3!}) proves Eq. (\ref{CERN-Thu-1}) for the $\frac{1}{5}$-BPS
attractor flow of the considered $\mathcal{N}=5$, $d=4$ supergravity. Such a
result was obtained by using Eq. (\ref{CERN-al-1}) and computing that
\begin{equation}
\begin{array}{l}
4G^{i\overline{j}}\left( \partial _{i}\mathcal{W}_{BPS}\right) \overline{%
\partial }_{\overline{j}}\mathcal{W}_{BPS}=4G^{i\overline{j}}\left( \partial
_{i}\mathcal{Z}_{1}\right) \overline{\partial }_{\overline{j}}\mathcal{Z}%
_{1}= \\
\\
=2G^{i\overline{j}}P_{,i}\overline{P}_{,\overline{j}}\left[ \frac{1}{2}Z_{AB}%
\overline{Z}^{AB}-\sqrt{Z_{AB}\overline{Z}^{BC}Z_{CD}\overline{Z}^{DA}-\frac{%
1}{4}\left( Z_{AB}\overline{Z}^{AB}\right) ^{2}}\right] = \\
\\
=\frac{1}{2}\left[ \frac{1}{2}Z_{AB}\overline{Z}^{AB}-\sqrt{Z_{AB}\overline{Z%
}^{BC}Z_{CD}\overline{Z}^{DA}-\frac{1}{4}\left( Z_{AB}\overline{Z}%
^{AB}\right) ^{2}}\right] =\mathcal{Z}_{2}^{2},
\end{array}
\end{equation}
where the relation
\begin{equation}
4G^{i\overline{j}}P_{,i}\overline{P}_{,\overline{j}}=1
\end{equation}
was used.

The considerations made at the end of Subsect. \ref
{N=2-d=4-quadratic-BPS-Attractor-Flow} hold also for the considered
attractor flow.

It is worth noticing out that Eq. (\ref{euro3!}) is consistent, because, as
pointed out above, the $\frac{1}{5}$-BPS-supporting BH charge configurations
in the considered theory is defined by the \textit{quartic} constraints $%
\mathcal{I}_{4}\left( p,q\right) >0$.

Furthermore, Eq. (\ref{Sun-1}) yields that the $\frac{1}{5}$-BPS attractor
flow of $\mathcal{N}=5$, $d=4$ supergravity does \textit{not }saturate the
\textit{marginal stability bound} (see \cite{Marginal-Refs} and \cite{GLS1};
see also the discussion at the end of Subsect. \ref
{N=2-d=4-quadratic-non-BPS-Z=0-Attractor-Flow}).\setcounter{equation}0

\section{\label{N=4,d=4-pure}$\mathcal{N}=4$ \textit{Pure} Supergravity
Revisited}

The treatment of $\mathcal{N}=4$, $d=4$ \textit{pure} supergravity is pretty
similar to the one given for $\mathcal{N}=5$, $d=4$ supergravity in Sect.
\ref{N=5,d=4}.

The (special K\"{a}hler) \textit{scalar manifold} is \cite{CSF}
\begin{equation}
\mathcal{M}_{\mathcal{N}=4,pure}=\frac{G_{\mathcal{N}=4,pure}}{H_{\mathcal{N}%
=4,pure}}=\frac{SU\left( 1,1\right) \times SU\left( 4\right) }{U\left(
1\right) \times SU\left( 4\right) }=\frac{SU\left( 1,1\right) }{U\left(
1\right) },~dim_{\mathbb{R}}=2,  \label{Sun-2}
\end{equation}
spanned by the complex scalar
\begin{equation}
s\equiv a+ie^{-2\varphi },~a,\varphi \in \mathbb{R},  \label{cocco-3}
\end{equation}
where $a$ and $\varphi $ are usually named \textit{axion} and \textit{dilaton%
}, respectively.

The $6$ vector field strengths and their duals, as well as their
asymptotical fluxes, sit in the \textit{bi-fundamental} irrepr. $\left(
\mathbf{2},\mathbf{6}\right) $ of the $U$-duality group $G_{\mathcal{N}%
=4,pure}=SU\left( 1,1\right) \times SO\left( 6\right) \sim SU\left(
1,1\right) \times SU\left( 4\right) $.

$Z_{AB}=Z_{\left[ AB\right] }$, $A,B=1,2,3,4=\mathcal{N}$ is the \textit{%
central charge matrix}. By means of a suitable transformation of the $%
\mathcal{R}$-symmetry $H_{\mathcal{N}=4,pure}=U\left( 1\right) \times
SO\left( 6\right) \sim U\left( 1\right) \times SU\left( 4\right) $, $Z_{AB}$
can be \textit{skew-diagonalized} by putting it in the \textit{normal form}
(see e.g. \cite{ADOT-1} and Refs. therein):
\begin{equation}
Z_{AB}=\left(
\begin{array}{cc}
\mathcal{Z}_{1}\epsilon &  \\
& \mathcal{Z}_{2}\epsilon
\end{array}
\right) ,
\end{equation}
where $\mathcal{Z}_{1},\mathcal{Z}_{1}\in \mathbb{R}_{0}^{+}$ are the $%
\mathcal{N}=4$ (moduli-dependent) \textit{skew-eigenvalues}, which can be
ordered as $\mathcal{Z}_{1}\geqslant \mathcal{Z}_{2}$ without any loss of
generality (up to renamings; see \textit{e.g.} \cite{ADOT-1}), and can be
formally expressed by the very same Eqs. (\ref{cuba-3})-(\ref{cuba-2}),
where now $I_{1}$ and $I_{2}$ are the two \textit{unique}\textbf{\ }%
(moduli-dependent) $H_{\mathcal{N}=4,pure}$-invariants.

The symplectic sections read as follows ($\Lambda \equiv \left[ AB\right]
=1,...,6$ throughout) \cite{ADF1,ADF2,ADFT}
\begin{equation}
f_{AB}^{\Lambda }=e^{\varphi }\delta _{AB}^{\Lambda },~~h_{\Lambda \mid
AB}=se^{\varphi }\delta _{\Lambda \mid AB}=\left( ae^{\varphi }+ie^{-\varphi
}\right) \delta _{\Lambda \mid AB},
\end{equation}
such that the kinetic vector matrix is given by (recall Eq. (\ref{Thu-aft-1}%
))
\begin{equation}
\mathcal{N}_{\Lambda \Sigma }=\left( \mathbf{hf}^{-1}\right) _{\Lambda
\Sigma }=s\delta _{\Lambda \Sigma }.
\end{equation}
By the general definition (\ref{Wed-8}), the c\textit{entral charge matrix}
is given by
\begin{equation}
Z_{AB}=f_{AB}^{\Lambda }q_{\Lambda }-h_{\Lambda \mid AB}p^{\Lambda }\
=e^{\varphi }\delta _{AB}^{\Lambda }q_{\Lambda }-se^{\varphi }\delta
_{\Lambda \mid AB}p^{\Lambda }=-e^{\varphi }\left( sp_{AB}-q_{AB}\right) .
\label{Thu-aft-2}
\end{equation}
Such an explicit expression allows one to elaborate Eqs. (\ref{cuba-3})-(\ref
{cuba-2}) further, obtaining
\begin{eqnarray}
I_{1} &\equiv &\frac{1}{2}Z_{AB}\overline{Z}^{AB}=\mathcal{Z}_{1}^{2}+%
\mathcal{Z}_{2}^{2}=\frac{1}{2}e^{2\varphi }\left( sp_{AB}-q_{AB}\right)
\left( \overline{s}p^{AB}-q^{AB}\right) =  \notag \\
&=&e^{2\varphi }\left( sp_{\Lambda }-q_{\Lambda }\right) \left( \overline{s}%
p^{\Lambda }-q^{\Lambda }\right) =  \notag \\
&=&\left( e^{2\varphi }a^{2}+e^{-2\varphi }\right) p^{2}+e^{2\varphi
}q^{2}-2ae^{2\varphi }p\cdot q;  \label{Thu-aft-3} \\
&&  \notag \\
I_{2} &\equiv &\frac{1}{2}Z_{AB}\overline{Z}^{BC}Z_{CD}\overline{Z}^{DA}=%
\mathcal{Z}_{1}^{4}+\mathcal{Z}_{2}^{4}=  \notag \\
&=&\frac{1}{2}e^{4\varphi }\left( sp_{AB}-q_{AB}\right) \left( \overline{s}%
p^{BC}-q^{BC}\right) \left( sp_{CD}-q_{CD}\right) \left( \overline{s}%
p^{DA}-q^{DA}\right) ,  \label{Thu-aft-4}
\end{eqnarray}
where $p^{2}\equiv \left( p^{1}\right) ^{2}+...+\left( p^{6}\right) ^{2}$, $%
q^{2}\equiv q_{1}^{2}+...+q_{6}^{2}$, and $p\cdot q\equiv p^{\Lambda
}q_{\Lambda }$ (see Eq. (7.1) of \cite{FHM}, fixing a typo in Eq. (225) of
\cite{ADFT}).

Only ($\frac{1}{4}$-)BPS attractor flow is \textit{non-degenerate} (\textit{%
i.e.} corresponding to \textit{large} BHs; see \textit{e.g.} the $n=0$ limit
of the discussion in \cite{ADFT}), and the corresponding (squared) \textit{%
first order fake superpotential} is identical to the one of the ($\frac{1}{5}
$-)BPS attractor flow in $\mathcal{N}=5$, $d=4$ supergravity \cite{ADOT-1},
given by Eq. (\ref{W-BPS}) above, which in the considered framework can be
further elaborated as follows:
\begin{eqnarray}
\mathcal{W}_{\left( \frac{1}{4}-\right) BPS}^{2} &=&\frac{1}{2}\left[ I_{1}+%
\sqrt{2I_{2}-I_{1}^{2}}\right] =\mathcal{Z}_{1}^{2}=  \notag \\
&&  \notag \\
&=&\frac{e^{2\varphi }}{4}\left[
\begin{array}{l}
\left( sp_{AB}-q_{AB}\right) \left( \overline{s}p^{AB}-q^{AB}\right) + \\
\\
+\sqrt{%
\begin{array}{l}
4\left( sp_{AB}-q_{AB}\right) \left( \overline{s}p^{BC}-q^{BC}\right) \left(
sp_{CD}-q_{CD}\right) \left( \overline{s}p^{DA}-q^{DA}\right) + \\
\\
-\left[ \left( sp_{AB}-q_{AB}\right) \left( \overline{s}p^{AB}-q^{AB}\right)
\right] ^{2}
\end{array}
}
\end{array}
\right] .  \notag \\
&&  \label{cocco-2}
\end{eqnarray}

\subsection{\label{N=4-d=4-pure-AEs}Attractor Equations and their Solutions :%
\newline
$\frac{1}{4}$-BPS Attractors and their Entropy}

Due to the absence of \textit{matter charges}, the \textit{BH effective
potential} $V_{BH}$ reads
\begin{eqnarray}
V_{BH} &=&\frac{1}{2}Z_{AB}\overline{Z}^{AB}=\mathcal{Z}_{1}^{2}+\mathcal{Z}%
_{2}^{2}=I_{1}=  \notag \\
&&  \notag \\
&=&\left( e^{2\varphi }a^{2}+e^{-2\varphi }\right) p^{2}+e^{2\varphi
}q^{2}-2ae^{2\varphi }p\cdot q,  \label{Thu-aft-5}
\end{eqnarray}
where in the second line we recalled Eq. (\ref{Thu-aft-3}) (see also the
treatments of \cite{ADFT} and \cite{FHM}). The \textit{complex} \textit{%
Attractor Eq.} of $\mathcal{N}=4$, $d=4$ \textit{pure} supergravity is
nothing but the \textit{criticality condition} for the $H_{\mathcal{N}%
=4,pure}$-invariant $I_{1}$. By using the relevant \textit{Maurer-Cartan Eqs.%
} (see \textit{e.g.} \cite{ADF-Duality-d=4,ADF1,ADF2}), such a \textit{%
complex} \textit{Attractor Eq.} can be written as follows ($\left| \varphi
\right| <\infty $):
\begin{equation}
\epsilon _{ABCD}\overline{Z}^{AB}\overline{Z}^{CD}\Leftrightarrow \epsilon
_{ABCD}\left( \overline{s}p^{AB}-q^{AB}\right) ^{\varphi }\left( \overline{s}%
p^{CD}-q^{CD}\right) =0.  \label{Thu-aft-6}
\end{equation}
An equivalent set of two \textit{real} Eqs. is given by the system
\begin{equation}
\frac{\partial V_{BH}}{\partial a}=0,~~\frac{\partial V_{BH}}{\partial
\varphi }=0,  \label{Thu-aft-7}
\end{equation}
with $V_{BH}$ given by Eqs. (\ref{Thu-aft-3}) or (\ref{Thu-aft-5}), yielding
Eqs. (7.2)-(7.3) of \cite{FHM}.

Thus, the criticality conditions (\ref{Thu-aft-6}), or equivalently (\ref
{Thu-aft-7}), are satisfied for a unique class of critical points:

($\frac{1}{4}$-)BPS:
\begin{equation}
\mathcal{Z}_{2}=0,~\mathcal{Z}_{1}>0,  \label{1/4-BPS}
\end{equation}
yielding Eqs. (7.2)-(7.3) of \cite{FHM}. Eqs. (\ref{Thu-aft-7}) are $2$
\textit{real} Eqs. in $2$ \textit{real} unknowns, namely the \textit{axion} $%
a$ and the \textit{dilaton} $\varphi $, which both are stabilized solely in
terms of the magnetic and electric BH charges. Thus no moduli space of $%
\frac{1}{4}$-BPS attractors in $\mathcal{N}=4$, $d=4$ \textit{pure}
supergravity exists at all \cite{ADFT,bellucci2,FHM}.

By recalling Eqs. (\ref{BH-entropy-I4}) and (\ref{Thu-aft-5}), and using Eq.
(\ref{1/4-BPS}), one achieves the following result \cite{N=4-pure-BH-entropy}%
:
\begin{equation}
\frac{S_{BH,BPS}}{\pi }=\frac{A_{H,BPS}}{4}=\left. V_{BH}\right| _{BPS}=%
\mathcal{Z}_{1,BPS}^{2}=\sqrt{\mathcal{I}_{4}}\ .
\end{equation}
$\mathcal{I}_{4}$ denotes the (unique) invariant of the \textit{%
bi-fundamental} representation $\left( \mathbf{2},\mathbf{6}\right) $ of the
$U$-duality group $G_{\mathcal{N}=4,pure}$. Such a representation is \textit{%
symplectic}\footnote{%
This fact has been observed above also for $\mathcal{N}=5$, $d=4$
supergravity.
\par
Even though in general the (unique) \textit{quartic} invariant of a
(semi-simple) Lie group can be built from a \textit{non-symplectic}
representation, this never happens for the $U$-duality groups of $\mathcal{N}%
=2$ symmetric and $\mathcal{N}>2$, $d=4$ theories. Thus, for all such
supergravities having a (unique) $U$-invariant \textit{quartic} in BH
charges, the relevant representation of the $U$-duality group is \textit{%
symplectic} (\textit{irreducible} to both $U$-duality and relevant
symplectic group).}, containing the singlet $\mathbf{1}_{a}$ in the tensor
product $\left( \mathbf{2},\mathbf{6}\right) \times \left( \mathbf{2},%
\mathbf{6}\right) $ \cite{Slansky} (and thus yielding a vanishing \textit{%
quadratic} $U$-invariant); consequently, it is \textit{irreducible} with
respect to both $G_{\mathcal{N}=4,pure}$ and $Sp\left( 12,\mathbb{R}\right) $%
). $\mathcal{I}_{4}$ is \textit{quartic} in BH charges (see Eq. (7.4), and
the related discussion, of \cite{FHM}) \cite{N=4-pure-BH-entropy}:
\begin{equation}
\mathcal{I}_{4}=4\left[ p^{2}q^{2}-\left( p\cdot q\right) ^{2}\right] .
\label{Wed-aft-8}
\end{equation}
In terms of the \textit{dressed} charges, \textit{i.e.} of the central
charge matrix $Z_{AB}$, $\mathcal{I}_{4}$ is formally given by the very same
Eqs. (\ref{cocco-1})-(\ref{Tr-A3}). $\mathcal{I}_{4}$ is the \textit{unique}
(moduli-independent) independent $G_{\mathcal{N}=4,pure}$-invariant
combination of (moduli-dependent) $H_{\mathcal{N}=4,pure}$-invariant
quantities (see \textit{e.g.} the $n=0$ limit of the discussion in \cite
{ADF-Duality-d=4} and \cite{ADFT}, and Refs. therein).

It is worth pointing out that only when $\Lambda =1,2$ (corresponding to the
truncation $\left( U\left( 1\right) \right) ^{6}\rightarrow \left( U\left(
1\right) \right) ^{2}$ of the gauge group) $\mathcal{I}_{4}$ is a \textit{%
perfect square}, thus reproducing the \textit{quadratic} invariant $\mathcal{%
I}_{2}$ of the ($1$-modulus, $n=1$ element of the) \textit{minimally coupled}
$\mathcal{N}=2$, $d=4$ sequence, given by Eq. (\ref{I2-N=2-quadr.-BH-charges}%
) (see \textit{e.g.} \cite{BFGM1} and \cite{FHM}, and Refs. therein).

\subsection{\label{N=4-d=4-pure-BPS}Black Hole Parameters for $\frac{1}{4}$%
-BPS Flow}

By using the \textit{Maurer-Cartan Eqs.} of $\mathcal{N}=4$, $d=4$ \textit{%
pure} supergravity (see \textit{e.g.} \cite{ADF-Duality-d=4,ADF1,ADF2}), one
gets \cite{ADOT-1}
\begin{eqnarray}
\partial _{s}\mathcal{Z}_{1} &=&\partial _{s}\mathcal{W}_{BPS}=  \notag \\
&=&\frac{P_{,s}}{\sqrt{2}}\sqrt{\frac{1}{2}Z_{AB}\overline{Z}^{AB}-\sqrt{%
Z_{AB}\overline{Z}^{BC}Z_{CD}\overline{Z}^{DA}-\frac{1}{4}\left( Z_{AB}%
\overline{Z}^{AB}\right) ^{2}}}=P_{,s}\mathcal{Z}_{2},
\label{CERN-al-1-N=4-pure}
\end{eqnarray}
where $P\equiv P_{,s}ds$ is the holomorphic Vielbein of $\mathcal{M}_{%
\mathcal{N}=5}$. Here, $\nabla $ denotes the $U\left( 1\right) $-K\"{a}hler
and $H_{\mathcal{N}=4,pure}$-covariant differential operator.

Thus, by using the explicit expressions of $\mathcal{W}_{BPS}^{2}$ given by
Eq. (\ref{cocco-2}), using the \textit{Maurer-Cartan Eqs.} of $\mathcal{N}=5$%
, $d=4$ supergravity (see \textit{e.g.} \cite{ADF-Duality-d=4,ADF1,ADF2}),
and exploiting the \textit{first order (fake supergravity) formalism}
discussed in Sect. \ref{First-Order}, one respectively obtains the following
expressions of the \textit{(square) ADM mass},\textit{\ axion-dilaton charge}
and \textit{(square) effective horizon radius} for the $\frac{1}{4}$-BPS
attractor flow:
\begin{eqnarray}
r_{H,BPS}^{2}\left( s_{\infty },\overline{s}_{\infty },p,q\right)
&=&M_{ADM,BPS}^{2}\left( s_{\infty },\overline{s}_{\infty },p,q\right) =%
\mathcal{W}_{BPS}^{2}\left( s_{\infty },\overline{s}_{\infty },p,q\right) =
\notag \\
&&  \notag \\
&=&\frac{1}{2}\lim_{\tau \rightarrow 0^{-}}\left[ \frac{1}{2}Z_{AB}\overline{%
Z}^{AB}+\sqrt{Z_{AB}\overline{Z}^{BC}Z_{CD}\overline{Z}^{DA}-\frac{1}{4}%
\left( Z_{AB}\overline{Z}^{AB}\right) ^{2}}\right] =  \notag \\
&&  \notag \\
&=&\left. \mathcal{Z}_{1}^{2}\right| _{\infty }=\frac{i}{2\left( s_{\infty }-%
\overline{s}_{\infty }\right) }\cdot  \notag \\
&&  \notag \\
&&\cdot \left[
\begin{array}{l}
\left( s_{\infty }p_{AB}-q_{AB}\right) \left( \overline{s}_{\infty
}p^{AB}-q^{AB}\right) + \\
\\
+\sqrt{%
\begin{array}{l}
4\left( s_{\infty }p_{AB}-q_{AB}\right) \left( \overline{s}_{\infty
}p^{BC}-q^{BC}\right) \cdot \\
\cdot \left( s_{\infty }p_{CD}-q_{CD}\right) \left( \overline{s}_{\infty
}p^{DA}-q^{DA}\right) + \\
\\
-\left[ \left( s_{\infty }p_{AB}-q_{AB}\right) \left( \overline{s}_{\infty
}p^{AB}-q^{AB}\right) \right] ^{2}
\end{array}
}
\end{array}
\right] ;  \label{cocco-6}
\end{eqnarray}
\begin{eqnarray}
\Sigma _{s,BPS}\left( s_{\infty },\overline{s}_{\infty },p,q\right) &\equiv
&2\lim_{\tau \rightarrow 0^{-}}\left( \partial _{s}\mathcal{W}_{BPS}\right)
\left( s\left( \tau \right) ,\overline{s}\left( \tau \right) ,p,q\right) =
\notag \\
&&  \notag \\
&=&\sqrt{2}\left[ P_{,s}\sqrt{\frac{1}{2}Z_{AB}\overline{Z}^{AB}-\sqrt{Z_{AB}%
\overline{Z}^{BC}Z_{CD}\overline{Z}^{DA}-\frac{1}{4}\left( Z_{AB}\overline{Z}%
^{AB}\right) ^{2}}}\right] _{\infty }=  \notag \\
&&  \notag \\
&=&2\left( P_{,s}\mathcal{Z}_{2}\right) _{\infty };  \label{cocco-5}
\end{eqnarray}
\begin{equation}
\begin{array}{l}
R_{H,BPS}^{2}=\mathcal{W}_{BPS}^{2}\left( z_{\infty },\overline{z}_{\infty
},p,q\right) + \\
\\
-4G^{i\overline{j}}\left( z_{\infty },\overline{z}_{\infty }\right) \left(
\partial _{i}\mathcal{W}_{BPS}\right) \left( z_{\infty },\overline{z}%
_{\infty },p,q\right) \left( \overline{\partial }_{\overline{j}}\mathcal{W}%
_{BPS}\right) \left( z_{\infty },\overline{z}_{\infty },p,q\right) = \\
\\
=\sqrt{Z_{AB}\overline{Z}^{BC}Z_{CD}\overline{Z}^{DA}-\frac{1}{4}\left(
Z_{AB}\overline{Z}^{AB}\right) ^{2}}= \\
\\
=\sqrt{2\mathcal{Z}_{1}^{4}+2\mathcal{Z}_{2}^{4}-\left( \mathcal{Z}_{1}^{2}+%
\mathcal{Z}_{2}^{2}\right) ^{2}}=\mathcal{Z}_{1}^{2}-\mathcal{Z}_{2}^{2}= \\
\\
=\sqrt{Tr\left( A^{2}\right) -\frac{1}{4}\left( Tr\left( A\right) \right)
^{2}}=\sqrt{\mathcal{I}_{4}\left( p,q\right) }>0,
\end{array}
\label{cocco-4}
\end{equation}
where, with suitable changes, the matrix $A_{A}^{~~B}$ and related
quantities are defined by Eqs. (\ref{Tr-A2}) and (\ref{Tr-A3}).

Eq. (\ref{cocco-4}) proves Eq. (\ref{CERN-Thu-1}) for the $\frac{1}{4}$-BPS
attractor flow of $\mathcal{N}=4$, $d=4$ \textit{pure} supergravity. Such a
result was obtained by using Eq. (\ref{CERN-al-1-N=4-pure}) and computing
that
\begin{equation}
\begin{array}{l}
4G^{s\overline{s}}\left( \partial _{s}\mathcal{W}_{BPS}\right) \overline{%
\partial }_{\overline{s}}\mathcal{W}_{BPS}=4G^{s\overline{s}}\left( \partial
_{s}\mathcal{Z}_{1}\right) \overline{\partial }_{\overline{s}}\mathcal{Z}%
_{1}= \\
\\
=2G^{s\overline{s}}P_{,s}\overline{P}_{,\overline{s}}\left[ \frac{1}{2}Z_{AB}%
\overline{Z}^{AB}-\sqrt{Z_{AB}\overline{Z}^{BC}Z_{CD}\overline{Z}^{DA}-\frac{%
1}{4}\left( Z_{AB}\overline{Z}^{AB}\right) ^{2}}\right] = \\
\\
=\frac{1}{2}\left[ \frac{1}{2}Z_{AB}\overline{Z}^{AB}-\sqrt{Z_{AB}\overline{Z%
}^{BC}Z_{CD}\overline{Z}^{DA}-\frac{1}{4}\left( Z_{AB}\overline{Z}%
^{AB}\right) ^{2}}\right] =\mathcal{Z}_{2}^{2},
\end{array}
\end{equation}
where the relation
\begin{equation}
4G^{s\overline{s}}P_{,s}\overline{P}_{,\overline{s}}=1
\end{equation}
was used.

From its very definition, by using Eqs. (\ref{cocco-2}) or (\ref{cocco-6}),
the \textit{axion-dilaton charge} $\Sigma _{s,BPS}$ can be explicitly
computed as follows:
\begin{eqnarray}
\Sigma _{s,BPS} &=&\frac{1}{M_{ADM,BPS}}\left( \partial _{s}\mathcal{W}%
_{BPS}^{2}\right) _{\infty }=\frac{1}{M_{ADM,BPS}}\cdot  \notag \\
&&  \notag \\
&&  \notag \\
&&\cdot \left\{
\begin{array}{l}
-\frac{1}{s_{\infty }-\overline{s}_{\infty }}M_{ADM,BPS}^{2}++\frac{i}{%
2\left( s_{\infty }-\overline{s}_{\infty }\right) }\cdot \\
\\
\cdot \left[
\begin{array}{l}
p_{AB}\left( \overline{s}_{\infty }p^{AB}-q^{AB}\right) + \\
\\
+\frac{1}{2}\left[
\begin{array}{l}
4\left( s_{\infty }p_{AB}-q_{AB}\right) \left( \overline{s}_{\infty
}p^{BC}-q^{BC}\right) \cdot \\
\cdot \left( s_{\infty }p_{CD}-q_{CD}\right) \left( \overline{s}_{\infty
}p^{DA}-q^{DA}\right) + \\
\\
-\left[ \left( s_{\infty }p_{AB}-q_{AB}\right) \left( \overline{s}_{\infty
}p^{AB}-q^{AB}\right) \right] ^{2}
\end{array}
\right] ^{-1/2}\cdot \\
\\
\cdot \left[
\begin{array}{l}
4p_{AB}\left( \overline{s}_{\infty }p^{BC}-q^{BC}\right) \left( s_{\infty
}p_{CD}-q_{CD}\right) \left( \overline{s}_{\infty }p^{DA}-q^{DA}\right) + \\
\\
+4\left( s_{\infty }p_{AB}-q_{AB}\right) \left( \overline{s}_{\infty
}p^{BC}-q^{BC}\right) p_{CD}\left( \overline{s}_{\infty
}p^{DA}-q^{DA}\right) + \\
\\
-2\left( s_{\infty }p_{AB}-q_{AB}\right) \left( \overline{s}_{\infty
}p^{AB}-q^{AB}\right) p_{CD}\left( \overline{s}_{\infty }p^{CD}-q^{CD}\right)
\end{array}
\right]
\end{array}
\right]
\end{array}
\right\} .  \notag \\
&&
\end{eqnarray}

Furthermore, from the definition (\ref{cocco-3}) and Eq. (\ref{cocco-5}), it
follows that
\begin{equation}
\Sigma _{s,BPS}\equiv 2\lim_{\tau \rightarrow 0^{-}}\left( \partial _{s}%
\mathcal{W}_{BPS}\right) =\Sigma _{a,BPS}+\frac{i}{2}e^{2\varphi _{\infty
}}\Sigma _{\varphi ,BPS},
\end{equation}
where
\begin{eqnarray}
\Sigma _{a,BPS} &\equiv &2\lim_{\tau \rightarrow 0^{-}}\left( \partial _{a}%
\mathcal{W}_{BPS}\right) =Re\left( \Sigma _{s,BPS}\right) ; \\
&&  \notag \\
\Sigma _{\varphi ,BPS} &\equiv &2\lim_{\tau \rightarrow 0^{-}}\left(
\partial _{\varphi }\mathcal{W}_{BPS}\right) =2e^{-2\varphi _{\infty
}}Im\left( \Sigma _{s,BPS}\right) =-i\left( s_{\infty }-\overline{s}_{\infty
}\right) Im\left( \Sigma _{s,BPS}\right)  \notag \\
&&
\end{eqnarray}
respectively are the \textit{axionic} and \textit{dilatonic} \textit{charges}
pertaining to the $\frac{1}{4}$-BPS attractor flow.

The considerations made at the end of Subsect. \ref
{N=2-d=4-quadratic-BPS-Attractor-Flow} hold also for the considered
attractor flow.

It is worth noticing out that Eq. (\ref{cocco-4}) is consistent, because, as
pointed out above, the $\frac{1}{4}$-BPS-supporting BH charge configurations
in the considered theory is defined by the \textit{quartic} constraints $%
\mathcal{I}_{4}\left( p,q\right) >0$.

Furthermore, Eq. (\ref{cocco-6}) yields that the $\frac{1}{4}$-BPS attractor
flow of $\mathcal{N}=4$, $d=4$ \textit{pure} supergravity does \textit{not }%
saturate the \textit{marginal stability bound} (see \cite{Marginal-Refs} and
\cite{GLS1}; see also the discussion at the end of Subsect. \ref
{N=2-d=4-quadratic-non-BPS-Z=0-Attractor-Flow}).\setcounter{equation}0

\section{\label{N=4,5,d=4}Peculiarity of \textit{Pure} $\mathcal{N}=4$ and $%
\mathcal{N}=5$ Supergravity}

By exploiting the \textit{first order (fake supergravity) formalism}
discussed in Sect. \ref{First-Order}, the expression of the \textit{squared
effective horizon radius (in the extremal case} $c=0$\textit{)} $R_{H}^{2}$
given by Eq. (\ref{CERN-night-3}) has been shown to hold for the following $%
d=4$ supergravity theories:

\begin{itemize}
\item  \textit{minimally coupled} $\mathcal{N}=2$ theory, whose scalar
manifold is given by the sequence $\frac{SU\left( 1,n\right) }{SU\left(
n\right) \times U(1)}$ (\cite{Luciani}, also named \textit{multi-dilaton
system} in \cite{FHM}; see Subsects. \ref
{N=2-d=4-quadratic-BPS-Attractor-Flow} and \ref
{N=2-d=4-quadratic-non-BPS-Z=0-Attractor-Flow});

\item  $\mathcal{N}=3$ (\cite{N=3-Ref}, see Subsects. \ref{N=3,d=4-BPS} and
\ref{N=3-d=4-non-BPS});

\item  $\mathcal{N}=5$ (\cite{N=5-Ref}, see Subsect. \ref{N=5,d=4,BPS});

\item  $\mathcal{N}=4$ \textit{pure} (\cite{CSF}, see Subsect. \ref
{N=4-d=4-pure-BPS}).
\end{itemize}

Such theories differ by a \textit{crucial} fact: whereas the $U$-invariant
of \textit{minimally coupled} $\mathcal{N}=2$ and $\mathcal{N}=3$
supergravity is \textit{quadratic}, the $U$-invariant of $\mathcal{N}=5$ and
\textit{pure }$\mathcal{N}=4$ theories is \textit{quartic} in BH charges.

Thus, among all $d=4$ supergravities with $U$-invariant \textit{quartic} in
BH charges, the cases $\mathcal{N}=5$ and \textit{pure }$\mathcal{N}=4$ turn
out to be peculiar ones.

Such a peculiarity can be traced back to the form of their \textit{Attractor
Eqs.}, which are \textit{structurally identical} to the ones of the \textit{%
minimally coupled} $\mathcal{N}=2$ and $\mathcal{N}=3$ cases (see \textit{%
e.g.} the treatments in \cite{ADFT} and \cite{ADOT-1}), and actually also to
the very structure of $\mathcal{W}_{BPS}^{2}$, as given by Eqs. (\ref{W-BPS}%
) and (\ref{cocco-2}), respectively.

This is ultimately due to a remarkable property, expressed by the last two
lines of Eqs. (\ref{euro3!}) and (\ref{cocco-4}): the (\textit{unique})
invariant $\mathcal{I}_{4}\left( p,q\right) $ of $G_{\mathcal{N}=5}$ and $G_{%
\mathcal{N}=4,pure}$, which is \textit{quartic} in the electric and magnetic
BH charges $\left( p,q\right) $, is a \textit{perfect square} of a quadratic
expression when written in terms of the \textit{moduli-dependent} \textit{%
skew-eigenvalues }$\mathcal{Z}_{1}$ and $\mathcal{Z}_{2}$:
\begin{eqnarray}
\mathcal{I}_{4}\left( p,q\right) &\equiv &Z_{AB}\overline{Z}^{BC}Z_{CD}%
\overline{Z}^{DA}-\frac{1}{4}\left( Z_{AB}\overline{Z}^{AB}\right)
^{2}=Tr\left( A^{2}\right) -\frac{1}{4}\left( Tr\left( A\right) \right)
^{2}=\left( \mathcal{Z}_{1}^{2}-\mathcal{Z}_{2}^{2}\right) ^{2}.  \notag \\
&&  \label{ostia-2}
\end{eqnarray}
Such a result, which is true in the \textit{whole} scalar manifolds $%
\mathcal{M}_{\mathcal{N}=5}$ and $\mathcal{M}_{\mathcal{N}=4,pure}$, does
\textit{not} generally hold for all other $\mathcal{N}>2$, $d=4$
supergravities with (unique) \textit{quartic} $U$-invariant, \textit{i.e.}
for $\mathcal{N}=4$ \textit{matter-coupled }and $\mathcal{N}=6,8$ theories,
as well as for $\mathcal{N}=2$ supergravity whose scalar manifold does
\textit{not} belong to the aforementioned sequence of complex Grassmannians $%
\frac{SU\left( 1,n\right) }{SU\left( n\right) \times U(1)}$.

This allows one to state that the relation (in the \textit{extremal case} $%
c=0$)\textit{\ }between the \textit{square effective horizon radius } $%
R_{H}^{2}$ and the \textit{square BH event horizon radius} $r_{H}^{2}$ for
the \textit{non-degenerate} attractor flows of such supergravities, \textit{%
if any}, is \textit{structurally different} from the one given by Eq. (\ref
{CERN-night-3}). Of course, in such theories one can still construct the
quantity $r_{H}^{2}\left( z_{\infty },\overline{z}_{\infty },p,q\right) -G_{i%
\overline{j}}\Sigma ^{i}\overline{\Sigma }^{\overline{j}}$ (eventually
within a real parametrization of the scalar fields), but, also in the
\textit{extremal} case, it will be \textit{moduli-dependent}, thus \textit{%
not} determining $R_{H}^{2}\left( p,q\right) $. \setcounter{equation}0

\section{\label{Relations}$\mathcal{N}\geqslant 2$ Supergravities with the
same Bosonic Sector\newline
and \textit{``Dualities''}}

In the present Section we consider $\mathcal{N}\geqslant 2$, $d=4$
supergravities\footnote{%
The relation between $\mathcal{N}=1$ and $\mathcal{N}=2$, $d=4$
supergravities and their attractor solutions is discussed in \cite{ADFT-2}.}
sharing the same bosonic sector, and thus with the same number of fermion
fields, but with \textit{different supersymmetric completions}.\medskip

\textbf{I)}

\begin{itemize}
\item  $\mathcal{N}=2$ (\textit{matter-coupled}) \textit{magic} supergravity
based on the degree $3$ complex Jordan algebra $J_{3}^{\mathbb{H}}$;

\item  $\mathcal{N}=6$ supergravity.
\end{itemize}

The scalar manifold of both such theories (upliftable to $d=5$) is $\frac{%
SO^{\ast }\left( 12\right) }{SU\left( 6\right) \times U\left( 1\right) }$ (%
\textit{rank-}$3$ homogeneous symmetric special K\"{a}hler space). In both
theories the $16$ vector field strengths and their duals, as well as their
asymptotical fluxes, sit in the left-handed spinor repr. $\mathbf{32}$ of
the $U$-duality group $SO^{\ast }\left( 12\right) $, which is \textit{%
symplectic}, containing the symmetric singlet $\mathbf{1}_{a}$ in the tensor
product $\mathbf{32}\times \mathbf{32}$, and thus irreducible with respect
to both $SO^{\ast }\left( 12\right) $ and $Sp\left( 32,\mathbb{R}\right) $.
For a discussion of the spin/field content, see \textit{e.g.} \cite
{ADF1,ADF2}.

The correspondences among the various classes of \textit{non-degenerate}
extremal BH attractors of such two theories have been studied in \cite{BFGM1}
(see \textit{e.g.} Table 9 therein).\medskip

\textbf{II)}
\begin{table}[t]
\begin{center}
\begin{tabular}{|c|c|c|}
\hline
$
\begin{array}{c}
\\
\text{\textit{Orbit}} \\
~
\end{array}
$ & $
\begin{array}{c}
\\
\mathcal{N}=2~\text{\textit{minimally~coupled}},~n_{V}=3 \\
~
\end{array}
$ & $
\begin{array}{c}
\\
\mathcal{N}=3,~m=1 \\
~
\end{array}
$ \\ \hline\hline
$
\begin{array}{c}
\\
\frac{SU(1,3)}{SU(3)} \\
~
\end{array}
$ & $
\begin{array}{c}
\mathcal{O}_{\frac{1}{2}-BPS}, \\
\ no~mod.~space, \\
\mathcal{I}_{2,\mathcal{N}=2}>0
\end{array}
$ & $
\begin{array}{c}
\mathcal{O}_{non-BPS,Z_{AB}=0}, \\
no~mod.~space, \\
\mathcal{I}_{2,\mathcal{N}=3}<0
\end{array}
$ \\ \hline
$
\begin{array}{c}
\\
\frac{SU(1,3)}{SU(1,2)} \\
~
\end{array}
$ & $
\begin{array}{c}
\mathcal{O}_{non-BPS,Z=0}, \\
mod.~space=\frac{SU(1,2)}{SU(2)\times U\left( 1\right) }, \\
\mathcal{I}_{2,\mathcal{N}=2}<0
\end{array}
$ & $
\begin{array}{c}
\mathcal{O}_{\frac{1}{3}-BPS}, \\
mod.~space=\frac{SU(1,2)}{SU(2)\times U\left( 1\right) }, \\
\mathcal{I}_{2,\mathcal{N}=3}>0
\end{array}
$ \\ \hline
\end{tabular}
\end{center}
\caption{$\mathcal{N}$\textbf{-dependent BPS-interpretations of the classes
of \textit{non-degenerate} orbits of the symmetric special K\"{a}hler
manifold $\frac{SU(1,3)}{SU(3)\times U(1)}$}}
\end{table}

\begin{itemize}
\item  $\mathcal{N}=2$ supergravity \textit{minimally coupled} to $n=n_{V}=3$
Abelian vector multiplets;

\item  $\mathcal{N}=3$ supergravity coupled to $m=1$ matter (Abelian vector)
multiplet.
\end{itemize}

All such theories (\textit{matter-coupled}, with \textit{quadratic} $U$%
-invariant, and not upliftable to $d=5$) share the same scalar manifold,
namely the \textit{rank-}$1$ symmetric special K\"{a}hler space $\frac{%
SU\left( 1,3\right) }{SU\left( 3\right) \times U\left( 1\right) }$.
Furthermore, in both such theories the $4$ vector field strengths and their
duals, as well as their asymptotical fluxes, sit in the \textit{fundamental}
$\mathbf{4}$ repr. of the $U$-duality group $SU\left( 3,1\right) $ (\textit{%
not} irreducible with respect to $SU\left( 3,1\right) $ itself, but only
with respect to $Sp\left( 8,\mathbb{R}\right) $).

By (local) supersymmetry, the number of fermion fields is the same in the
three theories, namely there are $8$ bosons and $8$ fermions, but with
\textit{different} relevant spin/field contents:
\begin{eqnarray}
&&
\begin{array}{lll}
\mathcal{N}=2~\text{\textit{minimally~coupled}},~n_{V}=3: &  & \left[
1\left( 2\right) ,2\left( \frac{3}{2}\right) ,1\left( 1\right) \right] ,3%
\left[ 1\left( 1\right) ,2\left( \frac{1}{2}\right) ,1_{\mathcal{C}}\left(
0\right) \right] ; \\
&  &  \\
\mathcal{N}=3,~m=1: &  & \left[ 1\left( 2\right) ,3\left( \frac{3}{2}\right)
,3\left( 1\right) ,1\left( \frac{1}{2}\right) \right] ,1\left[ 1\left(
1\right) ,4\left( \frac{1}{2}\right) ,3_{\mathcal{C}}\left( 0\right) \right]
.
\end{array}
\notag \\
&&
\end{eqnarray}
From this it follows that one can switch between such two theories by
transforming $1$\ \textit{gravitino} in $1$\ \textit{gaugino}, and \textit{%
vice versa}.

The relation among the various classes of \textit{non-degenerate} extremal
BH attractors of such three theories \cite{ADFT,bellucci2} is given in Table
1.

When switching between $\mathcal{N}=2$ and $\mathcal{N}=3$, the flip in sign
of the \textit{quadratic} $U$-invariant $\mathcal{I}_{2}=q^{2}+p^{2}$ can be
understood by recalling that $q^{2}\equiv \eta ^{\Lambda \Sigma }q_{\Lambda
}q_{\Sigma }$ and $p^{2}\equiv \eta _{\Lambda \Sigma }p^{\Lambda }p^{\Sigma
} $, with $\eta ^{\Lambda \Sigma }=\eta _{\Lambda \Sigma }=diag\left(
1,-1,-1,-1\right) $ in the case $\mathcal{N}=2$, and $\eta ^{\Lambda \Sigma
}=\eta _{\Lambda \Sigma }=diag\left( 1,1,1,-1\right) $ in the case $\mathcal{%
N}=3$ (recall Eq. (\ref{ven2})). It is here worth pointing out once again
that the positive signature pertains to the \textit{graviphoton charges},
while the negative signature corresponds to the charges given by the
asymptotical fluxes of the vector field strengths from the matter multiplets
(see also the discussion in Sect.\textbf{\ }\ref{Invariance-Props}). As
yielded by Table 1, the supersymmetry-preserving features of the attractor
solutions depend on the sign of $\mathcal{I}_{2}$.\medskip

\textbf{III)}
\begin{table}[t]
\begin{center}
\begin{tabular}{|c||c|c|}
\hline
$
\begin{array}{c}
\\
\text{\textit{Orbit}} \\
~
\end{array}
$ & $\mathcal{N}=2,~n_{V}=7$ & $\mathcal{N}=4,~n=2$ \\ \hline\hline
$
\begin{array}{c}
\\
\frac{SU(1,1)\times SO\left( 2,6\right) }{SO\left( 2\right) \times SO\left(
6\right) } \\
~
\end{array}
$ & $
\begin{array}{c}
\mathcal{O}_{\frac{1}{2}-BPS}, \\
\ no~mod.~space, \\
\mathcal{I}_{4,\mathcal{N}=2}>0
\end{array}
$ & $
\begin{array}{c}
\mathcal{O}_{non-BPS,Z_{AB}=0}, \\
\ no~mod.~space, \\
\mathcal{I}_{4,\mathcal{N}=4}>0
\end{array}
$ \\ \hline
$
\begin{array}{c}
\\
\frac{SU(1,1)\times SO\left( 2,6\right) }{SO\left( 2\right) \times SO\left(
2,4\right) } \\
~
\end{array}
$ & $
\begin{array}{c}
\mathcal{O}_{non-BPS,Z=0}, \\
\ mod.~space=\frac{SO\left( 2,4\right) }{SO\left( 2\right) \times SO\left(
4\right) } \\
\mathcal{I}_{4,\mathcal{N}=2}>0
\end{array}
$ & $
\begin{array}{c}
\mathcal{O}_{\frac{1}{4}-BPS}, \\
\ mod.~space=\frac{SO\left( 2,4\right) }{SO\left( 2\right) \times SO\left(
4\right) } \\
\mathcal{I}_{4,\mathcal{N}=4}>0
\end{array}
$ \\ \hline
$
\begin{array}{c}
\\
\frac{SU(1,1)\times SO\left( 2,6\right) }{SO\left( 1,1\right) \times
SO\left( 1,5\right) } \\
~
\end{array}
$ & $
\begin{array}{c}
\mathcal{O}_{non-BPS,Z\neq 0}, \\
\ mod.~space=SO\left( 1,1\right) \times \frac{SO\left( 1,5\right) }{SO\left(
5\right) } \\
\mathcal{I}_{4,\mathcal{N}=2}<0
\end{array}
$ & $
\begin{array}{c}
\mathcal{O}_{non-BPS,Z_{AB}\neq 0}, \\
\ mod.~space=SO\left( 1,1\right) \times \frac{SO\left( 1,5\right) }{SO\left(
5\right) } \\
\mathcal{I}_{4,\mathcal{N}=4}<0
\end{array}
$ \\ \hline
\end{tabular}
\end{center}
\caption{$\mathcal{N}$\textbf{-dependent BPS-interpretations of the classes
of \textit{non-degenerate} orbits of the reducible symmetric special
K\"{a}hler manifold $\frac{SU\left( 1,1\right) }{U\left( 1\right) }\times
\frac{SO(2,6)}{SO(2)\times SO(6)}$.} The structure of the\ \textit{%
``duality''} is analogous to the one pertaining to the manifold $\frac{%
SO^{\ast }\left( 12\right) }{SU\left( 6\right) \times U\left( 1\right) }$
(see point \textbf{I} above, as well as Table 9 of \protect\cite{BFGM1})}
\end{table}

\begin{itemize}
\item  $\mathcal{N}=2$\textbf{\ }supergravity coupled to $n_{V}=n+1=7$\
Abelian vector multiplets, with scalar manifold $\frac{SU\left( 1,1\right) }{%
U\left( 1\right) }\times \frac{SO\left( 2,6\right) }{SO\left( 2\right)
\times SO\left( 6\right) }$ (shortly named \textit{``cubic''}, $n_{V}=7$ in
Eq. (\ref{Tue-2}));

\item  $\mathcal{N}=4$\ supergravity coupled to $n_{m}=2$\ matter (Abelian
vector) multiplets.
\end{itemize}

The scalar manifold of both such theories (upliftable to $d=5$) is $\frac{%
SU\left( 1,1\right) }{U\left( 1\right) }\times \frac{SO\left( 2,6\right) }{%
SO\left( 2\right) \times SO\left( 6\right) }$ (homogeneous symmetric
reducible special K\"{a}hler, with \textit{rank }$3$). In both theories the $%
8$ vector field strengths and their duals, as well as their asymptotical
fluxes, sit in the (spinor/doublet)-vector (\textit{bi-fundamental}) repr. $%
\left( \mathbf{2},\mathbf{8}\right) $ of the $U$-duality group $SU\left(
1,1\right) \times SO(2,6)$, which is \textit{symplectic}, containing the
antisymmetric singlet $\mathbf{1}_{a}$ in the tensor product $\left( \mathbf{%
2},\mathbf{8}\right) \times \left( \mathbf{2},\mathbf{8}\right) $, and thus
irreducible with respect to both $SU\left( 1,1\right) \times SO(2,6)$ and $%
Sp\left( 16,\mathbb{R}\right) $.

Notice that, due to the isomorphism $\frak{so}\left( 6,2\right) \sim \frak{so%
}^{\ast }\left( 8\right) $ (see \textit{e.g.} \cite{Helgason}), the \textit{%
``dual'' }supersymmetric interpretation of the scalar manifold $\frac{%
SU\left( 1,1\right) }{U\left( 1\right) }\times \frac{SO\left( 2,6\right) }{%
SO\left( 2\right) \times SO\left( 6\right) }$ can be considered,
disregarding the \textit{axion-dilaton} sector $\frac{SU\left( 1,1\right) }{%
U\left( 1\right) }$, as a \textit{``subduality''} of the \textit{``duality''
}discussed in point \textbf{I}\textit{.}

By (local) supersymmetry, the number of fermion fields is the same in the
three theories, namely there are $16$ bosons and $16$ fermions, but with
\textit{different} relevant spin/field contents:
\begin{eqnarray}
&&
\begin{array}{lll}
\mathcal{N}=2~\text{\textit{``cubic''}},~n_{V}=7: &  & \left[ 1\left(
2\right) ,2\left( \frac{3}{2}\right) ,1\left( 1\right) \right] ,7\left[
1\left( 1\right) ,2\left( \frac{1}{2}\right) ,1_{\mathcal{C}}\left( 0\right)
\right] ; \\
&  &  \\
\mathcal{N}=4,n_{m}=2: &  & \left[ 1\left( 2\right) ,4\left( \frac{3}{2}%
\right) ,6\left( 1\right) ,4\left( \frac{1}{2}\right) ,1_{\mathcal{C}}\left(
0\right) \right] ,2\left[ 1\left( 1\right) ,4\left( \frac{1}{2}\right) ,3_{%
\mathcal{C}}\left( 0\right) \right] .
\end{array}
\notag \\
&&  \label{Tue-2}
\end{eqnarray}
From this it follows that one can switch between such two theories by
transforming $2$\ \textit{gravitinos} in $2$\ \textit{gauginos}, and \textit{%
vice versa}.

The correspondences among the various classes of \textit{non-degenerate}
extremal BH attractors of such two theories have been studied in \cite
{BFGM1,ferrara4,ADFT,bellucci2}, and it is given in Table 2.

As yielded by the comparison of Table 9 of \cite{BFGM1} and Table 2, such a
\textit{``duality''} is pretty similar to the \textit{``duality''} between $%
\mathcal{N}=2$ $J_{3}^{\mathbb{H}}$ and $\mathcal{N}=6$ considered at point
\textbf{I}, also because the sign of the \textit{quartic} $U$-invariant is
unchanged by the \textit{``duality''} relation (this is also consistent with
the \textit{``subduality''} relation mentioned above). In this sense, it
differs from the \textit{``duality''} between $\mathcal{N}=2$ \textit{%
minimally coupled}, $n_{V}=3$ and $\mathcal{N}=3$, $m=1$ considered at point
\textbf{II}, because in both such theories the $U$-invariant is \textit{%
quadratic}, and its sign is flipped by the \textit{``duality''} relation
(see Table 1). \bigskip

Points \textbf{I}-\textbf{III }present evidences against the conventional
wisdom that \textit{interacting} bosonic field theories have a unique
supersymmetric extension. The sharing of the same bosonic backgrounds with
different supersymmetric completions implies the \textit{``dual''}
interpretation with respect to the supersymmetry-preserving properties of
\textit{non-degenerate} extremal BH attractor solutions (see respectively
Table 9 of \cite{BFGM1}, Table 1 and Table 2).

\section{\label{Conclusion}Conclusion}

In the present investigation we have considered the class of $\mathcal{N}%
\geqslant 2$, $d=4$ \textit{ungauged} supergravity theories which do not
have a counterpart in $d=5$ space-time dimensions. For such theories, the
extremal BH parameters, namely, the \textit{ADM mass} $M_{ADM}$, the \textit{%
scalar charges} $\Sigma ^{a}$ and the \textit{effective horizon radius} $%
R_{H}$, pertaining to \textit{non-degenerate} attractor flows have a simple
formulation in the \textit{first order (fake supergravity) formalism}.

All such theories share the property that an \textit{effective radial
variable} $R$ can be defined, such that the \textit{effective BH (horizon)
area} $A$ is simply given by the surface of a sphere of radius $R_{H}\left(
p,q\right) $, where $R_{H}\left( p,q\right) $ is the \textit{%
moduli-independent effective horizon radius} of the extremal BH.

For $\mathcal{N}=2$ this holds for supergravity \textit{minimally coupled}
to Abelian vector multiplets, but it does \textit{not} hold for more general
matter couplings, such as symmetric spaces coming from degree three Jordan
Algebras \cite{GST} with cubic holomorphic prepotential. In this respect,
\textit{minimally coupled} $\mathcal{N}=2$ supergravity can be considered as
a \textit{multi-dilaton system} in that it generalizes the \textit{%
Maxwell-Einstein-axion-dilaton system}, studied in Refs. \cite{K3} and \cite
{Garfinkle} (see also the recent treatment in \cite{FHM}). \textit{%
Matter-coupled} supergravity with $\mathcal{N}=3$ shares similar properties.
Both such theories exhibit two class of \textit{non-degenerate} attractors
(BPS and non-BPS), and a Bekenstein-Hawking classical BH entropy \textit{%
quadratic} in the electric and magnetic BH charges. Furthermore, non-BPS ($%
Z=0$) $\mathcal{N}=2$ attractors and $\mathcal{N}=3$ (both ($\frac{1}{3}$%
-)BPS and non-BPS ($Z_{AB}=0$)) attractors yield a related moduli space of
solutions.

\textit{Pure} $\mathcal{N}=4$ and $\mathcal{N}=5$ supergravities have also
the same formula yielding to define $R_{H}\left( p,q\right) $, in spite of
the fact that the classical BH entropy is not \textit{quadratic}, but rather
the square root of a \textit{quartic} expression, in terms of the BH
charges. Such theories have only BPS attractors, and for $\mathcal{N}=5$ a
residual moduli space of solutions also exists.

It would be interesting to extend the notion of \textit{effective radius}
and of \textit{fake supergravity formalism} to other $d=4$ theories, such as
$\mathcal{N}=2$ \textit{not minimally coupled to} Abelian vector multiplets,
\textit{matter-coupled} $\mathcal{N}=4$, $\mathcal{N}=6$ and $\mathcal{N}=8$%
. In these cases, different formul\ae\ should occur to determine the BH
parameters, such as \textit{ADM mass} $M_{ADM}$ and \textit{scalar charges} $%
\Sigma ^{a}$, in terms of the geometry of the underlying (asymptotical)
scalar manifold.

Finally, one may wonder about a stringy realization of the theories
discussed in the present paper, and their extremal BH states. At the string
tree level, the massless spectrum of $\mathcal{N}=3$ and $\mathcal{N}=5$, $%
d=4$ supergravity can be obtained via asymmetric orbifolds of Type II
superstrings \cite{FK}--\nocite{FF,DH}\cite{KK}, or by orientifolds \cite
{Frey}. Furthermore, it should be remarked that in string theory the \textit{%
Attractor Mechanism} is essentially a non-perturbative phenomenon, either
because it fixes the dilaton, or because it involves non-perturbative string
states made out of $D$-brane bound states \cite{Bianchi}.

It is worth pointing out once again that the present analysis only covers
\textit{non-degenerate} extremal BH attractors, determining \textit{``large''%
} BH horizon geometries with \textit{non-vanishing} classical effective BH
horizon area in the field theory limit. For \textit{degenerate} extremal
BHs, having \textit{``small''} BH horizon geometries with \textit{vanishing}
classical effective BH horizon area, a departure from the Einsteinian
approximation, including higher curvature terms in the gravity sector, is
\textit{at least} required \cite{Wald}. Such corrections in supersymmetric
theories of gravity have been considered in \cite{Sen-old1} and \cite{MSW}--
\nocite{dW-Cardoso}\cite{Kraus-Larsen}.

\section*{Acknowledgments}

The authors gratefully acknowledge L. Andrianopoli, M. Bianchi, A. Ceresole,
R. D'Auria, F. Morales, M. Trigiante and A. Yeranyan for enlightening and
fruitful discussions.

We also would like to thank Mrs. Helen Webster for careful proofreading.

A. M. would like to thank the Department of Physics, Theory Unit Group at
CERN, where part of this work was done, for kind hospitality and stimulating
environment.

The work of S.F. has been supported in part by the European Community Human
Potential Programme under contract MRTN-CT-2004-503369 ``\textit{%
Constituents, Fundamental Forces and Symmetries of the Universe''} in
association with LNF-INFN and in part by DOE grant DE-FG03-91ER40662, Task C.

The work of A.M. has been supported by Museo Storico della Fisica e Centro
Studi e Ricerche ``\textit{Enrico Fermi''}, Rome, Italy, in association with
INFN-LNF. \newpage \appendix \setcounter{equation}0

\section{\label{N=4,d=4,n}Appendix I\newline
A Counterexample : $\mathcal{N}=4$ \textit{Matter-Coupled} Supergravity}

As an example of $d=4$ supergravity in which $\mathcal{I}_{4}$ is \textit{not%
} a perfect square of an expression quadratic in all the relevant
geometrical quantities (such as the \textit{skew-eigenvalues} of the \textit{%
central charge matrix} and the \textit{matter charges}), let us consider the
$\mathcal{N}=4$, $d=4$ \textit{matter-coupled} theory \cite{Bergshoeff,De
Roo}.

The real \textit{scalar manifold} is
\begin{equation}
\mathcal{M}_{\mathcal{N}=4}=\frac{G_{\mathcal{N}=4}}{H_{\mathcal{N}=4}}=%
\frac{SU\left( 1,1\right) }{U\left( 1\right) }\times \frac{SO\left(
6,n\right) }{SO\left( 6\right) \times SO\left( n\right) },~dim_{\mathbb{R}%
}=6n+2.  \label{Sun-3}
\end{equation}

The $6+n$ vector field strengths and their duals, as well as their
asymptotical fluxes, sit in the \textit{bi-fundamental} irrepr. $\left(
\mathbf{2},\left( \mathbf{6+n}\right) \right) $ of the $U$-duality group $G_{%
\mathcal{N}=4}=SU\left( 1,1\right) \times SO\left( 6,n\right) $.

$Z_{AB}=Z_{\left[ AB\right] }$, $A,B=1,2,3,4=\mathcal{N}$ is the \textit{%
central charge matrix}. As in the \textit{pure }supergravity treated in
Sect. \ref{N=4,d=4-pure} (obtained by setting $n=0$ in Eq. (\ref{Sun-3})),
by means of a suitable transformation of the $\mathcal{R}$-symmetry $U\left(
1\right) \times SO\left( 6\right) \sim U\left( 1\right) \times SU\left(
4\right) $, $Z_{AB}$ can be \textit{skew-diagonalized} by putting it in the
\textit{normal form} (see e.g. \cite{ADOT-1} and Refs. therein):
\begin{equation}
Z_{AB}=\left(
\begin{array}{cc}
\mathcal{Z}_{1}\epsilon &  \\
& \mathcal{Z}_{2}\epsilon
\end{array}
\right) ,
\end{equation}
where $\mathcal{Z}_{1},\mathcal{Z}_{2}\in \mathbb{R}_{0}^{+}$ are the $%
\mathcal{N}=4$ (moduli-dependent) \textit{skew-eigenvalues}, which can be
ordered as $\mathcal{Z}_{1}\geqslant \mathcal{Z}_{2}$ without any loss of
generality (up to renamings; see \textit{e.g.} \cite{ADOT-1}), and can be
formally expressed by the very same Eqs. (\ref{cuba-3})-(\ref{cuba-2}),
where $I_{1}$ and $I_{2}$ are still the two (moduli-dependent) $H_{\mathcal{N%
}=4,pure}$(and also $H_{\mathcal{N}=4}$)-invariants.

In this case, similarly to $\mathcal{N}=3$ supergravity, also the \textit{%
matter charges }$Z_{I}$ ($I=1,...,n$) enter the game, $n\in \mathbb{N}$
denoting the number of matter multiplets coupled to the gravity multiplet.
By a suitable rotation of $SO\left( n\right) $, the vector $Z_{I}$ can be
reduced in such a way that only one (strictly positive) \textit{real} and
one \textit{complex} matter charge are non-vanishing \cite{ADOT-1}:
\begin{equation}
Z_{I}=\left( Z_{1}=\rho _{1},Z_{2}=\rho _{2}e^{i\theta },Z_{\widehat{I}%
}=0\right) ,~~\rho _{1},\rho _{2}\in \mathbb{R}_{0}^{+},~\theta \in \left[
0,2\pi \right) ,~\widehat{I}=3,...,n.
\end{equation}
Thus, one can introduce the (moduli-dependent, unique) $SO\left( n\right) $%
(and also $H_{\mathcal{N}=4}$)-invariants
\begin{eqnarray}
I_{3} &\equiv &Z_{I}\overline{Z}^{I}=\rho _{1}^{2}+\rho _{2}^{2}; \\
&&  \notag \\
I_{4} &\equiv &Re\left( Z_{I}Z^{I}\right) =\rho _{1}^{2}+\rho
_{2}^{2}cos\left( 2\theta \right) .
\end{eqnarray}
Now, there are only three (moduli-dependent) $SO\left( 6,n\right) $%
-invariants, reading as follows \cite{ADF-Duality-d=4}:
\begin{eqnarray}
\mathbb{I}_{1} &\equiv &I_{1}-I_{3}=\mathcal{Z}_{1}^{2}+\mathcal{Z}%
_{2}^{2}-\rho _{1}^{2}-\rho _{2}^{2}; \\
&&  \notag \\
\mathbb{I}_{2} &\equiv &\frac{1}{4}\epsilon ^{ABCD}Z_{AB}Z_{CD}-\overline{Z}%
_{I}\overline{Z}^{I}=2\mathcal{Z}_{1}\mathcal{Z}_{2}-\rho _{1}^{2}-\rho
_{2}^{2}e^{-2i\theta }; \\
&&  \notag \\
\mathbb{I}_{3} &\equiv &\overline{\mathbb{I}_{2}}=2\mathcal{Z}_{1}\mathcal{Z}%
_{2}-\rho _{1}^{2}-\rho _{2}^{2}e^{2i\theta }.
\end{eqnarray}
The \textit{quartic} $G_{\mathcal{N}=4}$-invariant $\mathcal{I}_{4}$ of $%
\mathcal{N}=4$, $d=4$ supergravity is the following unique
(moduli-independent) $G_{\mathcal{N}=4}$-invariant combination of $\mathbb{I}%
_{1}$, $\mathbb{I}_{2}$ and $\mathbb{I}_{3}$ \cite{ADF-Duality-d=4}:
\begin{equation}
\begin{array}{l}
\mathcal{I}_{4}\equiv \mathbb{I}_{1}^{2}-\mathbb{I}_{2}\mathbb{I}_{3}=%
\mathbb{I}_{1}^{2}-\left| \mathbb{I}_{2}\right| ^{2}= \\
\\
=\left( \mathcal{Z}_{1}^{2}-\mathcal{Z}_{2}^{2}\right) ^{2}+\left( Z_{I}%
\overline{Z}^{I}\right) ^{2}-2\left( \mathcal{Z}_{1}^{2}+\mathcal{Z}%
_{2}^{2}\right) Z_{I}\overline{Z}^{I}+ \\
\\
+2\mathcal{Z}_{1}\mathcal{Z}_{2}\left( Z_{I}Z^{I}+\overline{Z}_{I}\overline{Z%
}^{I}\right) -\left| Z_{I}Z^{I}\right| ^{2}= \\
\\
=\left( \mathcal{Z}_{1}^{2}-\mathcal{Z}_{2}^{2}\right) ^{2}+\left( \rho
_{1}^{2}+\rho _{2}^{2}\right) ^{2}-2\left( \mathcal{Z}_{1}^{2}+\mathcal{Z}%
_{2}^{2}\right) \left( \rho _{1}^{2}+\rho _{2}^{2}\right) + \\
\\
+4\mathcal{Z}_{1}\mathcal{Z}_{2}\left[ \rho _{1}^{2}+\rho _{2}^{2}cos\left(
2\theta \right) \right] -\left[ \rho _{1}^{4}+\rho _{1}^{4}+2\rho
_{1}^{2}\rho _{2}^{2}cos\left( 2\theta \right) \right] .
\end{array}
\label{ostia-3}
\end{equation}
On the other hand, in terms of the BH charges $\left( q,p\right) $, $%
\mathcal{I}_{4}$ reads as follows (recall Eq. (\ref{Wed-aft-8})):
\begin{equation}
\mathcal{I}_{4}=4\left[ p^{2}q^{2}-\left( p\cdot q\right) ^{2}\right] ,
\label{Thu-11}
\end{equation}
where $p^{2}\equiv p^{\Lambda }p^{\Sigma }\eta _{\Lambda \Sigma }$, $%
q^{2}\equiv q_{\Lambda }q_{\Sigma }\eta ^{\Lambda \Sigma }$, with $\Lambda $
ranging $1,...,n+6$, and the scalar product $\cdot $ is defined by $\eta
_{\Lambda \Sigma }=\eta ^{\Lambda \Sigma }$, the Lorentzian metric with
signature $\left( n,6\right) $ (see \cite{FHM} and Refs. therein).

Looking at Eq. (\ref{ostia-3}), it is easy to realize that $\mathcal{I}_{4}$
is a non-trivial \textit{perfect square} of a function of degree $2$ of $%
\mathcal{Z}_{1}$, $\mathcal{Z}_{2}$, $\rho _{1}$, $\rho _{2}$ and $\theta $
\textit{only} in the \textit{pure }supergravity theory (obtained by setting $%
n=0$), \textit{i.e.} only in the case $\rho _{1}=\rho _{2}=0$. In such a
limit, Eq. (\ref{ostia-3}) consistently reduces to Eq. (\ref{ostia-2}).

As an example, we can workout the case $n=1$ (which uplifts to \textit{pure}
$\mathcal{N}=4$, $d=5$ supergravity). In this case, only a (strictly
positive) real \textit{matter charge} $Z_{1}=\rho _{1}\in \mathbb{R}_{0}^{+}$
is present, and the \textit{quartic} invariant $\mathcal{I}_{4}$ acquires
the following form:
\begin{eqnarray}
\mathcal{I}_{4} &=&\left( \mathcal{Z}_{1}^{2}-\mathcal{Z}_{2}^{2}\right)
^{2}-2\left( \mathcal{Z}_{1}^{2}+\mathcal{Z}_{2}^{2}\right) Z_{1}^{2}+4%
\mathcal{Z}_{1}\mathcal{Z}_{2}Z_{1}^{2}=  \notag \\
&=&\left( \mathcal{Z}_{1}-\mathcal{Z}_{2}\right) ^{2}\left[ \left( \mathcal{Z%
}_{1}+\mathcal{Z}_{2}\right) ^{2}-2Z_{1}^{2}\right] =  \notag \\
&=&\left( \mathcal{Z}_{1}-\mathcal{Z}_{2}\right) ^{2}\left( \mathcal{Z}_{1}+%
\mathcal{Z}_{2}+\sqrt{2}\rho _{1}\right) \left( \mathcal{Z}_{1}+\mathcal{Z}%
_{2}-\sqrt{2}\rho _{1}\right) ,
\end{eqnarray}
which is \textit{not} a non-trivial \textit{perfect square} of $\mathcal{Z}%
_{1}$, $\mathcal{Z}_{2}$ and $\rho _{1}$. \setcounter{equation}0
\pagebreak


\begin{thebibliography}{999}
\bibitem{ferrara1}  S. Ferrara, R. Kallosh, A. Strominger: $\mathcal{N}%
\mathit{=2}$\textit{\ extremal black holes}, Phys. Rev. \textbf{D52}, 5412
(1995).

\bibitem{ferrara2}  S. Ferrara, R. Kallosh: \textit{Supersymmetry and
attractors}, Phys. Rev. \textbf{D54}, 1514 (1996); S. Ferrara, R. Kallosh:
\textit{Universality of supersymmetric attractors}, Phys. Rev. \textbf{D54},
1525 (1996).

\bibitem{strominger2}  A. Strominger: \textit{Macroscopic entropy of }$%
\mathcal{N}\mathit{=2}$\textit{\ extremal black holes}, Phys. Lett. \textbf{%
B383}, 39 (1996).

\bibitem{FGK}  S.~Ferrara, G. W. Gibbons and R. Kallosh, \textit{Black Holes
and Critical Points in Moduli Space}, Nucl. Phys. \textbf{B500}, 75 (1997),
\texttt{hep-th/9702103}.

\bibitem{maldacena}  For reviews on black holes in superstring theory see
\textit{e.g.}: J. M. Maldacena: \textit{Black-Holes in String Theory},
\texttt{hep-th/9607235}; A. W. Peet: \textit{TASI lectures on black holes in
string theory}, \texttt{arXiv:hep-th/0008241}; B. Pioline: \textit{Lectures
on black holes, topological strings and quantum attractors}, Class. Quant.
Grav. \textbf{23}, S981 (2006); A. Dabholkar: \textit{Black hole entropy and
attractors}, Class. Quant. Grav. \textbf{23}, S957 (2006).

\bibitem{schwarz1}  For recent reviews see: J. H. Schwarz: \textit{Lectures
on superstring and }$\mathit{M}$\textit{-theory dualities}, Nucl. Phys.
Proc. Suppl. \textbf{B55}, 1 (1997); M. J. Duff: \textit{M-theory (the
theory formerly known as strings)}, Int. J. Mod. Phys. \textbf{A11}, 5623
(1996); A. Sen: \textit{Unification of string dualities}, Nucl. Phys. Proc.
Suppl. \textbf{58}, 5 (1997).

\bibitem{schwarz2}  J. H. Schwarz, A. Sen: \textit{Duality symmetries of }$%
\mathit{4D}$\textit{\ heterotic strings}, Phys. Lett. \textbf{B312}, 105
(1993); J. H. Schwarz, A. Sen: \textit{Duality Symmetrical Actions}, Nucl.
Phys. \textbf{B411}, 35 (1994).

\bibitem{gasperini}  M. Gasperini, J. Maharana, G. Veneziano: \textit{From
trivial to non-trivial conformal string backgrounds via }$\mathit{O(d,d)}$%
\textit{\ transformations}, Phys. Lett. \textbf{B272}, 277 (1991); J.
Maharana, J. H. Schwarz: \textit{Noncompact Symmetries in String Theory},
Nucl. Phys. \textbf{B390}, 3 (1993).

\bibitem{witten}  E. Witten: \textit{String Theory Dynamics in Various
Dimensions}, Nucl. Phys. \textbf{B443}, 85 (1995).

\bibitem{schwarz3}  J. H. Schwarz: $\mathit{M}$\textit{-theory extensions of
}$\mathit{T}$\textit{\ duality}, \texttt{arXiv:hep-th/9601077}; C. Vafa:
\textit{Evidence for }$\mathit{F}$\textit{-theory}, Nucl. Phys. \textbf{B469}%
, 403 (1996).

\bibitem{Sen-old1}  A. Sen, \textit{Black Hole Entropy Function and the
Attractor Mechanism in Higher Derivative Gravity}, JHEP \textbf{09}, 038
(2005), \texttt{hep-th/0506177}.

\bibitem{GIJT}  K. Goldstein, N. Iizuka, R. P. Jena and S. P. Trivedi,
\textit{Non-Supersymmetric Attractors}, Phys. Rev. \textbf{D72}, 124021
(2005), \texttt{hep-th/0507096}.

\bibitem{Sen-old2}  A. Sen, \textit{Entropy Function for Heterotic Black
Holes}, JHEP \textbf{03}, 008 (2006), \texttt{hep-th/0508042}.

\bibitem{K1}  R. Kallosh, \textit{New Attractors}, JHEP \textbf{0512}, 022
(2005), \texttt{hep-th/0510024}.

\bibitem{TT}  P. K. Tripathy and S. P. Trivedi, \textit{Non-Supersymmetric
Attractors in String Theory, }JHEP \textbf{0603}, 022 (2006),\textit{\ }%
\texttt{hep-th/0511117}.

\bibitem{G}  A. Giryavets, \textit{New Attractors and Area Codes}, JHEP
\textbf{0603}, 020 (2006), \texttt{hep-th/0511215}.

\bibitem{GJMT}  K. Goldstein, R. P. Jena, G. Mandal and S. P. Trivedi,
\textit{A }$\mathit{C}$\textit{-Function for Non-Supersymmetric Attractors, }%
JHEP \textbf{0602}, 053 (2006)\textit{, }\texttt{hep-th/0512138}.

\bibitem{Ebra1}  M. Alishahiha and H. Ebrahim, \textit{Non-supersymmetric
attractors and entropy function}, JHEP \textbf{0603}, 003 (2006), \texttt{%
hep-th/0601016}.

\bibitem{K2}  R. Kallosh, N. Sivanandam and M. Soroush, \textit{The Non-BPS
Black Hole Attractor Equation}, JHEP \textbf{0603}, 060 (2006), \texttt{%
hep-th/0602005}.

\bibitem{Ira1}  B. Chandrasekhar, S. Parvizi, A. Tavanfar and H. Yavartanoo,
\textit{Non-supersymmetric attractors in }$\mathcal{R}^{2}$\textit{\
gravities}, JHEP \textbf{0608}, 004 (2006), \texttt{hep-th/0602022}.

\bibitem{Tom}  J. P. Hsu, A. Maloney and A. Tomasiello, \textit{Black Hole
Attractors and Pure Spinors}, JHEP \textbf{0609}, 048 (2006), \texttt{%
hep-th/0602142}.

\bibitem{BFM}  S. Bellucci, S. Ferrara and A. Marrani, \textit{On some
properties of the Attractor Equations}, Phys. Lett. \textbf{B635}, 172
(2006), \texttt{hep-th/0602161}.

\bibitem{FKlast}  S. Ferrara and R. Kallosh, \textit{On }$\mathcal{N}\mathit{%
=8}$\textit{\ attractors}, Phys. Rev. D \textbf{73}, 125005 (2006), \texttt{%
hep-th/0603247}.

\bibitem{Ebra2}  M. Alishahiha and H. Ebrahim, \textit{New attractor,
Entropy Function and Black Hole Partition Function}, JHEP \textbf{0611}, 017
(2006), \texttt{hep-th/0605279}.

\bibitem{BFGM1}  S. Bellucci, S. Ferrara, M. G\"{u}naydin and A. Marrani,
\textit{Charge Orbits of Symmetric Special Geometries and Attractors}, Int.
J. Mod. Phys. \textbf{A21}, 5043 (2006), \texttt{hep-th/0606209}.

\bibitem{rotating-attr}  D. Astefanesei, K. Goldstein, R. P. Jena, A. Sen
and S. P. Trivedi, \textit{Rotating Attractors}, JHEP \textbf{0610}, 058
(2006), \texttt{hep-th/0606244}.

\bibitem{K2-bis}  R. Kallosh, N. Sivanandam and M. Soroush, \textit{Exact
Attractive non-BPS STU Black Holes}, Phys. Rev. \textbf{D74}, 065008 (2006),
\texttt{hep-th/0606263}.

\bibitem{Misra1}  P. Kaura and A. Misra, \textit{On the Existence of
Non-Supersymmetric Black Hole Attractors for Two-Parameter Calabi-Yau's and
Attractor Equations}, Fortsch. Phys. \textbf{54}, 1109 (2006), \texttt{%
hep-th/0607132}.

\bibitem{Lust2}  G. L. Cardoso, V. Grass, D. L\"{u}st and J. Perz, \textit{%
Extremal non-BPS Black Holes and Entropy Extremization}, JHEP \textbf{0609},
078 (2006), \texttt{hep-th/0607202}.

\bibitem{Morales}  J.~F.~Morales and H.~Samtleben, \textit{Entropy function
and attractors for AdS black holes}, JHEP \textbf{0610}, 074 (2006), \texttt{%
hep-th/0608044}.

\bibitem{Astefa}  D. Astefanesei, K. Goldstein and S. Mahapatra, \textit{%
Moduli and (un)attractor black hole thermodynamics}, \texttt{hep-th/0611140}.

\bibitem{CdWMa}  G.L. Cardoso, B. de Wit and S. Mahapatra, \textit{Black
hole entropy functions and attractor equations}, JHEP \textbf{0703}, 085
(2007) \texttt{hep-th/0612225}.

\bibitem{DFT07-1}  R. D'Auria, S. Ferrara and M. Trigiante, C\textit{ritical
points of the Black-Hole potential for homogeneous special geometries}, JHEP
\textbf{0703}, 097 (2007), \texttt{hep-th/0701090}.

\bibitem{BFM-SIGRAV06}  S. Bellucci, S. Ferrara and A. Marrani, {\textit{%
Attractor Horizon Geometries of Extremal Black Holes}}, contribution to the
Proceedings of the XVII SIGRAV Conference,4--7 September 2006, Turin, Italy,
\texttt{hep-th/0702019}.

\bibitem{Cer-Dal-1}  A. Ceresole and G. Dall'Agata, \textit{Flow Equations
for Non-BPS Extremal Black Holes}, JHEP \textbf{0703}, 110 (2007), \texttt{%
hep-th/0702088}.

\bibitem{ADFT-2}  L. Andrianopoli, R. D'Auria, S. Ferrara and M. Trigiante,
\textit{Black Hole Attractors in }$\mathcal{N}\mathit{=1}$\textit{\
Supergravity}, JHEP \textbf{0707}, 019 (2007), \texttt{hep-th/0703178}.

\bibitem{Saraikin-Vafa-1}  K. Saraikin and C. Vafa, \textit{%
Non-supersymmetric Black Holes and Topological Strings}, Class. Quant. Grav.
\textbf{25}, 095007 (2008), \texttt{hep-th/0703214}.

\bibitem{Ferrara-Marrani-1}  S. Ferrara and A. Marrani, $\mathcal{N}\mathit{%
=8}$\textit{\ non-BPS Attractors, Fixed Scalars and Magic Supergravities},
Nucl. Phys. \textbf{B788}, 63 (2008), \texttt{arXiV:0705.3866}.

\bibitem{TT2}  S. Nampuri, P. K. Tripathy and S. P. Trivedi, \textit{On The
Stability of Non-Supersymmetric Attractors in String Theory}, JHEP \textbf{%
0708}, 054 (2007), \texttt{arXiV:0705.4554}.

\bibitem{ADOT-1}  L. Andrianopoli, R. D'Auria, E. Orazi, M. Trigiante,
\textit{First Order Description of Black Holes in Moduli Space}, JHEP
\textbf{0711}, 032 (2007), \texttt{arXiV:0706.0712}.

\bibitem{ferrara4}  S. Ferrara and A. Marrani, \textit{On the Moduli Space
of non-BPS Attractors for }$\mathcal{N}\mathit{=2}$\textit{\ Symmetric
Manifolds}, Phys. Lett. \textbf{B652}, 111 (2007) , \texttt{arXiV:0706.1667}.

\bibitem{Astefanesei}  D.Astefanesei and H. Yavartanoo, \textit{Stationary
black holes and attractor mechanism}, Nucl. Phys. \textbf{B794}, 13 (2008),
\texttt{arXiv:0706.1847}.

\bibitem{CCDOP}  G. L. Cardoso, A. Ceresole, G. Dall'Agata, J. M.
Oberreuter, J. Perz, \textit{First-order flow equations for extremal black
holes in very special geometry}, JHEP \textbf{0710}, 063 (2007), \texttt{%
arXiV:0706.3373}.

\bibitem{Misra2}  A. Misra and P. Shukla, \textit{'Area codes', large volume
(non-)perturbative alpha-prime and instanton: Corrected non-supersymmetric
(A)dS minimum, the 'inverse problem' and 'fake superpotentials' for
multiple-singular-loci-two-parameter Calabi-Yau's}, \texttt{arXiV:0707.0105}.

\bibitem{Ceresole}  A. Ceresole, S. Ferrara and A. Marrani, \textit{4d/5d
Correspondence for the Black Hole Potential and its Critical Points}, Class.
Quant. Grav. \textbf{24}, 5651 (2007), \texttt{arXiV:0707.0964}.

\bibitem{Anber}  M. M. Anber and D. Kastor, \textit{The Attractor mechanism
in Gauss-Bonnet gravity}, JHEP \textbf{0710}, 084 (2007), \texttt{%
arXiv:0707.1464}.

\bibitem{Myung1}  Y. S. Myung, Y.-W. Kim and Y.-J. Park, \textit{New
attractor mechanism for spherically symmetric extremal black holes}, Phys.
Rev. \textbf{D76}, 104045 (2007), \texttt{arXiv:0707.1933}.

\bibitem{BMOS-1}  S. Bellucci, A. Marrani, E. Orazi and A. Shcherbakov,
\textit{Attractors with Vanishing Central Charge}, Phys. Lett. \textbf{B655}%
, 185 (2007), \texttt{arXiV:0707.2730}.

\bibitem{Hotta}  K. Hotta and T. Kubota, \textit{Exact Solutions and the
Attractor Mechanism in Non-BPS Black Holes}, Prog. Theor. Phys. \textbf{118N5%
}, 969 (2007), \texttt{arXiv:0707.4554}.

\bibitem{Gao}  X. Gao, \textit{Non-supersymmetric Attractors in Born-Infeld
Black Holes with a Cosmological Constant}, JHEP \textbf{0711}, 006 (2007),
\texttt{arXiv:0708.1226}.

\bibitem{Sen-review}  A. Sen, \textit{Black Hole Entropy Function,
Attractors and Precision Counting of Microstates}, \texttt{arXiv:0708.1270}.

\bibitem{Belhaj1}  A. Belhaj, L.B. Drissi, E.H. Saidi and A. Segui, $%
\mathcal{N}\mathit{=2}$\textit{\ Supersymmetric Black Attractors in Six and
Seven Dimensions}, Nucl. Phys. \textbf{B796}, 521 (2008), \texttt{%
arXiv:0709.0398}.

\bibitem{Gaiotto1}  D. Gaiotto, W. Li and M. Padi, \textit{%
Non-Supersymmetric Attractor Flow in Symmetric Spaces}, JHEP \textbf{0712},
093 (2007), \texttt{arXiv:0710.1638}.

\bibitem{GLS1}  E. G. Gimon, F. Larsen and J. Simon, \textit{Black Holes in
Supergravity: the non-BPS Branch}, JHEP \textbf{0801}, 040 (2008), \texttt{%
arXiv:0710.4967}.

\bibitem{ANYY1}  D. Astefanesei, H. Nastase, H. Yavartanoo and S. Yun,
\textit{Moduli flow and non-supersymmetric AdS attractors}, JHEP \textbf{0804%
}, 074 (2008), \texttt{arXiv:0711.0036}.

\bibitem{bellucci2}  S. Bellucci, S. Ferrara, R. Kallosh and A. Marrani,
\textit{Extremal Black Hole and Flux Vacua Attractors}, contribution to the
Proceedings of the Winter School on Attractor Mechanism 2006 (SAM2006),
20-24 March 2006, INFN-LNF, Frascati, Italy, \texttt{arXiv:0711.4547}.

\bibitem{Cai-Pang}  R.-G. Cai and D.-W. Pang, \textit{A Note on exact
solutions and attractor mechanism for non-BPS black holes}, JHEP \textbf{0801%
}, 046 (2008), \texttt{arXiv:0712.0217}.

\bibitem{Vaula}  M. Huebscher, P. Meessen, T. Ort\'{i}n and S. Vaul\`{a},
\textit{Supersymmetric }$\mathcal{N}\mathit{=2}$\textit{\
Einstein-Yang-Mills monopoles and covariant attractors}, \texttt{%
arXiv:0712.1530}.

\bibitem{Li}  W. Li: \textit{Non-Supersymmetric Attractors in Symmetric
Coset Spaces}, contribution to the Proceedings of 3rd School on Attractor
Mechanism (SAM 2007), Frascati, Italy, 18-22 Jun 2007, \texttt{%
arXiv:0801.2536}.

\bibitem{Saidi2}  E. H. Saidi, \textit{BPS and non BPS }$\mathit{7D}$\textit{%
\ Black Attractors in }$\mathit{M}$\textit{-Theory on }$\mathit{K3}$,
\texttt{arXiv:0802.0583}.

\bibitem{Saidi3}  E. H. Saidi, \textit{On Black Hole Effective Potential in }%
$\mathit{6D/7D}$\textit{\ }$\mathcal{N}\mathit{=2}$\textit{\ Supergravity},
\texttt{arXiv:0803.0827}.

\bibitem{Saidi4}  E. H. Saidi and A. Segui, \textit{Entropy of Pairs of Dual
Attractors in six and seven Dimensions}, \texttt{arXiv:0803.2945}.

\bibitem{FHM}  S. Ferrara, K. Hayakawa and A. Marrani, \textit{Erice
Lectures on Black Holes and Attractors}, contribution to the Proceedings of
the International School of Subnuclear Physics, 45th course \textit{``Search
for the ``Totally Unexpected'' in the LHC Era''}, 29 August--7 September
2007, Erice, Italy, \texttt{arXiv:0805.2498}.

\bibitem{Unattractor}  D. Astefanesei, N. Banerjee and S. Dutta, \textit{%
(Un)attractor black holes in higher derivative }$\mathit{AdS}$\textit{\
gravity}, \texttt{arXiv:0806.1334}.

\bibitem{Vaula2}  M. Huebscher, P. Meessen, T. Ort\'{i}n and S. Vaul\`{a}, $%
\mathcal{N}\mathit{=2}$\textit{\ Einstein-Yang-Mills's BPS solutions},
\texttt{arXiv:0806.1477}.

\bibitem{Trigiante}  E. Bergshoeff, W. Chemissany, A. Ploegh, M. Trigiante
and T. Van Riet, \textit{Generating Geodesic Flows and Supergravity Solutions%
}, \texttt{arXiv:0806.2310}.

\bibitem{OSV}  H. Ooguri, A. Strominger and C. Vafa: \textit{Black Hole
Attractors and the Topological String}, Phys. Rev. \textbf{D70}, 106007
(2004), \texttt{hep-th/0405146}.

\bibitem{OVV}  H. Ooguri, C. Vafa and E. Verlinde: \textit{Hartle-Hawking
wave-function for flux compactifications: the Entropic Principle}, Lett.
Math. Phys. \textbf{74}, 311 (2005), \texttt{hep-th/0502211}.

\bibitem{ANV}  M. Aganagic, A. Neitzke and C. Vafa: \textit{BPS microstates
and the open topological string wave function}, \texttt{hep-th/0504054}.

\bibitem{GSV}  S. Gukov, K. Saraikin and C. Vafa: \textit{The Entropic
Principle and Asymptotic Freedom}, Phys. Rev. \textbf{D73}, 066010 (2006),
\texttt{hep-th/0509109}.

\bibitem{ADFT}  L. Andrianopoli, R. D'Auria, S. Ferrara and M. Trigiante:
\textit{Extremal Black Holes in Supergravity}, in : \textit{``String Theory
and Fundamental Interactions''},M. Gasperini and J. Maharana eds. (LNP,
Springer, Berlin-Heidelberg, 2007), \texttt{hep-th/0611345}.

\bibitem{hawking2}  S. W. Hawking: \textit{Gravitational Radiation from
Colliding Black Holes}, Phys. Rev. Lett. \textbf{26}, 1344 (1971); J. D.
Bekenstein: \textit{Black Holes and Entropy}, Phys. Rev. \textbf{D7}, 2333
(1973).

\bibitem{CSF}  E. Cremmer, J. Scherk and S. Ferrara, $\mathit{SU(4)}$\textit{%
\ Invariant Supergravity Theory}, Phys. Lett. \textbf{B74}, 61 (1978).

\bibitem{N=5-Ref}  B. de Wit and H. Nicolai, \textit{Extended Supergravity
with Local }$\mathit{SO(5)}$\textit{\ Invariance}, Nucl. Phys. \textbf{B188}%
, 98 (1981).

\bibitem{Luciani}  J. F. Luciani: \textit{Coupling of }$\mathit{O(2)}$%
\textit{\ Supergravity with Several Vector Multiplets}, Nucl. Phys. \textbf{%
B132}, 325 (1978).

\bibitem{N=3-Ref}  L. Castellani, A. Ceresole, S. Ferrara, R. D'Auria, P.
Fr\'{e} and E. Maina, \textit{The Complete }$\mathcal{N}\mathit{=3}$\textit{%
\ Matter Coupled Supergravity}, Nucl. Phys. \textbf{B268}, 317 (1986).

\bibitem{K3}  R. Kallosh, A. D. Linde, T. Ort\'{i}n, A. W. Peet, A. Van
Proeyen: \textit{Supersymmetry as a cosmic censor}, Phys. Rev. \textbf{D46},
5278 (1992), \texttt{hep-th/9205027}; R. Kallosh, T. Ort\'{i}n, A. W. Peet:
\textit{Entropy and action of dilaton black holes}, Phys. Rev. \textbf{D47},
5400 (1993), \texttt{hep-th/9211015}; R. Kallosh, A. W. Peet: \textit{%
Dilaton black holes near the horizon}, Phys. Rev. \textbf{D46}, 5223 (1992),
\texttt{hep-th/9209116}.

\bibitem{Garfinkle}  D. Garfinkle, G. T. Horowitz and A. Strominger: \textit{%
Charged black holes in string theory}, Phys. Rev. \textbf{D43}, 3140 (1991)
[Erratum-ibid. \textbf{D45}, 3888 (1992)].

\bibitem{Fake-Refs}  D.Z. Freedman, C. Nunez, M. Schnabl and K. Skenderis,
\textit{Fake supergravity and domain wall stability}, Phys. Rev. \textbf{D69}%
, 104027 (2004), \texttt{hep-th/0312055}; A. Celi, A. Ceresole, G.
Dall'Agata, A. Van Proeyen and M. Zagermann, \textit{On the fakeness of fake
supergravity}, Phys. Rev. \textbf{D71}, 045009 (2005), \texttt{hep-th/0410126%
}; M. Zagermann, $\mathcal{N}\mathit{=4}$\textit{\ fake supergravity}, Phys.
Rev. \textbf{D71}, 125007 (2005), \texttt{hep-th/0412081}; K. Skenderis and
P. K. Townsend, \textit{Hidden supersymmetry of domain walls and cosmologies}%
, Phys. Rev. Lett. \textbf{96}, 191301 (2006), \texttt{hep-th/0602260}; D.
Bazeia, C.B. Gomes, L. Losano and R. Menezes, \textit{First-order formalism
and dark energy}, Phys. Lett. \textbf{B633}, 415 (2006), \texttt{%
astro-ph/0512197}; K. Skenderis and P. K. Townsend, \textit{%
Pseudo-Supersymmetry and the Domain-Wall/Cosmology Correspondence}, J. Phys.
\textbf{A40}, 6733 (2007), \texttt{hep-th/0610253}.

\bibitem{BPS}  G. W. Gibbons and C. M. Hull, \textit{A Bogomol'ny Bound for
General Relativity and Solitons in }$\mathit{N=2}$\textit{\ Supergravity},
Phys. Lett. \textbf{B109}, 190 (1982).

\bibitem{GST}  M. G\"{u}naydin, G. Sierra and P. K. Townsend, \textit{The
Geometry of }$\mathcal{N}\mathit{=2}$\textit{\ Maxwell-Einstein Supergravity
and Jordan Algebras}, Nucl. Phys. \textbf{B242}, 244 (1984).

\bibitem{arnowitt}  R. Arnowitt, S. Deser and C. W. Misner: \textit{%
Canonical Variables for General Relativity}, Phys. Rev. \textbf{117}, 1595
(1960).

\bibitem{Marginal-Refs}  J. Rahmfeld, \textit{Extremal black holes as bound
states}, Phys. Lett. \textbf{B372}, 198 (1996), \texttt{hep-th/9512089}; M.
J. Duff and J. Rahmfeld, \textit{Bound states of black holes and other }$%
\mathit{p}$\textit{-branes}, Nucl. Phys. \textbf{B481}, 332 (1996), \texttt{%
hep-th/9605085}; F. Denef, \textit{Supergravity flows and }$\mathit{D}$%
\textit{-brane stability}, JHEP \textbf{0008}, 050 (2000), \texttt{%
hep-th/0005049}; D. Gaiotto, A. Simons, A. Strominger and X. Yin, $\mathit{D0%
}$\textit{-branes in black hole attractors}, \texttt{hep-th/0412179}; A.
Ritz, M. A. Shifman, A. I. Vainshtein, M. B. Voloshin, \textit{Marginal
stability and the metamorphosis of BPS states}, Phys. Rev. \textbf{D63},
065018 (2001), \texttt{hep-th/0006028}; P. S. Aspinwall, A. Maloney, A.
Simons, \textit{Black hole entropy, marginal stability and mirror symmetry},
JHEP \textbf{0707}, 034 (2007), \texttt{hep-th/0610033}; A. Sen, \textit{%
Walls of Marginal Stability and Dyon Spectrum in }$\mathcal{N}\mathit{=4}$%
\textit{\ Supersymmetric String Theories}, JHEP \textbf{0705}, 039 (2007),
\texttt{hep-th/0702141}; F. Denef, D. Gaiotto, A. Strominger, D. Van den
Bleeken and X. Yin, \textit{Black Hole Deconstruction}, \texttt{%
hep-th/0703252}; A. Sen, $\mathcal{N}\mathit{=8}$\textit{\ Dyon Partition
Function and Walls of Marginal Stability}, \texttt{arXiv:0803.1014}.

\bibitem{4}  A. Ceresole, R. D'Auria and S. Ferrara: \textit{The Symplectic
Structure of }$\mathcal{N}\mathit{=2}$ \textit{Supergravity and Its Central
Extension}, Talk given at ICTP Trieste Conference on Physical and
Mathematical Implications of Mirror Symmetry in String Theory, Trieste,
Italy, 5-9 June 1995, Nucl. Phys. Proc. Suppl. \textbf{46} (1996), \texttt{%
hep-th/9509160}.

\bibitem{stu-Yeranyan}  S. Bellucci, S. Ferrara, A. Marrani and A. Yeranyan,
\textit{The }$\mathit{STU}$\textit{\ Unveiled}, to appear.

\bibitem{Helgason}  S. Helgason, \textit{Differential Geometry, Lie Groups
and Symmetric Spaces} (Academic Press, New York, 1978).

\bibitem{Gilmore}  R. Gilmore, \textit{Lie Groups, Lie Algebras, and Some of
Their Applications} (Dover Publications, 2006).

\bibitem{Gaillard-Zumino-1}  M. K. Gaillard and B. Zumino, \textit{Duality
Rotations for Interacting Fields}, Nucl. Phys. \textbf{B193}, 221 (1981).

\bibitem{ADF1}  L. Andrianopoli, R. D'Auria and S. Ferrara, \textit{Central
extension of extended supergravities in diverse dimensions}, Int. J. Mod.
Phys. \textbf{A12}, 3759 (1997), \texttt{hep-th/9608015}.

\bibitem{ADF2}  L. Andrianopoli, R. D'Auria and S. Ferrara, $\mathit{U}$%
\textit{\ duality and central charges in various dimensions revisited}, Int.
J. Mod. Phys. \textbf{A13}, 431 (1998), \texttt{hep-th/9612105}.

\bibitem{ADF-Duality-d=4}  L. Andrianopoli, R. D'Auria and S. Ferrara, $%
\mathit{U}$\textit{\ invariants, black hole entropy and fixed scalars},
Phys. Lett. \textbf{B403}, 12 (1997), \texttt{hep-th/9703156}.

\bibitem{FSZ}  S. Ferrara, J. Scherk and B. Zumino, \textit{Algebraic
Properties of Extended Supergravity Theories}, Nucl. Phys. \textbf{B121},
393 (1977).

\bibitem{DFL}  R. D'Auria, S. Ferrara and M. A. Lled\'{o}, \textit{On
central charges and Hamiltonians for }$\mathit{0}$\textit{-brane dynamics},
Phys. Rev. \textbf{D60}, 084007 (1999), \texttt{hep-th/9903089}.

\bibitem{DF-Fermion}  R. D'Auria and S. Ferrara, \textit{On fermion masses,
gradient flows and potential in supersymmetric theories}, JHEP \textbf{0105}%
, 034 (2001), \texttt{hep-th/0103153}.

\bibitem{dW-N=8}  B. de Wit, \textit{Properties Of }$\mathit{SO(8)}$\textit{%
\ Extended Supergravity}, Nucl. Phys. \textbf{B158}, 189 (1979).

\bibitem{Slansky}  R. Slansky, \textit{Group Theory for Unified Model
Building}, Phys. Rep. \textbf{79}, 1 (1981).

\bibitem{N=4-pure-BH-entropy}  M. Cvetic nd D. Youm, \textit{Dyonic BPS
saturated black holes of heterotic string on a six torus}, Phys. Rev.
\textbf{D53}, 584 (1996),\texttt{\ hep-th/9507090}; M. J. Duff, J. T. Liu
and J. Rahmfeld, \textit{Four-dimensional string-string-string triality},
Nucl. Phys. \textbf{B459}, 125 (1996), \texttt{hep-th/9508094}; M. Cvetic
and A. A. Tseytlin, \textit{Solitonic strings and BPS saturated dyonic black
holes}, Phys. Rev. \textbf{D53}, 5619 (1996); Erratum-ibid. \textbf{D55},
3907 (1997), \texttt{hep-th/9512031}.

\bibitem{Bergshoeff}  E. Bergshoeff, I. G. Koh and E. Sezgin, \textit{%
Coupling of Yang-Mills to }$\mathcal{N}\mathit{=4}$\textit{, }$\mathit{D=4}$%
\textit{\ Supergravity}, Phys. Lett. \textbf{B155}, 71 (1985).

\bibitem{De Roo}  M. de Roo and P. Wagemans, \textit{Gauge Matter Coupling
In }$\mathcal{N}\mathit{=4}$\textit{\ Supergravity}, Nucl. Phys. \textbf{B262%
}, 644 (1985).

\bibitem{FK}  S. Ferrara and C. Kounnas, \textit{Extended Supersymmetry In
Four-Dimensional Type II Strings}, Nucl. Phys. \textbf{B328}, 406 (1989).

\bibitem{FF}  S. Ferrara and P. Fr\'{e}, \textit{Type II Superstrings On
Twisted Group Manifolds And Their Heterotic Counterparts}, Int. J. Mod.
Phys. \textbf{A5}, 989 (1990).

\bibitem{DH}  A. Dabholkar and J. A. Harvey, \textit{String Islands}, JHEP
\textbf{9902}, 006 (1999), \texttt{hep-th/9809122}.

\bibitem{KK}  C. Kounnas and A. Kumar, \textit{BPS states in }$\mathcal{N}%
\mathit{=3}$\textit{\ Superstrings}, Nucl. Phys. \textbf{B511}, 216 (1998),
\texttt{hep-th/9709061}.

\bibitem{Frey}  A. R. Frey and J. Polchinski, $\mathcal{N}\mathit{=3}$%
\textit{\ Warped Compactifications}, Phys. Rev. \textbf{D65}, 126009 (2002),
\texttt{hep-th/0201029}.

\bibitem{Bianchi}  M. Bianchi, \textit{Bound-states of }$\mathit{D}$\textit{%
-branes in }$\mathit{L}$\textit{-}$\mathit{R}$\textit{\ asymmetric
superstring vacua}, \texttt{arXiv:0805.3276}.

\bibitem{Wald}  R. M. Wald, \textit{Black hole entropy in the Noether charge}%
, Phys. Rev. \textbf{D48}, 3427 (1993), \texttt{gr-qc/9307038}; T. Jacobson,
G. Kang and R. C. Myers, \textit{On Black Hole Entropy}, Phys. Rev. \textbf{%
D49}, 6587 (1994), \texttt{gr-qc/9312023}; V. Iyer and R. M. Wald, \textit{%
Some properties of Noether charge and a proposal for dynamical black hole
entropy}, Phys. Rev. \textbf{D50}, 846 (1994), \texttt{gr-qc/9403028}; T.
Jacobson, G. Kang and R. C. Myers, \textit{Black hole entropy in higher
curvature gravity}, \texttt{gr-qc/9502009}.

\bibitem{MSW}  J. M. Maldacena, A. Strominger and E. Witten, \textit{Black
Hole Entropy in }$\mathit{M}$\textit{\ Theory}, JHEP \textbf{9712}, 002
(1997), \texttt{hep-th/9711053}.

\bibitem{dW-Cardoso}  G. Lopes Cardoso, B. de Wit and T. Mohaupt, \textit{%
Corrections to Macroscopic Supersymmetric Black Hole Entropy}, Phys. Lett.
\textbf{B451}, 309 (1999), \texttt{hep-th/9812082}.

\bibitem{Kraus-Larsen}  P. Kraus and F. Larsen, \textit{Microscopic Black
Hole Entropy in Theories with Higher Derivatives}, JHEP \textbf{0509}, 034
(2005), \texttt{hep-th/0506176}.
\end{thebibliography}
\end{document}